\documentclass[a4paper,10pt]{article}
\usepackage{graphicx}
\usepackage{subcaption}
\usepackage{pstricks}
\usepackage{pstricks-add}
\usepackage{tikz}
\usepackage{verbatim}
\usetikzlibrary{decorations.pathreplacing}
\usetikzlibrary{decorations.markings}

\usepackage{amsmath}
\usepackage{amsfonts}
\usepackage[title]{appendix}
\usepackage[margin=2cm]{geometry}

\newcommand{\beq}{\begin{equation}}

\newcommand{\eeq}{\end{equation}}
\newcommand{\no}{\noindent}

\newcommand{\q}{\mathfrak{q}}

\usepackage{color}

\newcommand{\BX}{\mathcal{X}}
\newcommand{\BW}{\mathcal{W}}
\newcommand{\StTL}[1]{\mathcal{W}_{#1}}
\newcommand{\IrrTL}[1]{\mathcal{X}_{#1}}
\newcommand{\TL}[1]{\mathsf{TL}_{#1}}

\begin{document}

\title{Conformally invariant boundary conditions in the antiferromagnetic Potts model and the $SL(2,\mathbb{R})/U(1)$ sigma model }
\date{}

\maketitle

\begin{center}

\vskip 1cm

{\large Niall F.\ Robertson$^{1}$,  Jesper Lykke Jacobsen$^{1,2,3}$,  and Hubert Saleur$^{1,4}$}

\vspace{1.0cm}
{\sl\small $^1$ Institut de Physique Th\'eorique, Universit\'e Paris Saclay, CEA, CNRS, F-91191 Gif-sur-Yvette,France\\}
{\sl\small $^2$ Laboratoire de Physique de l'Ecole Normale Sup\'erieure, ENS, Universit\'e PSL, CNRS,
Sorbonne Universit\'e, Universit\'e de Paris, Paris, France\\}
{\sl\small $^3$ Sorbonne Universit\'e, \'Ecole Normale Sup\'erieure, CNRS, Laboratoire de Physique (LPENS), 75005 Paris, France\\}
{\sl\small $^4$ Department of Physics and Astronomy,
University of Southern California,
Los Angeles, CA 90089, USA\\}

\end{center}

\vskip 2cm

\begin{abstract}

We initiate a study of the boundary version of the square-lattice $Q$-state Potts antiferromagnet, with $Q \in [0,4]$ real, motivated by the fact that the continuum limit of the corresponding bulk
model is  a non-compact CFT,  closely related with the $SL(2,\mathbb{R})_k/U(1)$ Euclidian black-hole coset model.

While various types of conformal boundary conditions (discrete and continuous branes) have been formally identified for the the $SL(2,\mathbb{R})_k/U(1)$ coset CFT,  we are only able in this work to identify conformal boundary conditions  (CBC) leading to a discrete boundary spectrum.

The CBC we find are of two types. The first is free boundary Potts spins, for which we  confirm an old conjecture for the generating functions of conformal levels,
and show them to be related to characters in a non-linear deformation of the $W_\infty$ algebra.

The second type of CBC - which  corresponds to  restricting the values of the Potts spins  to a subset of size $Q_1$, or its complement of size $Q-Q_1$, at alternating sites along the  boundary - is new, and turns out to be conformal in the antiferromagnetic case only. 
Using algebraic and numerical techniques, we show that the corresponding spectrum generating functions produce {\sl all the characters of discrete representations for the coset CFT}. The normalizability bounds of the associated discrete states in the coset CFT are found to have a simple interpretation in terms of boundary phase transitions in the lattice model. 
In the two-boundary case, with two distinct alt conditions, we obtain similar results, at least in the case when the corresponding boundary condition changing operator also
inserts a number of defect lines.

For $\sqrt{Q} = 2 \cos \frac{\pi}{k}$, with $k \ge 3$ integer, we show also how our boundary conditions can be reformulated in terms of a RSOS height model.
The spectrum generating functions are then identified with string functions of  the compact $SU(2)_{k-2} / U(1)$ parafermion theory  (with symmetry $Z_{k-2}$). The new alt conditions are needed to cover all the string functions. We provide an algebraic proof that the two-boundary alt conditions correctly produce
the fusion rules of string functions. We expose in detail the special case of $Q=3$ and its link with three-colourings of the square lattice and a corresponding
boundary six-vertex model.

Finally, we discuss the case of an odd number of sites (in the loop model) and the relation with wired boundary conditions (in the spin model). In this case
the RSOS restriction produces the disorder operators of the parafermion theory.

\end{abstract}

\section{Introduction}

The  critical antiferromagnetic Potts model on the square lattice enjoys remarkable  properties which are not fully understood, despite years of work on the topic \cite{B-AF,S-AF,JS-AF}. The most remarkable aspect of this model is that it is described, in the continuum limit, by a conformal field theory (CFT) with a continuous spectrum of critical exponents \cite{IJS2008,IJS-AFlett,CanduIkhlef,BKKL}, closely related---after parametrising $\sqrt{Q}=2\cos\gamma$, where $\gamma={\pi\over k}$ and $k\geq 2$---with the $SL(2,\mathbb{R})_k/U(1)$ coset ``Euclidian black-hole'' CFT  \cite{Witten,DVV}. While this kind of relationship between a compact (albeit non-unitary) lattice model and a non-compact  CFT has been extended to other cases---for instance,  two coupled antiferromagnetic Potts models \cite{VJS}, polymers at the theta-point \cite{VJS0,VJS1}, or truncations of the Chalker-Coddington model \cite{CVJS}---it has only been studied so far in the bulk. It is natural to wonder how the relation between models persists in the boundary case, and what this might tell us about  issues ranging, for instance,  from the microscopic interpretation of the non-compact degrees of freedom in the continuum limit of the antiferromagnetic (AF) Potts model to the possible conformally invariant boundary conditions in the Euclidian black hole CFT \cite{RS}. 

This paper is intended as the first in a series devoted to  study this question. We will mostly focus here on the Potts model per se, and discuss in a subsequent work what happens for the underlying staggered six-vertex model---or to the cognate staggered (``alternating'') XXZ spin chain \cite{IJSstag1,IJSstag2}. 

In section \ref{PottsQgen} we review well-known facts about the Potts model, its clusters and loops representations in the bulk, and its known critical lines. In section \ref{BDRPotts} we review aspects of the boundary Potts model, together with known conformal boundary conditions for the ferromagnetic case. In  section \ref{KnownPottsBC} we review  the  known case of free boundary conditions for the AF Potts model---which turns out to be conformal indeed---pointing out in particular the presence  of an  underlying (deformation of) $W_\infty$ symmetry.

Section \ref{NewPottsBC} is the centrepiece of the present paper. There, we identify new conformal boundary conditions specific to the AF Potts model---which we nickname ``alt'' because they involve the {\em alternation} of certain variables---together with the associated partition functions, which turn out to be combinations of discrete characters of the $SL(2,\mathbb{R})/U(1)$ coset model. Many issues are discussed, in particular those concerning normalisability. Section \ref{RSOSkint} delineates the relationship with parafermions \cite{paraf} when $k$  is an integer, and leads to the definition of a new staggered restricted solid-on-solid (RSOS) model \cite{PasquierRSOS} in the $Z_k$ universality class. Section \ref{3statepotts} examines the special case of the two and three-state Potts models and interprets the ``alt'' boundary conditions in terms of these well-studied models. Finally, in section~\ref{oddsites}, we study the RSOS and loop models with an odd number of sites, which we find to give the so-called ``disorder operators'' arising in parafermion theories, which were first studied in \cite{Disorderparaf}.

\bigskip

For the reader's convenience, we here give a list of notations, consistent with our earlier works on related topics: 

\begin{itemize}

\item $\BW^{b/u}_{j}$ ---  standard modules over the blob algebra,
\item  $\BX^{b/u}_{j}$  --- simple modules over the blob algebra,
\item$\StTL{j}$ --- standard modules over $\TL{N}$,
\item$\IrrTL{j}$ --- simple modules over $\TL{N}$,
\item$j$ --- the $U_qsl(2)$ spin, with $l=2j$ the number of through-lines,
\item$J$ --- the $sl(2,\mathbb{R})$ spin,
\item$L$ --- number of Potts spins in a horizontal row of the lattice,
\item$N$ --- number of strands in the loop model/the number of sites in the spin chain ($N=2L$),
\item$\lambda^d_{J,M}$ --- discrete character of the $SL(2,\mathbb{R})/U(1)$ coset model,
\item$\lambda_{r,j}$  --- generating function of levels in the loop model with ``alt" boundary conditions,
\item$c^m_l$ --- string function, i.e., the generating function of levels in the $Z_{k-2}$ parafermion CFT.

\end{itemize}

\bigskip

\section{The (bulk) $Q$-state Potts Model on the square lattice}\label{PottsQgen}

\subsection{Clusters and loops}

We will concern ourselves here with the $Q$-state Potts model on the square lattice defined by the Hamiltonian
\beq\label{PottsHamAn}
\mathcal{H}=-K_1\sum\limits_{\langle ij \rangle_1}\delta_{\sigma_i,\sigma_j}-K_2\sum\limits_{\langle ij \rangle_2}\delta_{\sigma_i,\sigma_j} \,,
\eeq
where $\langle ij \rangle_1$ and $\langle ij \rangle_2$ denote respectively the set of horizontal and vertical nearest neighbours, while $K_1$ and $K_2$ are the corresponding coupling constants. Eq.~(\ref{PottsHamAn}) gives the partition function
\beq\label{Zan}
\mathcal{Z}=\sum\limits_{\{\sigma\}}\prod\limits_{\langle ij \rangle_1}\exp(K_1\delta_{\sigma_i,\sigma_j})\prod\limits_{\langle ij \rangle_2}\exp(K_2\delta_{\sigma_i,\sigma_j}) \,,
\eeq 
where the sum is over all configurations of the Potts spins $\sigma$. Here each spin can take the integer values $\sigma_i = 1,2,\ldots,Q$, and
$\{ \sigma \}$ denotes the collection of all spins.

We will be particularly interested in the isotropic model, i.e. $K_1=K_2=K$. In this case, eqs.~(\ref{PottsHamAn})--(\ref{Zan}) become
\beq\label{PottsHam}
\mathcal{H}=-K\sum\limits_{\langle ij \rangle}\delta_{\sigma_i,\sigma_j}
\eeq
and
\beq\label{Pottspart}
\mathcal{Z}=\sum\limits_{\{\sigma\}}\exp(-\mathcal{H})=\sum\limits_{\{\sigma\}}\prod\limits_{\langle ij \rangle}\exp(K\delta_{\sigma_i,\sigma_j})
\eeq
respectively. It is known that the model admits a duality transformation, under which the variable $v \equiv \exp(K)-1$ is replaced by its dual value $v^* \equiv Q/v$.
The (positive) fixed point of this relation, namely $v_{\rm c} = \sqrt{Q}$, is known to correspond to the ferromagnetic critical point \cite{BaxterFerroPotts}. Notice that under duality, a horizontal pair of nearest neighbours is replaced by a dual vertical pair, and vice versa.

Very remarkably the isotropic square-lattice Potts model is also critical at another value, $v_{\rm AF} = -2+\sqrt{4-Q}$, the antiferromagnetic critical point \cite{B-AF}.
Since this is not a fixed point under duality, the dual value $v^*_{\rm AF} \equiv Q / v_{\rm AF} = -2-\sqrt{4-Q}$ provides a second AF critical point in the same universality class as $v_{\rm AF}$.

Returning to the anisotropic model, we can write
\beq
\begin{aligned}
&\exp(K_1\delta_{\sigma_i,\sigma_j})=1+v_1\delta_{\sigma_i,\sigma_j}\\
&\exp(K_2\delta_{\sigma_i,\sigma_j})=1+v_2\delta_{\sigma_i,\sigma_j} \label{FKtrick}
\end{aligned}
\eeq
and evaluating these identities for $\delta_{\sigma_i,\sigma_j} = 0$ or $1$, we see that they are satisfied provided we set
\beq
\begin{aligned}
&v_1=\exp(K_1)-1\\
&v_2=\exp(K_2)-1 \,.
\end{aligned}
\eeq
Following the strategy of Fortuin and Kasteleyn (FK) \cite{FK72}, we insert (\ref{FKtrick}) into (\ref{Zan}) and expand out the products. For each term
in the expansion, we draw a line between neighbouring Potts spins $i$ and $j$ provided it corresponds to picking the second term
$v_k \delta_{\sigma_i,\sigma_j}$  in (\ref{FKtrick}), and no line if the first term $1$ is taken. Making this choice for each nearest neighbour pair
defines a graph $G$ of clusters (connected components) containing $|G_1|$ horizontal and $|G_2|$ vertical lines.
We can then write the partition function as a sum over all possible clusters:
\beq\label{zcluster1}
\mathcal{Z}=\sum\limits_{G}v_1^{|G_1|}v_2^{|G_2|}\sum\limits_{\{\sigma\}}\delta_{\sigma_i,\sigma_j} \,,
\eeq
where $\sum\limits_{G}$ is a sum over all clusters. We denote moreover the total number of lines in $G$ as
\beq
|G|=|G_1|+|G_2| \,.
\eeq
If we define the quantity $C(G)$ as the number of connected components in $G$, the partition function (\ref{Zan}) can be rewritten as \cite{FK72}
\beq\label{zcluster2}
\mathcal{Z}=\sum\limits_{G}v_1^{|G_1|}v_2^{|G_2|}Q^{C(G)} \,.
\eeq

\begin{figure}
	\centering
\begin{tikzpicture}[scale=1.2]
\draw[black,line width = 1pt](1.5,2.5)--(2,3);
\draw[black,line width = 1pt](1.5,2.5)--(1,3);
\draw[black,line width = 1pt](1,3)--(1.5,3.5);
\draw[black,line width = 1pt](2,3)--(1.5,3.5);
\draw[black,line width = 1pt](1.5,2.5)--(1.5,3.5);

\filldraw[black] (1.5,2.5) circle (2pt);
\filldraw[black] (1.5,3.5) circle (2pt);

\node at (1.5,2) {(a)};

\draw[black,line width = 1pt] (1.25,2.75) .. controls (1.375,3) .. (1.25,3.25);
\draw[black,line width = 1pt] (1.75,2.75) .. controls (1.625,3) .. (1.75,3.25);

\draw[black,line width = 1pt](3.5,2.5)--(4,3);
\draw[black,line width = 1pt](3.5,2.5)--(3,3);
\draw[black,line width = 1pt](3,3)--(3.5,3.5);
\draw[black,line width = 1pt](4,3)--(3.5,3.5);
\draw[black,line width = 1pt](3,3)--(4,3);

\filldraw[black] (3,3) circle (2pt);
\filldraw[black] (4,3) circle (2pt);

\node at (3.5,2) {(b)};

\draw[black,line width = 1pt] (3.25,2.75) .. controls (3.5,2.875) .. (3.75,2.75);
\draw[black,line width = 1pt] (3.25,3.25) .. controls (3.5,3.125) .. (3.75,3.25);

\draw[black,line width = 1pt](5.5,2.5)--(6,3);
\draw[black,line width = 1pt](5.5,2.5)--(5,3);
\draw[black,line width = 1pt](5,3)--(5.5,3.5);
\draw[black,line width = 1pt](6,3)--(5.5,3.5);

\filldraw[black] (5.5,2.5) circle (2pt);
\filldraw[black] (5.5,3.5) circle (2pt);

\draw[black,line width = 1pt] (5.25,2.75) .. controls (5.5,2.875) .. (5.75,2.75);
\draw[black,line width = 1pt] (5.25,3.25) .. controls (5.5,3.125) .. (5.75,3.25);

\node at (5.5,2) {(c)};

\draw[black,line width = 1pt](7.5,2.5)--(8,3);
\draw[black,line width = 1pt](7.5,2.5)--(7,3);
\draw[black,line width = 1pt](7,3)--(7.5,3.5);
\draw[black,line width = 1pt](8,3)--(7.5,3.5);

\filldraw[black] (7,3) circle (2pt);
\filldraw[black] (8,3) circle (2pt);

\draw[black,line width = 1pt] (7.25,2.75) .. controls (7.375,3) .. (7.25,3.25);
\draw[black,line width = 1pt] (7.75,2.75) .. controls (7.625,3) .. (7.75,3.25);

\node at (7.5,2) {(d)};

\end{tikzpicture}

\caption{Each tile on the lattice will take one of the forms shown in panels (a), (b), (c) and (d). The black circles represent the points where the Potts spins lie. The vertices with no black circles are points on the dual lattice.}\label{fourtiles}
\end{figure}
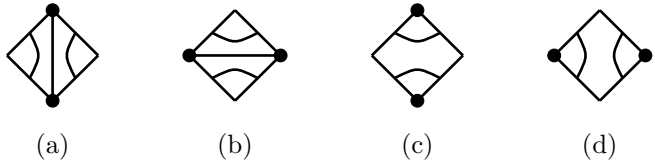

\begin{figure}
	\centering
\begin{tikzpicture}[scale=0.8]
\draw[black,line width = 1pt](1,1)--(9,9);
\draw[black,line width = 1pt](3,1)--(9,7);
\draw[black,line width = 1pt](5,1)--(9,5);
\draw[black,line width = 1pt](7,1)--(9,3);

\draw[black,line width = 1pt](1,3)--(7,9);
\draw[black,line width = 1pt](1,5)--(5,9);
\draw[black,line width = 1pt](1,7)--(3,9);

\draw[black,line width = 1pt](9,1)--(1,9);
\draw[black,line width = 1pt](7,1)--(1,7);
\draw[black,line width = 1pt](5,1)--(1,5);
\draw[black,line width = 1pt](3,1)--(1,3);

\draw[black,line width = 1pt](9,3)--(3,9);
\draw[black,line width = 1pt](9,5)--(5,9);
\draw[black,line width = 1pt](9,7)--(7,9);

\filldraw[black] (2,2) circle (4pt);
\filldraw[black] (4,2) circle (4pt);
\filldraw[black] (6,2) circle (4pt);
\filldraw[black] (8,2) circle (4pt);

\filldraw[black] (2,4) circle (4pt);
\filldraw[black] (4,4) circle (4pt);
\filldraw[black] (6,4) circle (4pt);
\filldraw[black] (8,4) circle (4pt);

\filldraw[black] (2,6) circle (4pt);
\filldraw[black] (4,6) circle (4pt);
\filldraw[black] (6,6) circle (4pt);
\filldraw[black] (8,6) circle (4pt);

\filldraw[black] (2,8) circle (4pt);
\filldraw[black] (4,8) circle (4pt);
\filldraw[black] (6,8) circle (4pt);
\filldraw[black] (8,8) circle (4pt);

\draw[black,line width = 1pt] (1.5,1.5) .. controls (1.25,2) .. (1.5,2.5);
\draw[black,line width = 1pt] (1.5,3.5) .. controls (1.25,4) .. (1.5,4.5);
\draw[black,line width = 1pt] (1.5,5.5) .. controls (1.25,6) .. (1.5,6.5);
\draw[black,line width = 1pt] (1.5,7.5) .. controls (1.25,8) .. (1.5,8.5);

\draw[black,line width = 1pt] (8.5,1.5) .. controls (8.75,2) .. (8.5,2.5);
\draw[black,line width = 1pt] (8.5,3.5) .. controls (8.75,4) .. (8.5,4.5);
\draw[black,line width = 1pt] (8.5,5.5) .. controls (8.75,6) .. (8.5,6.5);
\draw[black,line width = 1pt] (8.5,7.5) .. controls (8.75,8) .. (8.5,8.5);

\draw[black,line width = 2pt](2,2)--(2,4);
\draw[black,line width = 2pt](2,2)--(4,2);
\draw[black,line width = 2pt](2,4)--(6,4);
\draw[black,line width = 2pt](4,4)--(4,2);
\draw[black,line width = 2pt](6,4)--(6,2);
\draw[black,line width = 2pt](2,6)--(4,6);
\draw[black,line width = 2pt](2,8)--(4,8);
\draw[black,line width = 2pt](8,2)--(8,8);

\draw[black,line width = 1pt] (1.5,1.5) .. controls (2,1.25) .. (2.5,1.5);
\draw[black,line width = 1pt] (3.5,1.5) .. controls (4,1.25) .. (4.5,1.5);
\draw[black,line width = 1pt] (5.5,1.5) .. controls (6,1.25) .. (6.5,1.5);
\draw[black,line width = 1pt] (7.5,1.5) .. controls (8,1.25) .. (8.5,1.5);

\draw[black,line width = 1pt] (2.5,1.5) .. controls (3,1.75) .. (3.5,1.5);
\draw[black,line width = 1pt] (2.5,2.5) .. controls (3,2.25) .. (3.5,2.5);
\draw[black,line width = 1pt] (4.5,1.5) .. controls (4.75,2) .. (4.5,2.5);
\draw[black,line width = 1pt] (5.5,1.5) .. controls (5.25,2) .. (5.5,2.5);
\draw[black,line width = 1pt] (6.5,1.5) .. controls (6.75,2) .. (6.5,2.5);
\draw[black,line width = 1pt] (7.5,1.5) .. controls (7.25,2) .. (7.5,2.5);

\draw[black,line width = 1pt] (1.5,2.5) .. controls (1.75,3) .. (1.5,3.5);
\draw[black,line width = 1pt] (2.5,2.5) .. controls (2.25,3) .. (2.5,3.5);
\draw[black,line width = 1pt] (3.5,2.5) .. controls (3.75,3) .. (3.5,3.5);
\draw[black,line width = 1pt] (4.5,2.5) .. controls (4.25,3) .. (4.5,3.5);
\draw[black,line width = 1pt] (5.5,2.5) .. controls (5.75,3) .. (5.5,3.5);
\draw[black,line width = 1pt] (6.5,2.5) .. controls (6.25,3) .. (6.5,3.5);
\draw[black,line width = 1pt] (7.5,2.5) .. controls (7.75,3) .. (7.5,3.5);
\draw[black,line width = 1pt] (8.5,2.5) .. controls (8.25,3) .. (8.5,3.5);

\draw[black,line width = 1pt] (2.5,3.5) .. controls (3,3.75) .. (3.5,3.5);
\draw[black,line width = 1pt] (2.5,4.5) .. controls (3,4.25) .. (3.5,4.5);
\draw[black,line width = 1pt] (4.5,3.5) .. controls (5,3.75) .. (5.5,3.5);
\draw[black,line width = 1pt] (4.5,4.5) .. controls (5,4.25) .. (5.5,4.5);
\draw[black,line width = 1pt] (6.5,3.5) .. controls (6.75,4) .. (6.5,4.5);
\draw[black,line width = 1pt] (7.5,3.5) .. controls (7.25,4) .. (7.5,4.5);

\draw[black,line width = 1pt] (1.5,4.5) .. controls (2,4.75) .. (2.5,4.5);
\draw[black,line width = 1pt] (1.5,5.5) .. controls (2,5.25) .. (2.5,5.5);
\draw[black,line width = 1pt] (3.5,4.5) .. controls (4,4.75) .. (4.5,4.5);
\draw[black,line width = 1pt] (3.5,5.5) .. controls (4,5.25) .. (4.5,5.5);
\draw[black,line width = 1pt] (5.5,4.5) .. controls (6,4.75) .. (6.5,4.5);
\draw[black,line width = 1pt] (5.5,5.5) .. controls (6,5.25) .. (6.5,5.5);
\draw[black,line width = 1pt] (7.5,4.5) .. controls (7.75,5) .. (7.5,5.5);
\draw[black,line width = 1pt] (8.5,4.5) .. controls (8.25,5) .. (8.5,5.5);

\draw[black,line width = 1pt] (2.5,5.5) .. controls (3,5.75) .. (3.5,5.5);
\draw[black,line width = 1pt] (2.5,6.5) .. controls (3,6.25) .. (3.5,6.5);
\draw[black,line width = 1pt] (4.5,5.5) .. controls (4.75,6) .. (4.5,6.5);
\draw[black,line width = 1pt] (5.5,5.5) .. controls (5.25,6) .. (5.5,6.5);
\draw[black,line width = 1pt] (6.5,5.5) .. controls (6.75,6) .. (6.5,6.5);
\draw[black,line width = 1pt] (7.5,5.5) .. controls (7.25,6) .. (7.5,6.5);

\draw[black,line width = 1pt] (1.5,6.5) .. controls (2,6.75) .. (2.5,6.5);
\draw[black,line width = 1pt] (1.5,7.5) .. controls (2,7.25) .. (2.5,7.5);
\draw[black,line width = 1pt] (3.5,6.5) .. controls (4,6.75) .. (4.5,6.5);
\draw[black,line width = 1pt] (3.5,7.5) .. controls (4,7.25) .. (4.5,7.5);
\draw[black,line width = 1pt] (5.5,6.5) .. controls (6,6.75) .. (6.5,6.5);
\draw[black,line width = 1pt] (5.5,7.5) .. controls (6,7.25) .. (6.5,7.5);
\draw[black,line width = 1pt] (7.5,6.5) .. controls (7.75,7) .. (7.5,7.5);
\draw[black,line width = 1pt] (8.5,6.5) .. controls (8.25,7) .. (8.5,7.5);

\draw[black,line width = 1pt] (2.5,7.5) .. controls (3,7.75) .. (3.5,7.5);
\draw[black,line width = 1pt] (2.5,8.5) .. controls (3,8.25) .. (3.5,8.5);
\draw[black,line width = 1pt] (4.5,7.5) .. controls (4.75,8) .. (4.5,8.5);
\draw[black,line width = 1pt] (5.5,7.5) .. controls (5.25,8) .. (5.5,8.5);
\draw[black,line width = 1pt] (6.5,7.5) .. controls (6.75,8) .. (6.5,8.5);
\draw[black,line width = 1pt] (7.5,7.5) .. controls (7.25,8) .. (7.5,8.5);

\draw[black,line width = 1pt] (1.5,8.5) .. controls (2,8.75) .. (2.5,8.5);
\draw[black,line width = 1pt] (3.5,8.5) .. controls (4,8.75) .. (4.5,8.5);
\draw[black,line width = 1pt] (5.5,8.5) .. controls (6,8.75) .. (6.5,8.5);
\draw[black,line width = 1pt] (7.5,8.5) .. controls (8,8.75) .. (8.5,8.5);

\end{tikzpicture}

\caption{The one-to-one mapping between cluster configurations and loop configurations.}\label{fulllattice}
\end{figure}
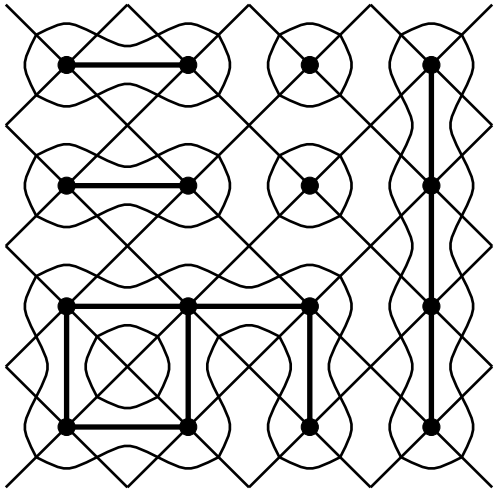

Following Baxter, Kelland and Wu \cite{BKW76}, we can next transform this cluster model into a loop model, i.e., rewrite the partition function as a sum over loops. 
The graph of clusters $G$ is made up of four different types of tiles, shown in Figure~\ref{fourtiles}, defined by the following two binary choices:
1) For tiles of type (a) and (c), the Potts spins (resp.\ dual Potts spins) stand at the top and bottom vertices (resp.\ at the left and right vertices) of the tile,
while it is the other way around for tiles of type (b) and (d). 
2) For tiles of type (a) and (b), the FK expansion contains a line between the two Potts spins, while for tiles of type (c) and (d) there is no such line.
A loop representation equivalent to the FK one is now defined by assigning to each tile two quarter-turn loop segments, as shown in Figure~\ref{fourtiles}.
Doing this for each tile of the lattice (with appropriate boundary conditions), we obtain an ensemble of closed loops, as shown in Figure~\ref{fulllattice}.
Notice in particular that the loops surround the FK clusters and are themselves surrounded by the cycles of the FK clusters.

It follows from this property of surrounding that the number of loops $\ell$ in any given configuration is the sum of the number of connected components $C$
and the number of independent cycles $S$ in the corresponding FK cluster configuration:
\beq\label{topid1}
\ell=C+S
\eeq
We furthermore have another easily verified topological identity (Euler relation) for each cluster configuration:
\beq\label{topid2}
C=|V|-|G|+S \,,
\eeq
where $|V|$ is the number of vertices on the lattice. Combining eqs.~(\ref{topid1})--(\ref{topid2}) gives us
\beq
C=\frac{1}{2}(|V|-|G|+\ell) \,,
\eeq
and inserting this into (\ref{zcluster2}) gives us the partition function expressed in terms of the loop-related quantities \cite{BKW76}
\beq\label{looppartition}
\mathcal{Z}=Q^{\frac{|V|}{2}}\sum\limits_{\text{loops}}x_1^{|G_1|}x_2^{|G_2|}Q^{\frac{\ell}{2}} \,,
\eeq
where we have defined $x_1 \equiv \frac{\exp K_1-1}{\sqrt{Q}}$ and $x_2 \equiv \frac{\exp K_2-1}{\sqrt{Q}}$.
The isotropic case corresponds to $x_1=x_2=x$. In other words, apart from an unimportant overall factor, ${\mathcal Z}$
consists of local weights $x_1, x_2$ depending on the choice of tiles, and a non-local weight of $\sqrt{Q}$ per loop.

\subsection{Critical lines}

There are two well-known critical lines of physical interest for the $Q$-state Potts model on the square lattice. The ferromagnetic self-dual line is obtained for $x_1x_2=1$, with $x_i>0$. The isotropic case corresponds to $x_1=x_2=1$ and thus 
\begin{equation}
e^{K_1}=e^{K_2}=1+\sqrt{Q}
\end{equation}
The corresponding loop model is then purely ``topological'' in the sense that the weight of configurations only depends on the number of loops,
each of which comes with a fugacity $\sqrt{Q}$. The continuum limit of this model is reviewed in many places. Using the parametrisation $\sqrt{Q}=2\cos\gamma$, with $\gamma={\pi\over k}$ and $k\geq 2$,
the central charge in particular is known to be 
\begin{equation}
c=1-{6\over k(k-1)} \,.
\end{equation}
The critical exponents and three point-functions are well known, and closely related with Liouville CFT at $c\leq 1$ (sometimes called time-like Liouville)
\cite{DelfinoViti3pt,PiccoSantachiaraVitiDelfino,IkhlefJacobsenSaleur}.
The question of higher correlation functions in the model remains still partly open, but has recently witnessed important progress \cite{Ribault,JS4point}.
The antiferromagnetic (AF) critical line is not self-dual,%
\footnote{As we have seen above, its image under the duality transformation gives rise to a non-physical regime with complex Boltzmann weights.}
and given by
\begin{equation}
\left(e^{K_1}+1\right)\left(e^{K_2}+1\right)=4-Q \,.
\end{equation}
The isotropic point corresponds to 
\begin{equation}
e^{K_1}=e^{K_2}=-1+\sqrt{4-Q} \,. \label{isoafcouplings}
\end{equation}
Note that the corresponding loop model has weights that depend also on $|G|$, the number of lines in the graphical expansion.
Note that for the model to be ``physical'' one would in general require $e^{K}\geq 0$ that is $Q\leq 3$.
We will, however, consider all values of $Q\in [0,4]$ in what follows, since many of the parameters characterising the corresponding
CFT turn out to depend continuously on $Q$ throughout this range.

The continuum limit of this model is quite intricate, and the present paper considers but one of its many aspects which has not yet
been brought under sufficient control. The central charge along the AF line is however well established to be \cite{S-AF,JS-AF}
\begin{equation}
c_{\rm PF}=2-{6\over k}\label{cafpotts} \,,
\end{equation}
where the subscript PF abbreviates ``parafermion'' for reasons that will be exposed in details below.
Moreover, the exponents are closely related \cite{IJS2008,IJS-AFlett,CanduIkhlef,BKKL} with the spectrum of the $SL(2,\mathbb{R})_k/U(1)$ coset
model \cite{Witten,DVV}, henceforth simply referred to as the black-hole (BH) theory because of its string-theory origins. Recall that this CFT has central charge 
 \begin{equation}\label{cbh}
 c_{\rm BH}=2+{6\over k-2} \,,
 \end{equation}
 while the conformal weights read:
 \begin{equation}
 h_{\rm BH}=-{J(J-1)\over k-2}+{(n\pm wk)^2\over 4k}\label{confw} \,,
 \end{equation}
where $J$ is an $sl(2,\mathbb{R})$ spin.

Since the target of this CFT is non-compact, care must be taken with issues of normalisability. It is known, in particular, that the ground state (corresponding to $J=0$) is non-normalisable. Normalisable states come in a few discrete representations (see below), and in the continuum representations $J={1\over 2}+is$, with $s\in \mathbb{R}$. The central charge of the antiferromagnetic Potts model (\ref{cafpotts}) is obtained from (\ref{cbh}) by a ``twist'' of the compact-boson on top of the $J={1\over 2}$ state in (\ref{confw})---see \cite{IJS2008} for a detailed discussion of this point. 

An intriguing aspect of the antiferromagnetic Potts model meanwhile is that is also exhibits, for $k$ integer, a close relationship with {\sl compact}  parafermions $Z_{k-2}$. The latter can be considered as the coset theory $SU(2)_{k-2}/U(1)$, and are well known to have the central charge (\ref{cafpotts}). It is important to stress that we are not talking here about the non-compact parafermions naturally associated with the $SL(2,\mathbb{R}))_k/U(1)$ theory. These correspond to a symmetry $\hat{W}_\infty(k)$, and correspond formally to $SU(2)_{-k}/U(1)$, since 
\begin{equation}
2+{6\over k-2}=2-{6\over (-k+2)}
\end{equation}
Instead, we really mean here a symmetry $W_{k-2}$, ``hidden'', as it were, within $\hat{W}_\infty(k)$, under some sort of ``symmetry'' $k\to 2-k$. We will shed some more light on this question in the subsequent sections. 

In general, we shall denote exponents in the BH theory by $h$ and those in the AF Potts model by $\Delta$.

\section{Boundary Potts model}\label{BDRPotts}

In the following it will be useful to turn to a more algebraic framework whose basic pieces we now recall. We will exclusively consider the Potts model on a strip of the square lattice of width $L$ (the number of Potts spins), with the transfer matrix ``propagating'' in the vertical direction. 

\subsection{Free boundary conditions and the Temperley-Lieb algebra}\label{secTL}

The simplest boundary conditions are obtained when the Potts spins on the boundary are ``free''. In this case, the cluster expansion of section \ref{PottsQgen} goes through as is. Meanwhile, it is well known that the transfer matrix can be expressed as a product of elementary edge generators:
\beq\label{TMat}
T = (x_1+e_1)(x_1+e_3)\cdots(x_1+e_{2L-1})(1+x_2e_2)(1+x_2e_4)\cdots(1+x_2e_{2L-2}) \,.
\eeq
In the factors of the form $(x_1+e_{2k-1})$ the first term corresponds to a tile of type (a) in Figure~\ref{fourtiles}, while the second term corresponds
to a tile of type (c). Similarly, in the factors of the form $(1+x_2 e_{2k})$ the first (resp.\ second) term corresponds to a tile of type (d) (resp.\ type (b)).
In other words, the former factors add a row of vertical edges between the Potts spins, while the latter ones add a row of horizontal edges.
In these expressions the $e_i$ are operators acting on the Potts spins Hilbert space will well-known expressions.%
\footnote{These expressions depend on the type of representation chosen: original Potts spins $\sigma_i$, FK clusters, loops, six-vertex model, etc. Their
form for the last two representations will be discussed in some detail below.}
More relevant to us is the fact that the $e_i$ obey  the defining relations of the Temperley-Lieb (TL) algebra \cite{TL71}
\beq \label{TLrelations}
\begin{aligned}
e_i^2 &= \sqrt{Q} e_i \,, \\
e_ie_{i\pm1}e_i &= e_i \,, \\
e_ie_j &= e_je_i \text{ for } |i-j|\geq2 \,. \\
\end{aligned}
\eeq
While the integer-$Q$ Potts model provides one representation of this algebra, the FK cluster and loop models correspond to other representations
(essentially the same ones in these latter two cases), which now makes sense for all values of $Q$. In the loop language, we can 
interpret the TL relations graphically by associating a tile with the loop configurations shown in Figure \ref{fourtiles}(d) to the identity operator,
and the tile in Figure \ref{fourtiles}(c) to the TL operator $e_i$. The corresponding tiles (a) and (b) have the same interpretation, except
that they include the factors $x_{1,2}$ in order to account for the weighting of the lines in the expansion.

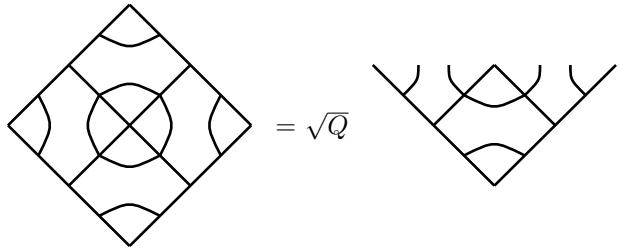
\begin{figure}
	\centering
\begin{tikzpicture}[scale=0.8]
	
\draw[black,line width = 1pt](0,2)--(2,0);
\draw[black,line width = 1pt](2,0)--(4,2);
\draw[black,line width = 1pt](4,2)--(2,4);
\draw[black,line width = 1pt](2,4)--(0,2);
\draw[black,line width = 1pt](1,1)--(3,3);
\draw[black,line width = 1pt](3,1)--(1,3);

\draw[black,line width = 1pt] (1.5,1.5) .. controls (2,1.25) .. (2.5,1.5);
\draw[black,line width = 1pt] (1.5,0.5) .. controls (2,0.75) .. (2.5,0.5);

\draw[black,line width = 1pt] (1.5,3.5) .. controls (2,3.25) .. (2.5,3.5);
\draw[black,line width = 1pt] (1.5,2.5) .. controls (2,2.75) .. (2.5,2.5);

\draw[black,line width = 1pt] (0.5,1.5) .. controls (0.75,2) .. (0.5,2.5);
\draw[black,line width = 1pt] (1.5,1.5) .. controls (1.25,2) .. (1.5,2.5);

\draw[black,line width = 1pt] (2.5,1.5) .. controls (2.75,2) .. (2.5,2.5);
\draw[black,line width = 1pt] (3.5,1.5) .. controls (3.25,2) .. (3.5,2.5);

\node at (5,2) {=   $\sqrt{Q}$};

\draw[black,line width = 1pt](6,3)--(8,1);
\draw[black,line width = 1pt](8,1)--(10,3);
\draw[black,line width = 1pt](7,2)--(8,3);
\draw[black,line width = 1pt](9,2)--(8,3);

\draw[black,line width = 1pt] (7.5,1.5) .. controls (8,1.75) .. (8.5,1.5);
\draw[black,line width = 1pt] (7.5,2.5) .. controls (8,2.25) .. (8.5,2.5);

\draw[black,line width = 1pt] (9.5,2.5) .. controls (9.25,2.75) .. (9.25,3);
\draw[black,line width = 1pt] (8.5,2.5) .. controls (8.75,2.75) .. (8.75,3);

\draw[black,line width = 1pt] (7.5,2.5) .. controls (7.25,2.75) .. (7.25,3);
\draw[black,line width = 1pt] (6.5,2.5) .. controls (6.75,2.75) .. (6.75,3);

\end{tikzpicture}
\caption{Graphical interpretation of the Temperley Lieb algebra $e_i^2=\sqrt{Q}e_i$. Multiplying Temperley Lieb operators corresponds to stacking tiles vertically. Stacking the tiles in the left-hand part of the figure corresponds to the following string of Temperley Lieb operators: $e_iI_{i-1}I_{i+1}e_i=e_ie_i=\sqrt{Q}e_i$. }\label{TL1}
\end{figure}

\begin{figure}
	\centering
\begin{tikzpicture}[scale=0.8]
	
\draw[black,line width = 1pt](0,2)--(2,0);
\draw[black,line width = 1pt](2,0)--(4,2);
\draw[black,line width = 1pt](4,2)--(2,4);
\draw[black,line width = 1pt](2,4)--(0,2);
\draw[black,line width = 1pt](1,1)--(3,3);
\draw[black,line width = 1pt](3,1)--(1,3);

\draw[black,line width = 1pt] (1.5,1.5) .. controls (2,1.25) .. (2.5,1.5);
\draw[black,line width = 1pt] (1.5,0.5) .. controls (2,0.75) .. (2.5,0.5);

\draw[black,line width = 1pt] (1.5,3.5) .. controls (2,3.25) .. (2.5,3.5);
\draw[black,line width = 1pt] (1.5,2.5) .. controls (2,2.75) .. (2.5,2.5);

\draw[black,line width = 1pt] (0.5,1.5) .. controls (1,1.75) .. (1.5,1.5);
\draw[black,line width = 1pt] (0.5,2.5) .. controls (1,2.25) .. (1.5,2.5);

\draw[black,line width = 1pt] (2.5,1.5) .. controls (2.75,2) .. (2.5,2.5);
\draw[black,line width = 1pt] (3.5,1.5) .. controls (3.25,2) .. (3.5,2.5);

\node at (5,2) {=};

\draw[black,line width = 1pt](6,3)--(8,1);
\draw[black,line width = 1pt](8,1)--(10,3);
\draw[black,line width = 1pt](7,2)--(8,3);
\draw[black,line width = 1pt](9,2)--(8,3);

\draw[black,line width = 1pt] (7.5,1.5) .. controls (8,1.75) .. (8.5,1.5);
\draw[black,line width = 1pt] (7.5,2.5) .. controls (8,2.25) .. (8.5,2.5);

\draw[black,line width = 1pt] (9.5,2.5) .. controls (9.25,2.75) .. (9.25,3);
\draw[black,line width = 1pt] (8.5,2.5) .. controls (8.75,2.75) .. (8.75,3);

\draw[black,line width = 1pt] (7.5,2.5) .. controls (7.25,2.75) .. (7.25,3);
\draw[black,line width = 1pt] (6.5,2.5) .. controls (6.75,2.75) .. (6.75,3);

\end{tikzpicture}
\caption{Graphical interpretation of the Temperley Lieb algebra $e_ie_{i-1}e_i=e_i$. Stacking the tiles in the left-hand part of the figure corresponds to the following string of Temperley Lieb operators: $e_iI_{i+1}e_{i-1}e_i=e_ie_{i-1}e_i$}\label{TL2}
\end{figure}
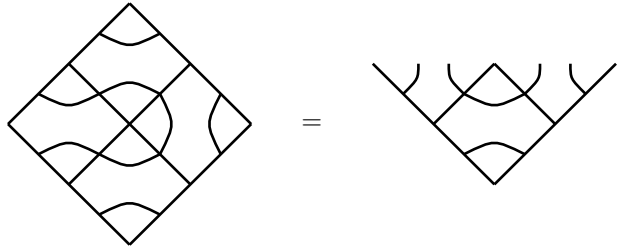

Multiplying TL generators then corresponds to stacking tiles vertically. The graphical interpretation of the relation $e_i^2=\sqrt{Q} e_i$ is shown in Figure \ref{TL1}. We see that stacking an $e_i$ tile on top of another $e_i$ tile creates a loop, and this loop gets the Boltzmann weight $\sqrt{Q}$. Similarly, the relation $e_ie_{i-1}e_i$ is illustrated in Figure \ref{TL2}.

Note that, for the time being, the boundary conditions imposed in eq.~(\ref{TMat}) are free on both sides of the lattice. This corresponds to the fact that
the loop segments that touch the left and right sides of Figure~\ref{fulllattice} are simply reflected back into the system. We can see that the transfer
matrix (\ref{TMat}) creates all the possible loop configurations with such boundary conditions (such as the one in Figure \ref{fulllattice}),
and eq.~(\ref{TLrelations}) ensures that the corresponding loops get the correct Boltzmann weights. In other words, repeated applications of 
$T$ in eq.~(\ref{TMat}) builds up the partition function ${\mathcal Z}$ in eq.~(\ref{looppartition}).%
\footnote{Specifying the boundary conditions between the top and bottom layers would require a more detailed discussion.}

Another useful representation of the TL algebra is obtained by turning to the six-vertex model, where
\beq\label{envert}
e_n=I^{\otimes n-1}\otimes
\begin{bmatrix}
	0 & 0 & 0 & 0\\
	0 & e^{-i\gamma} & 1 & 0\\
	0 & 1 & e^{i\gamma} & 0\\
	0 & 0 & 0 & 0\\
\end{bmatrix}
\otimes
I^{\otimes 2L-n-1} \,.
\eeq
 The operator $e_n$ can also be written in terms of Pauli matrices
 \beq\label{enpauli}
e_n=\frac{1}{2}\left[\sigma^x_n\sigma^x_{n+1}+\sigma^y_n\sigma^y_{n+1}-\cos\gamma(\sigma^z_n\sigma^z_{n+1}-I)-i\sin\gamma(\sigma^z_n-\sigma^z_{n+1})\right] \,.
\eeq
In the Euclidian description of the model, the vertex representation of the Temperley-Lieb algebra leads to a reformulation of eq.~(\ref{looppartition}) as
a staggered six-vertex model \cite{IJSstag1,IJSstag2}. With free boundary conditions for the Potts model, the transfer matrix of that vertex model  commutes with the quantum group $U_\q sl(2)$.

\subsection{Possible other boundary conditions and the blob algebra}

Another type of boundary conditions---often referred to as ``blob'' for reasons to be discussed below---consists in restricting the 
Potts spins on the boundary to take values in $\{1,2,\ldots,Q_1\}$, that is, in a subset of the original range $\{1,2,\ldots,Q\}$.
Note that for the original $Q$-integer Potts model,  this only makes sense if $Q_1\leq Q$ is also integer. However, once the definition of the model
has been extended to all real values of $Q$---e.g., by going to the loop or six-vertex representation---it is no longer necessary to impose this restriction on
$Q_1$, which can hence be taken arbitrary real as well.

By following the steps of the mapping onto clusters and loops, this blob boundary condition corresponds to giving clusters touching the boundary the weight $Q_1$ instead of $Q$ and to loops touching the boundary the weight $Q_1/\sqrt{Q}$ \cite{JS-Blob}. Note that in this definition, the blob boundary conditions is only applied to one side---conventionally the left one---, while the other side retains free boundary conditions. It is also possible to choose blob boundary conditions on both sides, as we discuss below. 

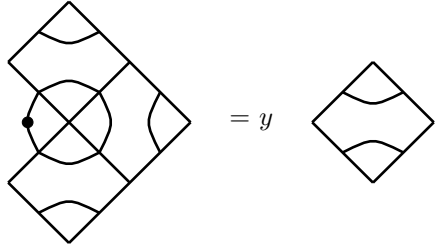
\begin{figure}
	\centering
\begin{tikzpicture}[scale=0.8]
	
\draw[black,line width = 1pt](1,1)--(2,0);
\draw[black,line width = 1pt](2,0)--(4,2);
\draw[black,line width = 1pt](4,2)--(2,4);
\draw[black,line width = 1pt](2,4)--(1,3);
\draw[black,line width = 1pt](1,1)--(3,3);
\draw[black,line width = 1pt](3,1)--(1,3);

\draw[black,line width = 1pt] (1.5,1.5) .. controls (2,1.25) .. (2.5,1.5);
\draw[black,line width = 1pt] (1.5,0.5) .. controls (2,0.75) .. (2.5,0.5);

\draw[black,line width = 1pt] (1.5,3.5) .. controls (2,3.25) .. (2.5,3.5);
\draw[black,line width = 1pt] (1.5,2.5) .. controls (2,2.75) .. (2.5,2.5);

\draw[black,line width = 1pt] (1.5,1.5) .. controls (1.25,2) .. (1.5,2.5);

\draw[black,line width = 1pt] (2.5,1.5) .. controls (2.75,2) .. (2.5,2.5);
\draw[black,line width = 1pt] (3.5,1.5) .. controls (3.25,2) .. (3.5,2.5);

\filldraw[black] (1.32,2) circle (2.5pt);

\node at (5,2) {=   $y$};

\draw[black,line width = 1pt](6,2)--(7,1);
\draw[black,line width = 1pt](7,1)--(8,2);
\draw[black,line width = 1pt](6,2)--(7,3);
\draw[black,line width = 1pt](8,2)--(7,3);

\draw[black,line width = 1pt] (6.5,1.5) .. controls (7,1.75) .. (7.5,1.5);
\draw[black,line width = 1pt] (6.5,2.5) .. controls (7,2.25) .. (7.5,2.5);

\end{tikzpicture}
\caption{Graphical interpretation of the blob algebra relation $e_1 b e_1=y e_1$}\label{blob}
\end{figure}

The blob boundary conditions are easily implemented by introducing the ``blob algebra'' \cite{MartinSaleurBlob, JS-Blob,CBLM}
which is defined by supplementing (\ref{TLrelations}) by an extra generator $b$ subject to the new relations:
\beq\label{blobrelations}
\begin{aligned}
e_1 b e_1 &=ye_1 \,, \\
b^2 &=b \,, \\
e_i b &= b e_i \text{ for } i > 1 \,.
\end{aligned}
\eeq
The graphical interpretation of the first of these relations is illustrated in Figure~\ref{blob}. The blob operator $b$ adds a ``blob'' to the left most loop strand and closed ``blobbed loops'' then get the modified Boltzmann weight $y$. The relation $b^2=b$ describes the property that loops which touch the left boundary more than once get the same weight as loops that touch the left boundary exactly once.

\begin{figure}
	\centering
\begin{tikzpicture}[scale=0.8]
\draw[black,line width = 1pt](1,1)--(9,9);
\draw[black,line width = 1pt](3,1)--(9,7);
\draw[black,line width = 1pt](5,1)--(9,5);
\draw[black,line width = 1pt](7,1)--(9,3);

\draw[black,line width = 1pt](1,3)--(7,9);
\draw[black,line width = 1pt](1,5)--(5,9);
\draw[black,line width = 1pt](1,7)--(3,9);

\draw[black,line width = 1pt](9,1)--(1,9);
\draw[black,line width = 1pt](7,1)--(1,7);
\draw[black,line width = 1pt](5,1)--(1,5);
\draw[black,line width = 1pt](3,1)--(1,3);

\draw[black,line width = 1pt](9,3)--(3,9);
\draw[black,line width = 1pt](9,5)--(5,9);
\draw[black,line width = 1pt](9,7)--(7,9);

\filldraw[black] (1.32,2) circle (2.5pt);
\filldraw[black] (1.32,4) circle (2.5pt);
\filldraw[black] (1.32,6) circle (2.5pt);
\filldraw[black] (1.32,8) circle (2.5pt);

\filldraw[black] (2,2) circle (4pt);
\filldraw[black] (4,2) circle (4pt);
\filldraw[black] (6,2) circle (4pt);
\filldraw[black] (8,2) circle (4pt);

\filldraw[black] (2,4) circle (4pt);
\filldraw[black] (4,4) circle (4pt);
\filldraw[black] (6,4) circle (4pt);
\filldraw[black] (8,4) circle (4pt);

\filldraw[black] (2,6) circle (4pt);
\filldraw[black] (4,6) circle (4pt);
\filldraw[black] (6,6) circle (4pt);
\filldraw[black] (8,6) circle (4pt);

\filldraw[black] (2,8) circle (4pt);
\filldraw[black] (4,8) circle (4pt);
\filldraw[black] (6,8) circle (4pt);
\filldraw[black] (8,8) circle (4pt);

\draw[black,line width = 1pt] (1.5,1.5) .. controls (1.25,2) .. (1.5,2.5);
\draw[black,line width = 1pt] (1.5,3.5) .. controls (1.25,4) .. (1.5,4.5);
\draw[black,line width = 1pt] (1.5,5.5) .. controls (1.25,6) .. (1.5,6.5);
\draw[black,line width = 1pt] (1.5,7.5) .. controls (1.25,8) .. (1.5,8.5);

\draw[black,line width = 1pt] (8.5,1.5) .. controls (8.75,2) .. (8.5,2.5);
\draw[black,line width = 1pt] (8.5,3.5) .. controls (8.75,4) .. (8.5,4.5);
\draw[black,line width = 1pt] (8.5,5.5) .. controls (8.75,6) .. (8.5,6.5);
\draw[black,line width = 1pt] (8.5,7.5) .. controls (8.75,8) .. (8.5,8.5);

\draw[black,line width = 2pt](2,2)--(2,4);
\draw[black,line width = 2pt](2,2)--(4,2);
\draw[black,line width = 2pt](2,4)--(6,4);
\draw[black,line width = 2pt](4,4)--(4,2);
\draw[black,line width = 2pt](6,4)--(6,2);
\draw[black,line width = 2pt](2,6)--(4,6);
\draw[black,line width = 2pt](2,8)--(4,8);
\draw[black,line width = 2pt](8,2)--(8,8);

\draw[black,line width = 1pt] (1.5,1.5) .. controls (2,1.25) .. (2.5,1.5);
\draw[black,line width = 1pt] (3.5,1.5) .. controls (4,1.25) .. (4.5,1.5);
\draw[black,line width = 1pt] (5.5,1.5) .. controls (6,1.25) .. (6.5,1.5);
\draw[black,line width = 1pt] (7.5,1.5) .. controls (8,1.25) .. (8.5,1.5);

\draw[black,line width = 1pt] (2.5,1.5) .. controls (3,1.75) .. (3.5,1.5);
\draw[black,line width = 1pt] (2.5,2.5) .. controls (3,2.25) .. (3.5,2.5);
\draw[black,line width = 1pt] (4.5,1.5) .. controls (4.75,2) .. (4.5,2.5);
\draw[black,line width = 1pt] (5.5,1.5) .. controls (5.25,2) .. (5.5,2.5);
\draw[black,line width = 1pt] (6.5,1.5) .. controls (6.75,2) .. (6.5,2.5);
\draw[black,line width = 1pt] (7.5,1.5) .. controls (7.25,2) .. (7.5,2.5);
\draw[black,line width = 1pt] (1.5,2.5) .. controls (1.75,3) .. (1.5,3.5);
\draw[black,line width = 1pt] (2.5,2.5) .. controls (2.25,3) .. (2.5,3.5);
\draw[black,line width = 1pt] (3.5,2.5) .. controls (3.75,3) .. (3.5,3.5);
\draw[black,line width = 1pt] (4.5,2.5) .. controls (4.25,3) .. (4.5,3.5);
\draw[black,line width = 1pt] (5.5,2.5) .. controls (5.75,3) .. (5.5,3.5);
\draw[black,line width = 1pt] (6.5,2.5) .. controls (6.25,3) .. (6.5,3.5);
\draw[black,line width = 1pt] (7.5,2.5) .. controls (7.75,3) .. (7.5,3.5);
\draw[black,line width = 1pt] (8.5,2.5) .. controls (8.25,3) .. (8.5,3.5);

\draw[black,line width = 1pt] (2.5,3.5) .. controls (3,3.75) .. (3.5,3.5);
\draw[black,line width = 1pt] (2.5,4.5) .. controls (3,4.25) .. (3.5,4.5);
\draw[black,line width = 1pt] (4.5,3.5) .. controls (5,3.75) .. (5.5,3.5);
\draw[black,line width = 1pt] (4.5,4.5) .. controls (5,4.25) .. (5.5,4.5);
\draw[black,line width = 1pt] (6.5,3.5) .. controls (6.75,4) .. (6.5,4.5);
\draw[black,line width = 1pt] (7.5,3.5) .. controls (7.25,4) .. (7.5,4.5);

\draw[black,line width = 1pt] (1.5,4.5) .. controls (2,4.75) .. (2.5,4.5);
\draw[black,line width = 1pt] (1.5,5.5) .. controls (2,5.25) .. (2.5,5.5);
\draw[black,line width = 1pt] (3.5,4.5) .. controls (4,4.75) .. (4.5,4.5);
\draw[black,line width = 1pt] (3.5,5.5) .. controls (4,5.25) .. (4.5,5.5);
\draw[black,line width = 1pt] (5.5,4.5) .. controls (6,4.75) .. (6.5,4.5);
\draw[black,line width = 1pt] (5.5,5.5) .. controls (6,5.25) .. (6.5,5.5);
\draw[black,line width = 1pt] (7.5,4.5) .. controls (7.75,5) .. (7.5,5.5);
\draw[black,line width = 1pt] (8.5,4.5) .. controls (8.25,5) .. (8.5,5.5);

\draw[black,line width = 1pt] (2.5,5.5) .. controls (3,5.75) .. (3.5,5.5);
\draw[black,line width = 1pt] (2.5,6.5) .. controls (3,6.25) .. (3.5,6.5);
\draw[black,line width = 1pt] (4.5,5.5) .. controls (4.75,6) .. (4.5,6.5);
\draw[black,line width = 1pt] (5.5,5.5) .. controls (5.25,6) .. (5.5,6.5);
\draw[black,line width = 1pt] (6.5,5.5) .. controls (6.75,6) .. (6.5,6.5);
\draw[black,line width = 1pt] (7.5,5.5) .. controls (7.25,6) .. (7.5,6.5);

\draw[black,line width = 1pt] (1.5,6.5) .. controls (2,6.75) .. (2.5,6.5);
\draw[black,line width = 1pt] (1.5,7.5) .. controls (2,7.25) .. (2.5,7.5);
\draw[black,line width = 1pt] (3.5,6.5) .. controls (4,6.75) .. (4.5,6.5);
\draw[black,line width = 1pt] (3.5,7.5) .. controls (4,7.25) .. (4.5,7.5);
\draw[black,line width = 1pt] (5.5,6.5) .. controls (6,6.75) .. (6.5,6.5);
\draw[black,line width = 1pt] (5.5,7.5) .. controls (6,7.25) .. (6.5,7.5);
\draw[black,line width = 1pt] (7.5,6.5) .. controls (7.75,7) .. (7.5,7.5);
\draw[black,line width = 1pt] (8.5,6.5) .. controls (8.25,7) .. (8.5,7.5);

\draw[black,line width = 1pt] (2.5,7.5) .. controls (3,7.75) .. (3.5,7.5);
\draw[black,line width = 1pt] (2.5,8.5) .. controls (3,8.25) .. (3.5,8.5);
\draw[black,line width = 1pt] (4.5,7.5) .. controls (4.75,8) .. (4.5,8.5);
\draw[black,line width = 1pt] (5.5,7.5) .. controls (5.25,8) .. (5.5,8.5);
\draw[black,line width = 1pt] (6.5,7.5) .. controls (6.75,8) .. (6.5,8.5);
\draw[black,line width = 1pt] (7.5,7.5) .. controls (7.25,8) .. (7.5,8.5);

\draw[black,line width = 1pt] (1.5,8.5) .. controls (2,8.75) .. (2.5,8.5);
\draw[black,line width = 1pt] (3.5,8.5) .. controls (4,8.75) .. (4.5,8.5);
\draw[black,line width = 1pt] (5.5,8.5) .. controls (6,8.75) .. (6.5,8.5);
\draw[black,line width = 1pt] (7.5,8.5) .. controls (8,8.75) .. (8.5,8.5);

\end{tikzpicture}

\caption{The blobbed loop model where loops that touch the left boundary at least once (blobbed loops) get a modified Boltzmann weight.}
\label{blobfulllattice}
\end{figure}
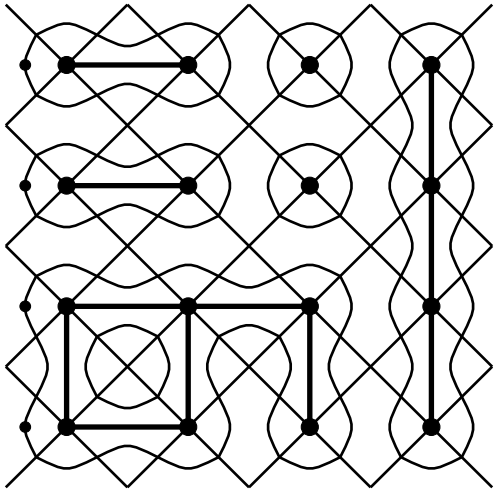

\subsection{The ferromagnetic Potts model}\label{ferrpotts}

In the case of the ferromagnetic Potts model, ``blob'' boundary conditions are conformally invariant for all values of $y$ \cite{CBLM},
provided of course that the bulk theory is critical ($0 \le Q \le 4$).
A configuration of the model with these boundary conditions is illustrated in Figure~\ref{blobfulllattice}. Note that the
general blob boundary conditions include the case of free boundary conditions ($Q_1 = Q$), as well as, for $Q \in \{1,2,3,4\}$ integer,
the case of fixed ($Q_1 = 1$) or ``mixed'' ($1 < Q_1 < Q$ integer) boundary conditions \cite{AOS}.

It is sometimes useful to discuss boundary conditions in terms of the dual model as well. Consider for instance a square lattice, like
in Figure~\ref{blobfulllattice}, with the top and bottom sides identified, so that the lattice has the topology of the annulus with the two
rims being the left and right sides. With free boundary conditions along both rims, the duality transformation produces another square-lattice
annulus, but with different boundary conditions. Indeed there is now a single dual spin to the left of the first column of original Potts spins,
connected to each of the dual spins between the first and second columns.%
\footnote{The situation along the right rim is of course analogous.}
This is sometimes called ``wired'' boundary conditions. 
Notice that this dual lattice can also be seen as a regular square lattice with an extra column on the left, all of whose spins have been contracted to form
a single one. It follows that free boundary conditions are dual to fixed boundary conditions.%
\footnote{The situation becomes slightly more subtle if we play with the difference between the values of the contracted spins on the left and right rims.
This can be conveniently discussed in terms of the two-boundary TL algebra \cite{DJS}. We shall come back to boundary conditions that distinguish both rims
later in this paper.}
The opposite is of course also true: if we give the
same value to all the Potts spins in the first column, we can contract them, so the dual lattice is a regular square lattice, hence sustains free
boundary conditions.

The new boundary condition discussed in \cite{AOS} is dual to the mixed boundary condition $Q=3$, $Q_1=2$. It appears however less
obvious how to discuss it directly in terms of the original spins, i.e., without invoking duality.

\section{Free boundary conditions in the AF Potts model, and $W_\infty$ algebra}\label{KnownPottsBC}

Free boundary conditions are known to be conformally invariant in for the antiferromagnetic Potts model, just like they were for the ferromagnetic one.
The transfer matrix is given by eq.~(\ref{TMat}), which reads in the isotropic case
\beq\label{tmataf}
T=(x+e_1)\cdots(x+e_{2L-1})(1+xe_2)\cdots(1+xe_{2L-2}) \,,
\eeq
but now with the AF choice (\ref{isoafcouplings}) of the coupling constants, viz.
\beq
 x = \frac{e^K - 1}{\sqrt{Q}} = \frac{-2 + \sqrt{4-Q}}{\sqrt{Q}} \,.
\eeq
This corresponds, in the six-vertex model representation of the TL algebra, to 
a staggered six-vertex model with full $U_\q qsl(2)$ symmetry: the transfer matrix (\ref{tmataf}) commutes
with the generators of the quantum group \cite{PasquierSaleurQG}.

The continuum limit of the AF Potts model with these boundary conditions was elucidated as early as \cite{S-AF}, and further studied and confirmed in \cite{CJSSS-Forests,JS-Forests,JS-AF}. Free boundary conditions for the Potts model translate into an even number of spins, $N=2L$, for the vertex model/spin chain, and thus into an integer value of the $U_qsl(2)$ spin $j$.  It is convenient in what follows to introduce the (even) integer $l=2j$. The generating function of conformal levels was then proposed to be \cite{S-AF}
\begin{equation}
K_l=\hbox{Tr}_{\StTL{j=l/2}}q^{L_0-c/24}={q^{(l+1)^2/4k}\over \eta(q)^2}\left[1+2\sum_{n=1}^\infty (-1)^n q^{n(n+l+1)/2}\right]\label{Kfct} \,,
\end{equation}
where we have parameterised $\sqrt{Q}=\q+\q^{-1}$,  with $\q\equiv e^{i\gamma}\equiv e^{i\pi/k}$.
In the loop representation, the number $l$ can be interpreted geometrically as the number of defect lines (unpaired loop
strands) that run from the bottom to the top of the lattice. Such lines are also called through-lines and have a well-known algebraic interpretation
in terms of the TL algebra.

Assuming that the case $l=0$---which by the above remark is tantamount to imposing free boundary conditions on both sides of the lattice (or
free-free boundary conditions, for short)---corresponds to the identity sector of the theory, this allows one to identify the  central charge
\begin{equation}
c=2-{6\over k}
\end{equation}
and coincides indeed with the central charge of the $SU(2)_{k-2}/U(1)$ coset theory (formally  extended to $k$ real). Meanwhile, the leading critical exponent in $K_l$ is 
\begin{equation}
\Delta_l= {l(l+2)\over 4k}\label{PFexp} \,,
\end{equation}
which can be identified \cite{S-AF,JS-AF} as one of a more general family of exponents for the $SU(2)_{k-2}/U(1)$ theory
\begin{equation}
\Delta^{m}_l={l(l+2)\over 4k}-{m^2\over 4(k-2)} \,.
\end{equation}
In particular, we have $\Delta_l=\Delta_l^0$. 

While the initial identification of the $K_l$ generating functions was restricted to a few levels in \cite{S-AF}, we have carefully checked the validity of the expansion (\ref{Kfct}) for a large number of levels. For $K_0$  for instance we have 
\beq\label{Kj0}
K_0=q^{-\frac{c_{\rm AF}}{24}}(1+q^2+2q^3+4q^4+6q^5+11q^6+ \cdots) \,. 
\eeq
This has been observed on the lattice by the analysis of the first 40 eigenvalues of the transfer matrix (\ref{tmataf}). 
To calculate the central charge and leading exponent corresponding to the transfer matrix in (\ref{tmataf}) we use the well-known formula relating the finite-size free-energy density fo the critical exponents $h$ and the central charge $c$:
\beq\label{flscaling}
f_L=f_0+\frac{f_s}{L}-\frac{\pi(\frac{c}{24}-h)}{L^2}+\mathcal{O}\left( \frac{1}{L^3} \right) \,,
\eeq
where the free-energy density $f_L$ is related to the leading transfer matrix eigenvalue $\lambda_0$ by%
\footnote{The factor of $2$ appearing here comes from the fact that (\ref{tmataf}) is a double-row transfer matrix. We will later consider an example of a four-row transfer matrix---see eq.\ (\ref{altblobfull})---where this factor of $2$ will be replaced by $4$ when computing the exponents.} 
\beq
f_L=\frac{\log\lambda_0}{2L} \,.
\eeq 
We compute $f_L$ explicitly by exact numerical diagonalisation methods, then extract $h$ and $c$ by studying the scaling of $f_L$ and comparing with (\ref{flscaling}). We find the remaining exponents in the model by extrapolating to the limit $\frac{1}{L}\rightarrow 0$ the quantities
\beq\label{gapformula}
h^{(L)}=-\frac{L}{\pi}\log \left| \frac{\lambda^i}{\lambda_0} \right| \,,
\eeq
where $\lambda^i$ denotes the $i$-th leading transfer matrix eigenvalue.

By considering sizes up to $N=24$ we can see the first six levels (with multiplicities) of the spectrum generating function, as is written in equation (\ref{Kj0}). For concreteness, we have focussed on the case $k=4.2$, but other values confirm this result. Similarly, we have observed the following levels for $j=1$:
\beq\label{Kj1}
K_2=q^{\Delta_2-\frac{c_{\rm AF}}{24}}(1+2q+3q^2+6q^3+10q^4+\cdots) \,.
\eeq

The $K_l$ are interesting objects for several reasons. If we recall the expression of characters of the $W_\infty$ algebra \cite{Odake}
\begin{equation}
\hbox{Ch}_n^{W_\infty}={1\over \eta(q)^2}\sum_{m=1}^\infty (-1)^m q^{{m(m-1)\over 2}+mn}(q^m-1)
\end{equation}
(this result is valid for $n \in \mathbb{Z}$), we see first of all that 
\begin{equation}
K_0=q^{1\over 4k}\hbox{Ch}_0^{W_\infty}\label{KWid1} \,.
\end{equation}
While the power of $q$ corresponds to the difference between the central charge  $c=2$ of the $W_\infty$ theory and  $c_{\rm AF}$, the appearance of the same character $\hbox{Ch}_0^{W_\infty}$ suggests that the  antiferromagnetic Potts  model exhibits a symmetry that should be some sort of deformation of the $W_\infty$ algebra. Recall that for the latter, one has an  infinity of generators $W^s$ with spin $s$ and  linear commutation relations
\begin{equation}
[W_m^s,W_n^{s'}]=[(s'-1)m-(s-1)n]W_{m+n}^{s+s'-2} \,.
\end{equation}

It is believed that the $SL(2,\mathbb{R})/U(1)$ theory admits a symmetry which is a non-linear deformation of $W_\infty$ that depends on $k$, dubbed $\widehat{W}_\infty(k)$ in \cite{BK91}. The deformation is, for $k$ generic, not expected to change the characters, since it does not affect the spin of the generators. Indeed, it is easy to match the first few terms in the expansion of (\ref{KWid1})---i.e., those shown explicitly in (\ref{Kj0})---with elementary counting  where, for every integer spin $s \ge 2$, a new independent field of conformal weight $n$ appears:%
\footnote{Note that there is no field with $h=1$ in this list, hence no $U(1)$ current.}
\beq
\begin{aligned}
h=2: & \ T\nonumber\\
h=3: & \ \partial T, W_3\nonumber\\
h=4: & \ \partial^2 T, T^2, \partial W_3,W_4 \\
h=5: & \ \partial^3T, T\partial T, \partial^2 W_3, TW_3,\partial W_4,W_5\nonumber\\
h=6: & \ \partial^4 T,T\partial^2T, W_3\partial T, W_3^2,T\partial W_3,T^3,TW_4,\partial^3 W_3,\partial^2 W_4,\partial W_5,W_6\nonumber\\
\ldots&
\end{aligned}
\eeq

The relevance of $\hbox{Ch}_n^{W_\infty}$ characters to the $SL(2,\mathbb{R})/U(1)$ model can be seen in the expression for the 
{\sl discrete, trivial representation} character of the latter theory, given in \cite{BK91}
\begin{equation}
\hat{c}_{m}^{j=0}=q^{|m|+{m^2\over k}}{{q^{-6\over k-2}}\over \eta(q)^2} \left[1+\sum_{n=1}^\infty (-1)^n q^{{n^2+n(2|m|+1)\over 2}}(1+q^{-|m|})\right]=q^{{m^2\over k}+{6\over k-2}}~ \hbox{Ch}_m^{W_\infty}\label{identcharac}
\end{equation}
It is also expected \cite{BK91} that the $\widehat{W}_\infty(k)$ algebra truncates when $k=-k'$, $k'\in \mathbb{N}$, to the symmetry algebra of the $Z_{k'}$ parafermionic theory. Since our model only corresponds to $k\geq 2$, this observation will not be directly relevant to us.

The emergence of $W_\infty$ (or $\widehat{W}_\infty(k)$) extends to all the generating functions in this sector, through the following easily established identity
\begin{equation}
K_{l}=q^{(l+1)^2/4k}\left( \hbox{Ch}_{n=0}^{W_\infty}+2 \hbox{Ch}_{n=1}^{W_\infty}+\ldots +2\hbox{Ch}_{n=l/2}^{W_\infty}\right) \,.
\end{equation}
Note that the number of terms in this expression is $2\times {l\over 2}+1=2j+1$ (recalling that $l=2j$). 

In conclusion,  it is difficult to interpret the Potts model partition functions with free boundary conditions  directly in terms of the $SL(2,\mathbb{R})/U(1)$ sigma model, essentially because of the shift of ground-state energy, which does not seem to have a natural interpretation in terms of discrete spins in (\ref{identcharac}). On the other hand, the appearance of the spectrum of the $W_\infty$  identity representation as excitations over the AF Potts ground state is a strong indication that this model enjoys a deformed version of the $W_\infty$ symmetry. 

\section{New boundary conditions for the AF Potts model}\label{NewPottsBC}

We start with the observation that the usual ``blobbed'' boundary conditions (which, we recall, correspond to fixing the values of the spin in the $Q$-state Potts model to a subset $\{1,2,...,Q_1\}$ on the boundary) do not seem, in general to be conformally invariant in the critical antiferromagnetic case.%
\footnote{There does not seem much point in providing data  to justify this statement: what is observed is simply that the scaled gaps, while converging to fixed (real) values for large systems as they should in a scaling theory, do not reproduce any of the features expected from a CFT. Most noticeably, they do not form conformal towers with integer-spaced scaling levels characteristic of descendent states.}
Meanwhile, we have, by trial and error---and inspired by the Ising case, as discussed below---identified a whole new family of boundary conditions which are conformally invariant in the AF Potts model, but are not in the ferromagnetic one. We refer to these boundary conditions as ``alt'', for reasons which will become clear below. 

\subsection{The alt boundary conditions}\label{altbcsec}

Instead of fixing the spins to a subset on one of the boundaries---by convention the left one---, we now fix them to two complementary, alternating subsets. In other words,
we decide that Potts spins on even (say) boundary sites can only take a particular set of $Q_{1}\leq Q$ values, while spins on odd boundary sites can only take  value in the complementary set of $Q_{2}=Q-Q_1$ values. No spin can be in both sets.

As before, we can make sense of this definition for all real values of $Q$, $Q_1$ by going to the loop or vertex representation. In particular, we no longer require $0 \le Q_1 \le Q$. The partition function of this model in the loop representation is:
\beq\label{zpotts}
\mathcal{Z}^{\rm Alt}_{\rm Potts}=Q^{\frac{V}{2}}\sum\limits_{\rm loops}x^{|E|}Q^{\frac{\ell}{2}}\left(\frac{Q_{1}}{Q}\right)^{\ell_1}\left(\frac{Q_{2}}{Q}\right)^{\ell_2} \,,
\eeq
where the sum is over all loops that do not touch the boundary at both even and odd sites, and $\ell_1$ and $\ell_2$ are the number of loops that touch the boundary at exclusively even and exclusively odd sites, respectively. This describes what we may call the partition function of an alternating boundary loop model.
Eliminating the nugatory overall factor, and matching the notation of the usual blobbed loop model, we define
\beq\label{zalt}
\mathcal{Z}^{\rm Alt}_{\rm Loop}=\sum\limits_{\rm loops}x^{|E|}Q^{\ell-\ell_1-\ell_2\over 2}(y_1)^{\ell_1}(y_2)^{\ell_2} \,,
\eeq
where $y_1$ and $y_2$ are the Boltzmann weights of loops that touch the boundary at exclusively even and exclusively odd sites respectively, and $n_0=\sqrt{Q}$ is the weight of any of the $\ell-\ell_1-\ell_2$ loops in the bulk (i.e., loops that do not touch any site on the left boundary). Once again the sum is over loops that do not touch the boundary at both even and odd sites. We can rewrite the foregoing expression as
\beq\label{zalt2}
\mathcal{Z}^{\rm Alt}_{\rm Loop}=\sum\limits_{\rm loops}x^{|E|}Q^{\ell \over 2}\left(\frac{y_1}{n_0}\right)^{\ell_1}\left(\frac{y_2}{n_0}\right)^{\ell_2} \,.
\eeq

The algebraic framework that permits us to analyse this object is now the blob algebra instead of the Temperley-Lieb algebra. The corresponding blob parameter is  $y_1=Q_1/\sqrt{Q}$, which we parameterise as 
\begin{equation}\label{blobloopweight}
y_1={Q_1\over\sqrt{Q}}\equiv {\sin(r+1)\gamma\over\sin r\gamma} \,,
\end{equation}
while 
\begin{equation}\label{y1y2}
y_2={Q-Q_1\over\sqrt{Q}}=\sqrt{Q}-y_1={\sin(r-1)\gamma\over\sin r\gamma} \,,
\end{equation}
where we recall that we have $n_0 = \sqrt{Q}=2\cos\gamma$. The model is described by a transfer matrix $T$ written entirely in terms of blob algebra generators:
\beq\label{altblobfull}
T=t_1t_2 \,,
\eeq
where
\begin{eqnarray} \label{altblob1}
t_1 &=& b(e_1)(x+e_3)(x+e_5)\cdots(x+e_{2L-1})(1+xe_2)(1+xe_4)\cdots(1+xe_{2L-2}) \,, \\
\label{altblob2}
t_2 &=& (1-b)(e_1)(x+e_3)(x+e_5)\cdots(x+e_{2L-1})(1+xe_2)(1+xe_4)\cdots(1+xe_{2L-2}) \,.
\end{eqnarray}

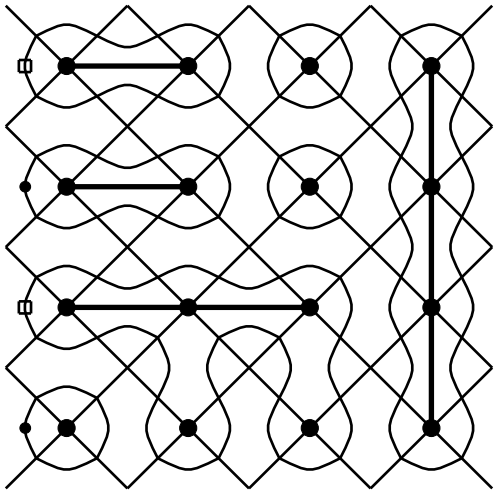
\begin{figure}
	\centering
\begin{tikzpicture}[scale=0.8]
\draw[black,line width = 1pt](1,1)--(9,9);
\draw[black,line width = 1pt](3,1)--(9,7);
\draw[black,line width = 1pt](5,1)--(9,5);
\draw[black,line width = 1pt](7,1)--(9,3);

\draw[black,line width = 1pt](1,3)--(7,9);
\draw[black,line width = 1pt](1,5)--(5,9);
\draw[black,line width = 1pt](1,7)--(3,9);

\draw[black,line width = 1pt](9,1)--(1,9);
\draw[black,line width = 1pt](7,1)--(1,7);
\draw[black,line width = 1pt](5,1)--(1,5);
\draw[black,line width = 1pt](3,1)--(1,3);

\draw[black,line width = 1pt](9,3)--(3,9);
\draw[black,line width = 1pt](9,5)--(5,9);
\draw[black,line width = 1pt](9,7)--(7,9);

\filldraw[black] (1.32,2) circle (2.5pt);

\draw[black,line width = 1pt](1.215,3.9)--(1.415,3.9);
\draw[black,line width = 1pt](1.215,3.9)--(1.215,4.1);
\draw[black,line width = 1pt](1.415,3.9)--(1.415,4.1);
\draw[black,line width = 1pt](1.215,4.1)--(1.415,4.1);

\filldraw[black] (1.32,6) circle (2.5pt);

\draw[black,line width = 1pt](1.215,7.9)--(1.415,7.9);
\draw[black,line width = 1pt](1.215,7.9)--(1.215,8.1);
\draw[black,line width = 1pt](1.415,7.9)--(1.415,8.1);
\draw[black,line width = 1pt](1.215,8.1)--(1.415,8.1);

\filldraw[black] (2,2) circle (4pt);
\filldraw[black] (4,2) circle (4pt);
\filldraw[black] (6,2) circle (4pt);
\filldraw[black] (8,2) circle (4pt);

\filldraw[black] (2,4) circle (4pt);
\filldraw[black] (4,4) circle (4pt);
\filldraw[black] (6,4) circle (4pt);
\filldraw[black] (8,4) circle (4pt);

\filldraw[black] (2,6) circle (4pt);
\filldraw[black] (4,6) circle (4pt);
\filldraw[black] (6,6) circle (4pt);
\filldraw[black] (8,6) circle (4pt);

\filldraw[black] (2,8) circle (4pt);
\filldraw[black] (4,8) circle (4pt);
\filldraw[black] (6,8) circle (4pt);
\filldraw[black] (8,8) circle (4pt);

\draw[black,line width = 1pt] (1.5,1.5) .. controls (1.25,2) .. (1.5,2.5);
\draw[black,line width = 1pt] (1.5,3.5) .. controls (1.25,4) .. (1.5,4.5);
\draw[black,line width = 1pt] (1.5,5.5) .. controls (1.25,6) .. (1.5,6.5);
\draw[black,line width = 1pt] (1.5,7.5) .. controls (1.25,8) .. (1.5,8.5);

\draw[black,line width = 1pt] (8.5,1.5) .. controls (8.75,2) .. (8.5,2.5);
\draw[black,line width = 1pt] (8.5,3.5) .. controls (8.75,4) .. (8.5,4.5);
\draw[black,line width = 1pt] (8.5,5.5) .. controls (8.75,6) .. (8.5,6.5);
\draw[black,line width = 1pt] (8.5,7.5) .. controls (8.75,8) .. (8.5,8.5);

\draw[black,line width = 2pt](2,4)--(6,4);
\draw[black,line width = 2pt](2,6)--(4,6);
\draw[black,line width = 2pt](2,8)--(4,8);
\draw[black,line width = 2pt](8,2)--(8,8);

\draw[black,line width = 1pt] (1.5,1.5) .. controls (2,1.25) .. (2.5,1.5);
\draw[black,line width = 1pt] (3.5,1.5) .. controls (4,1.25) .. (4.5,1.5);
\draw[black,line width = 1pt] (5.5,1.5) .. controls (6,1.25) .. (6.5,1.5);
\draw[black,line width = 1pt] (7.5,1.5) .. controls (8,1.25) .. (8.5,1.5);

\draw[black,line width = 1pt] (2.5,1.5) .. controls (2.75,2) .. (2.5,2.5);
\draw[black,line width = 1pt] (3.5,1.5) .. controls (3.25,2) .. (3.5,2.5);
\draw[black,line width = 1pt] (4.5,1.5) .. controls (4.75,2) .. (4.5,2.5);
\draw[black,line width = 1pt] (5.5,1.5) .. controls (5.25,2) .. (5.5,2.5);
\draw[black,line width = 1pt] (6.5,1.5) .. controls (6.75,2) .. (6.5,2.5);
\draw[black,line width = 1pt] (7.5,1.5) .. controls (7.25,2) .. (7.5,2.5);

\draw[black,line width = 1pt] (1.5,2.5) .. controls (2,2.75) .. (2.5,2.5);
\draw[black,line width = 1pt] (1.5,3.5) .. controls (2,3.25) .. (2.5,3.5);
\draw[black,line width = 1pt] (3.5,2.5) .. controls (3.75,3) .. (3.5,3.5);
\draw[black,line width = 1pt] (4.5,2.5) .. controls (4.25,3) .. (4.5,3.5);
\draw[black,line width = 1pt] (5.5,2.5) .. controls (5.75,3) .. (5.5,3.5);
\draw[black,line width = 1pt] (6.5,2.5) .. controls (6.25,3) .. (6.5,3.5);
\draw[black,line width = 1pt] (7.5,2.5) .. controls (7.75,3) .. (7.5,3.5);
\draw[black,line width = 1pt] (8.5,2.5) .. controls (8.25,3) .. (8.5,3.5);

\draw[black,line width = 1pt] (2.5,3.5) .. controls (3,3.75) .. (3.5,3.5);
\draw[black,line width = 1pt] (2.5,4.5) .. controls (3,4.25) .. (3.5,4.5);
\draw[black,line width = 1pt] (4.5,3.5) .. controls (5,3.75) .. (5.5,3.5);
\draw[black,line width = 1pt] (4.5,4.5) .. controls (5,4.25) .. (5.5,4.5);
\draw[black,line width = 1pt] (6.5,3.5) .. controls (6.75,4) .. (6.5,4.5);
\draw[black,line width = 1pt] (7.5,3.5) .. controls (7.25,4) .. (7.5,4.5);

\draw[black,line width = 1pt] (1.5,4.5) .. controls (2,4.75) .. (2.5,4.5);
\draw[black,line width = 1pt] (1.5,5.5) .. controls (2,5.25) .. (2.5,5.5);
\draw[black,line width = 1pt] (3.5,4.5) .. controls (4,4.75) .. (4.5,4.5);
\draw[black,line width = 1pt] (3.5,5.5) .. controls (4,5.25) .. (4.5,5.5);
\draw[black,line width = 1pt] (5.5,4.5) .. controls (6,4.75) .. (6.5,4.5);
\draw[black,line width = 1pt] (5.5,5.5) .. controls (6,5.25) .. (6.5,5.5);
\draw[black,line width = 1pt] (7.5,4.5) .. controls (7.75,5) .. (7.5,5.5);
\draw[black,line width = 1pt] (8.5,4.5) .. controls (8.25,5) .. (8.5,5.5);

\draw[black,line width = 1pt] (2.5,5.5) .. controls (3,5.75) .. (3.5,5.5);
\draw[black,line width = 1pt] (2.5,6.5) .. controls (3,6.25) .. (3.5,6.5);
\draw[black,line width = 1pt] (4.5,5.5) .. controls (4.75,6) .. (4.5,6.5);
\draw[black,line width = 1pt] (5.5,5.5) .. controls (5.25,6) .. (5.5,6.5);
\draw[black,line width = 1pt] (6.5,5.5) .. controls (6.75,6) .. (6.5,6.5);
\draw[black,line width = 1pt] (7.5,5.5) .. controls (7.25,6) .. (7.5,6.5);

\draw[black,line width = 1pt] (1.5,6.5) .. controls (2,6.75) .. (2.5,6.5);
\draw[black,line width = 1pt] (1.5,7.5) .. controls (2,7.25) .. (2.5,7.5);
\draw[black,line width = 1pt] (3.5,6.5) .. controls (4,6.75) .. (4.5,6.5);
\draw[black,line width = 1pt] (3.5,7.5) .. controls (4,7.25) .. (4.5,7.5);
\draw[black,line width = 1pt] (5.5,6.5) .. controls (6,6.75) .. (6.5,6.5);
\draw[black,line width = 1pt] (5.5,7.5) .. controls (6,7.25) .. (6.5,7.5);
\draw[black,line width = 1pt] (7.5,6.5) .. controls (7.75,7) .. (7.5,7.5);
\draw[black,line width = 1pt] (8.5,6.5) .. controls (8.25,7) .. (8.5,7.5);

\draw[black,line width = 1pt] (2.5,7.5) .. controls (3,7.75) .. (3.5,7.5);
\draw[black,line width = 1pt] (2.5,8.5) .. controls (3,8.25) .. (3.5,8.5);
\draw[black,line width = 1pt] (4.5,7.5) .. controls (4.75,8) .. (4.5,8.5);
\draw[black,line width = 1pt] (5.5,7.5) .. controls (5.25,8) .. (5.5,8.5);
\draw[black,line width = 1pt] (6.5,7.5) .. controls (6.75,8) .. (6.5,8.5);
\draw[black,line width = 1pt] (7.5,7.5) .. controls (7.25,8) .. (7.5,8.5);

\draw[black,line width = 1pt] (1.5,8.5) .. controls (2,8.75) .. (2.5,8.5);
\draw[black,line width = 1pt] (3.5,8.5) .. controls (4,8.75) .. (4.5,8.5);
\draw[black,line width = 1pt] (5.5,8.5) .. controls (6,8.75) .. (6.5,8.5);
\draw[black,line width = 1pt] (7.5,8.5) .. controls (8,8.75) .. (8.5,8.5);

\end{tikzpicture}

\caption{The ``alternating loop model'': The ``blobs" correspond to the operator $b$ and the squares correspond to the operator $u=1-b$.}\label{blobaltfulllattice}
\end{figure}

We stress that the new boundary condition---which we shall call ``alt'' in the following---is for the time being imposed on only one side of the system. Note also that under $r\to -r$,  $y_1$ and $y_2$ are swapped, which corresponds simply to swapping  the odd and even sites (or odd and even  loops), and does not change any of the properties of the system. In what follows, we can therefore assume without loss of generality that $r\geq 0$. We can furthermore define the operator $u \equiv 1-b$ such that $b,u$ are orthogonal projectors:
\beq
\begin{aligned}
&b^2=b\\
&u^2=u\\
&b u =u b=0\\
&b+u=1
\end{aligned}
\eeq
and we can rewrite eq.~(\ref{altblob2}) as 
\beq\label{altunblob}
t_2=u (e_1)(x+e_3)(x+e_5)\cdots(x+e_{2L-1})(1+xe_2)(1+xe_4)\cdots(1+xe_{2L-2}) \,.
\eeq
We shall represent the operator $u$ as a square; closed loops with a square then get the modified Boltzmann weight $y_2$. A configuration of the alternating loop model is shown in Figure~\ref{blobaltfulllattice}. 

\subsection{Discrete characters for $sl(2,\mathbb{R})$}

\begin{figure}
	\centering
	\includegraphics[scale=0.35]{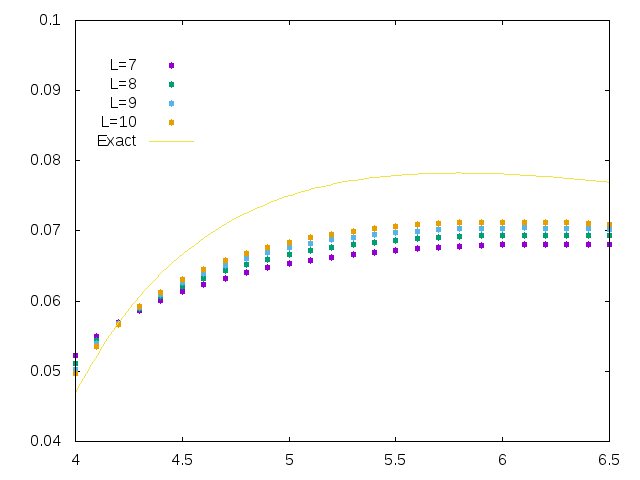}
	\caption{The critical exponent plotted vs $k$, using the values $r=2.5$ and $j=0$. It can be seen that the finite-size values for the critical exponents converge (slowly) to the exact value.}\label{hvskr25l0}
\end{figure}

\begin{figure}
	\captionsetup{width=0.3\textwidth}
\centering
\begin{minipage}{.35\textwidth}
\includegraphics[width=.9\linewidth]{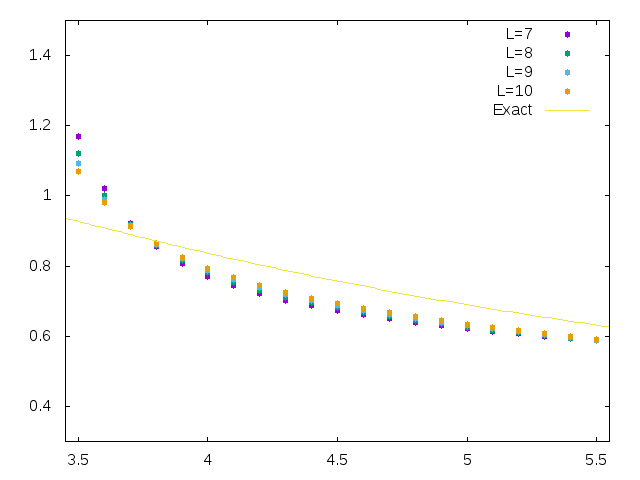}
 \captionof{figure}{The critical exponent plotted vs $k$, using the values $r=2.1$ and $j=1$ in the blob sector.}
 \label{hvskl2blob}
 \end{minipage}%
\begin{minipage}{.35\textwidth}
\includegraphics[width=.9\linewidth]{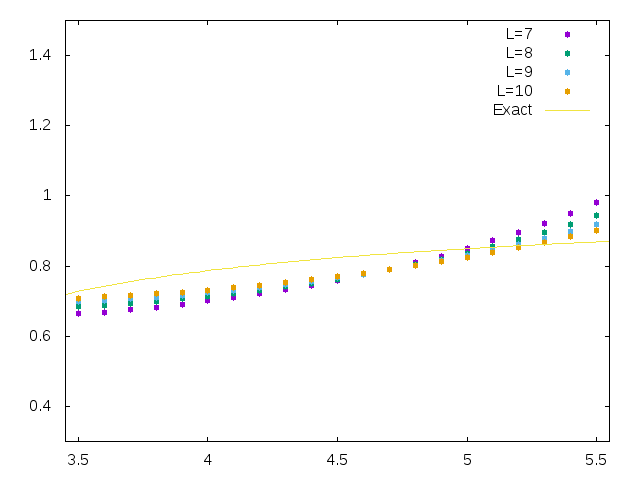}
\captionof{figure}{The critical exponent plotted vs $k$, using the values $r=2.1$ and $j=1$ in the unblob sector.}
\label{hvskl2unblob}
\end{minipage}%
\end{figure}

While the blob algebra is reminiscent of the Temperley-Lieb algebra, the two have interesting differences. First of all, blob standard modules in general come in two types, depending on whether the leftmost through-line is blobbed or unblobbed. The two types have the same dimension
 \begin{equation} \label{eqDimStdblob}
\mathrm{dim}\; \BW^{b/u}_{j} = \binom{N}{N/2 - j} \ .
\end{equation}
where as usual $2j$ is the number of through-lines.  The generating functions of the levels for the alt transfer matrix 
 in each of the two sectors are found numerically (see Table~\ref {altgenfunctab}) to have the following scaling limit:
\begin{eqnarray}
\label{blobsector}
\hbox{Tr}_{\BW_j^b}q^{L_0-c/24} &\mapsto& \lambda_{r,j} \,, \\
\label{unblobsector}
\hbox{Tr}_{\BW_j^u}q^{L_0-c/24} &\mapsto& \lambda_{k-r,j} \,,
\end{eqnarray}
where $r$ parameterises the blob parameter according to eqs.~(\ref{blobloopweight})--(\ref{y1y2}). Note that the result (\ref{unblobsector}) follows from (\ref{blobsector}), since $r\to k-r$ exchanges  $y_1$ and $y_2$ (using also periodicity of  $y_1,y_2$ under $r\to r+k$), while exchanging $y_1$ and $y_2$ obviously also exchanges the roles of $b$ and $u$. The quantities $\lambda_{r,j}$ entering eqs.~(\ref{blobsector})--(\ref{unblobsector}) are defined as
\beq
\lambda_{r,j}=\frac{1}{\eta(q)^2}q^{\frac{(r+2j)^2}{4k}-\frac{(r-1)^2}{4(k-2)}}S_j \,,
\eeq
where 
\beq\label{Sjdef}
S_j=\sum\limits_{n=0}^{\infty}(-1)^n q^{\frac{n^2}{2}+\frac{n}{2}(2j+1)}
\eeq
and $\eta(q) = q^{1/24} \prod_{n=1}^\infty (1-q^n)$ is Dedekind's eta function.

Remarkably, these quantities are formally identical with  characters  of discrete representations ${\cal D}^d_{J,M}$ for the $SL(2,\mathbb{R})/U(1)$ theory. Indeed from \cite{Troostetal,RS} these characters are, for level $k$ and $M\geq 0$,
\begin{equation}
\lambda^d_{J,M}=\eta(q)^{-2} q^{{(J+M)^2\over k}}q^{-{(J-1/2)^2\over k-2}}S_M,~M\geq 0\label{discCharcpos} \,,
\end{equation}
where ${1\over 2}<J<{k-1\over 2}$, $2J\in\mathbb{N}$  is the spin of the discrete $SL(2,\mathbb{R})$ representation and $J+M$ are the $U(1)$ charges, with $M\in\mathbb{Z}$. We see that the correspondence with our case must be
\begin{eqnarray}
J&=&{r\over 2} \,, \nonumber\\
M&=&j \,, \nonumber \\
\lambda_{r,j}&=&\lambda^d_{J,M} \,. \label{paramcorr} 
\end{eqnarray}
Equally remarkably, we see that the ``spin'' in $U_qsl(2)$ corresponds to the $U(1)$ number in the coset theory, while it is the $r$-number (which can be interpreted in terms of a boundary spin \cite{JS-Blob}) that corresponds to the $SL(2,\mathbb{R})$ spin. Note that since we deal with systems of even length, $j$ is integer and thus so is $M$. Of course the parameter $r$ in the lattice model does not have to be integer; but since $2J\in \mathbb{N}$, it is only this case that lends itself to an interpretation in terms of  discrete representations. Note, however, that even when $r$ is an integer, $k-r$ is not an integer for $k$ generic. The  generating function in the unblobbed sector can nevertheless be interpreted in terms of $SL(2,\mathbb{R})$ as well. Indeed, for $M\leq -1$, the discrete characters are usually expressed slightly differently  \cite{RS}\footnote{In fact, it can be shown that expression (\ref{discCharcpos}) also holds for $M\leq -1$ thanks to the identity $S_{-j}+S_{j-1}=1$ \cite{Israel}. Note also that $q^{-j}S_{-j}=S_j$, so we have the algebraic identity $\lambda_{r,j}=\lambda_{k-r,-j}$.}\footnote{For comparison with \cite{RS} we have $\lambda^d_{J,M}=\chi^d_{-J,M+J}$. In other words, the labels $j,l$ in this reference are given by $j=-J,l=M$.}
\begin{equation}
\lambda^d_{J,M}=\eta(q)^{-2} q^{{(J-|M|)^2\over k}}q^{-{(J-1/2)^2\over k-2}} q^{|M|}S_{|M|} \,, \mbox{ for } M\leq -1 \,,
\end{equation}
and we find correspondingly
\begin{eqnarray}
J&=&{r\over 2} \,, \nonumber\\
M&=&-j \,, \nonumber\\
\lambda_{k-r,j}&=&\lambda^d_{J,M} \,. \label{paramcorr1}
\end{eqnarray}

In Table~\ref{altgenfunctab} we compare the generating functions of the alternating loop model observed on the lattice with the quantites defined in equations (\ref{blobsector}) and (\ref{unblobsector}). In Figures~\ref{hvskr25l0}, \ref{hvskl2blob} and \ref{hvskl2unblob} we plot the critical exponent as a function of $k$, and three different spin sectors: $j=0$, $j=1$ in the blob sector, and $j=1$ in the unblob sector.

\begin{table}
\begin{center}
\begin{tabular}{l | l | l | l | l}
$k$ & Sector & $r$ & Exponent & Generating function\\
\hline
4.5 & $j=0$ & $2$ & $0.0667$ & $1+q+3q^2+6q^3+...$ \\[1.1ex]
\hline
4.5 & $j=1$ blob & $2$ & $0.733$ & $1+2q+4q^2+8q^3+...$ \\[1.1ex]
\hline
4.5 & $j=1$ unblob & $2$ & $0.844$ & $1+2q+4q^2+8q^3+...$ \\[1.1ex]
\hline
4.5 & $j=2$ unblob& $2$ & $2.067$ & $1+2q+5q^2+...$ \\[1.1ex]
\hline
4.5 & $j=2$  blob& $2$ & $1.844$ & $1+2q+5q^2+...$ \\[1.1ex]
\hline
5.1 & $j=0$ & $2.2$ & $0.0664$ & $1+q+3q^2+6q^3+...$\\[1.1ex]
\hline
5.1 & $j=1$  blob & $2.2$ & $0.655$ & $1+2q+4q^2+8q^3+...$ \\[1.1ex]
\hline
5.1 & $j=1$  unblob & $2.2$ & $0.837$ & $1+2q+4q^2+8q^3+...$ \\[1.1ex]
\hline
5.1 & $j=2$ unblob & $2.2$ & $2.066$ & $1+2q+5q^2+...$ \\[1.1ex]
\hline
5.1 & $j=2$  blob & $2.2$ & $1.635$ & $1+2q+5q^2+...$ \\[1.1ex]
\end{tabular}
\caption{Some examples of the generating functions and critical exponents observed numerically on the lattice for two different values of $k$ and $r$, in both the blobbed and the unblobbed sectors. The generating functions are written to the order to which they can be clearly observed on the lattice by diagonalising the transfer matrix for the alternating loop model. Up to this order they agree with the generating functions defined in eqs.~(\ref{blobsector})--(\ref{unblobsector}).}\label{altgenfunctab}
\end {center}
\end{table}

Since $M=\pm j$ depending on the sector, we see that all allowed values of $M$ in the discrete characters are observed in the lattice model. We will discuss the values of $J$ in section \ref{secnormalizeissues}. Note that the characters $K_l$ from section~\ref{KnownPottsBC} can also be expressed in terms of discrete characters. We find indeed the simple identities
\begin{equation}
K_l=\lambda^d_{{1\over 2},{l\over 2}}-\lambda^d_{{1\over 2},-{l+2\over 2}}\label{Kdisexp} \,.
\end{equation}
Since the $K_l$ are obtained by focussing only on the Temperley-Lieb subalgebra of the blob algebra---that is, in fact, the top module of the blob algebra for the degenerate case $r=1$---it is tempting to think that the first discrete character in (\ref{Kdisexp}) would be obtained by taking the trace over the full module $\BX_{r=1,j=l/2}^{b}$.%
\footnote{A similar feature is known to occur in the ferromagnetic case, where the trace over the full module gives the character of the Verma module $q^{h_{1,1+l}-c/24}/P(q)$ while the trace over the top module gives the character of the `Kac' module $q^{h_{1,1+l}-c/24}(1-q^{1+l}) /P(q)$.}
But this is not at all what happens---extending the transfer matrix from the Temperley-Lieb to the full blob algebra in the case $r=1$ leads to a complete change of thermodynamic properties, and is connected with normalisability issues we discuss next. We also emphasise that, while the alt boundary conditions are also described with a label $r$, there is no way to go from them to the boundary conditions of section~\ref{KnownPottsBC} by sending $r\to 1$. 

\subsection{Normalisability issues}\label{secnormalizeissues}
The inequality ${1\over 2}<J<{k-1\over 2}$, where $2J\in\mathbb{N}$, for normalisable states in the CFT suggests that the identification of the generating function of levels in the lattice model  with discrete characters must break down at some point  for $r<1$ or $r>k-1$. Where exactly it breaks down is not so clear, since $J$ in the CFT is necessarily integer while  our variable $r$ is continuous. We find in fact that the identification breaks down for $r<r_c$ where $1<r_c<2$ is some critical value of $r$ dependent on $k$. The same phenomenon must then happen for $r>k-r_c$, because of the $r\to k-r$ symmetry.

We find numerically that the analytical continuation of the levels contributing to the discrete character within the interval $[r_c,k-r_c]$ correspond, outside this interval, to highly excited states. In other words, there are numerous level crossings at $r_c$ and $k-r_c$. Outside this interval, the true ground state of the theory does not follow analytically from the ground state within the interval. This is related to the behaviour of the boundary energy in eq.~(\ref{flscaling}) as will now be discussed.

\medskip

Recall that to calculate the leading critical exponent from the finite-size scaling of the lattice model we use eq.~(\ref{flscaling}), while to find the descendant states within a given sector we use eq.~(\ref{gapformula}). It is found, however, that there are in fact two different types of states in the spectrum, with two different boundary free energies $f_s$ in eq.~(\ref{flscaling}). We can see from (\ref{flscaling}) that only states with the lower value of $f_s$ will contribute to the low-energy spectrum in the thermodynamic limit. The values of these two boundary free energies, $f_s^1$ and $f_s^2$, however depend on $r$, and it is found that there exists a critical value $1<r_c<2$ such that they cross. Accordingly, when $r<r_c$ (or $r>k-r_c$) the low-energy part of the spectrum in the thermodynamic limit is no longer described by the generating functions in (\ref{blobsector})--(\ref{unblobsector}).

\begin{figure}
	\captionsetup{width=0.3\textwidth}
\centering
\begin{minipage}{.35\textwidth}
\includegraphics[width=.9\linewidth]{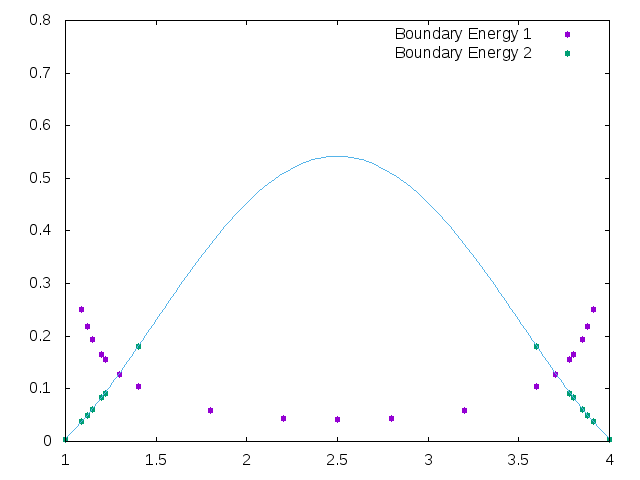}
 \captionof{figure}{The two boundary energies $f_s^1$ and $f_s^2$ plotted vs $r$, for $k=5$ in the $j=0$ sector. The critical values of $r_c$ are the points of intersection of the two curves.}
 \label{fsvsrk5}
 \end{minipage}%
\begin{minipage}{.35\textwidth}
\includegraphics[width=.9\linewidth]{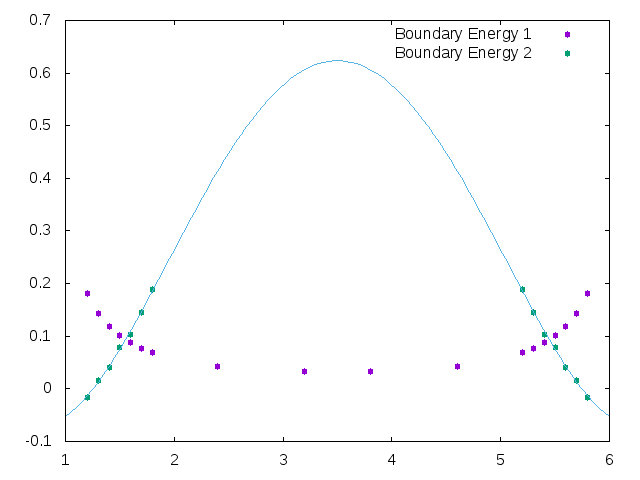}
\captionof{figure}{The two boundary energies $f_s^1$ and $f_s^2$ plotted vs $r$, for $k=7$ in the $j=0$ sector. The critical values of $r_c$ are the points of intersection of the two curves.}
\label{fsvsrk7}
\end{minipage}%
\end{figure}

We can see this phenomenon illustrated quite clearly in Figures \ref{fsvsrk5}--\ref{fsvsrk7}. When the green line is below the purple line, the $r \in [r_c,k-r_c]$ regime, the spectrum is described by equations (\ref{blobsector}) and (\ref{unblobsector}). When the green line is above the purple line, the $r \notin [r_c,k-r_c]$ regime, the states corresponding to the green line no longer affect the low-energy part of the spectrum in the thermodynamic limit.

\begin{figure}
	\captionsetup{width=0.3\textwidth}
\centering
\begin{minipage}{.35\textwidth}
\includegraphics[width=.9\linewidth]{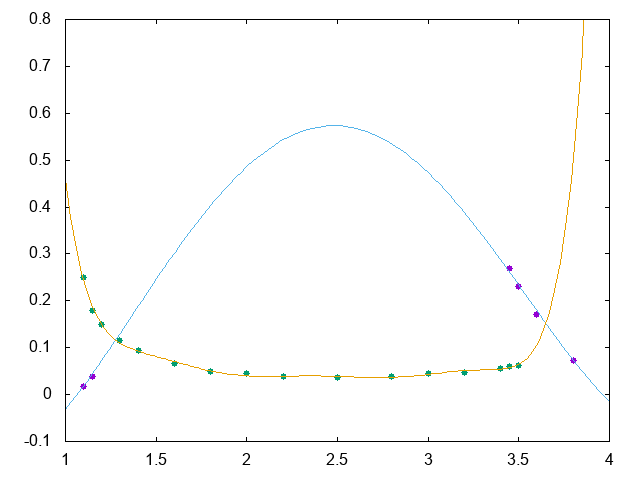}
 \captionof{figure}{The two boundary energies $f_s^1$ and $f_s^2$ plotted vs $r$, for $k=5$ in the $j=3$ blob sector. The critical values of $r_c$ are the points of intersection of the two curves.}
 \label{fsvsrblob}
 \end{minipage}%
\begin{minipage}{.35\textwidth}
\includegraphics[width=.9\linewidth]{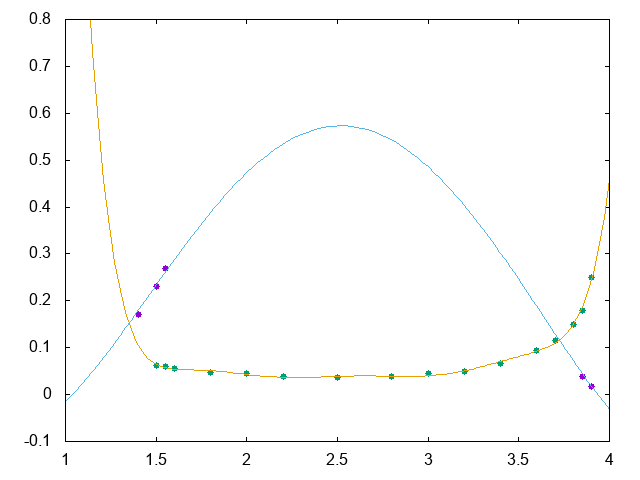}
\captionof{figure}{The two boundary energies $f_s^1$ and $f_s^2$ plotted vs $r$, for $k=5$ in the $j=3$ unblob sector. The critical values of $r_c$ are the points of intersection of the two curves.}
\label{fsvsrunblob}
\end{minipage}%
\end{figure}

Certainly the states whose boundary energy is the continuation of the boundary energy within the interval do not contribute, outside the interval, to partition functions in the scaling limit. This is because the BCFT is obtained after subtracting the non-universal ground state energy $f_s$ associated with a given boundary condition:%
\footnote{It is important to recall here that $f_s$ does in general depend on the boundary condition (while it is always set equal to zero in the BCFT). The point is that, in this particular case, we have two families of states, with the same boundary conditions, and  with different values of $f_s$.} states with a higher ground-state energy  are exponentially suppressed as $e^{-L(f_s'-f_s)}$ in the partition function.  It is not clear what these states possibly encode from a field theoretic point of view---that is, how the low-lying excitations above the ``wrong'' ground state might scale. We are not sure in particular  whether or not we would find the continuation of the generating functions observed for $r\in [r_c,k-r_c]$.  It seems nevertheless worth pointing out that there is a known case \cite{VernierColourings} of a {\em bulk} model possessing a continuous parameter, which can be adjusted so as to make two different {\em bulk} free energies cross; and on either side of this transition both the true ground state and the ``wrong'' ground state obtained by analytic continuation from the other side of the transition---each together with their low-energy excitations---behave as two different fully-fledged conformal field theories. We do not wish to rule out that the present model might provide a boundary analogue of this scenario.

Note that in Figures \ref{fsvsrblob}--\ref{fsvsrunblob} we see the same crossing phenomenon occurring when there are defect lines present in the system. In particular, Figures \ref{fsvsrblob}---\ref{fsvsrunblob} illustrate this for $j=3$. This is the scenario that we would expect from the connection between level-crossing and normalisability; the inequality ${1\over 2}<J<{k-1\over 2}$ with $2J\in\mathbb{N}$---recall the associations made in (\ref{paramcorr1})---for normalisable states suggests that the breakdown of the correspondence between the continuum limit of the lattice model and the discrete character is independent of $j$ and therefore should also occur for $j\neq 0$. Figures \ref{fsvsrblob}--\ref{fsvsrunblob} show that this indeed the case.

Meanwhile  we  note that we have not been able to interpret in a satisfactory way the structure of the excitations above the true ground state for $r\not\in[r_c,k-r_c]$ either: it is not clear whether they have anything to do with a boundary CFT any longer.

\subsection{A first-order boundary phase transition}

The level crossing observed at $r_c$ and $k-r_c$ can be interpreted as a first-order boundary phase transition, since it corresponds to a discontinuity of the derivative of the boundary free energy. Defining $f_s=-\frac{\log\mathcal{Z}}{M}$ where $M$ is the number of sites in the vertical direction of the lattice and $\mathcal{Z}$ is the partition function defined in (\ref{zalt2}), an easy calculation shows that

\beq\label{dfdr}
\frac{\partial f_s}{\partial r}=\frac{\gamma\sin\gamma}{M\sin(r\gamma)}\left[\frac{\langle \ell_1 \rangle}{\sin((r+1)\gamma)}-\frac{\langle \ell_2 \rangle}{\sin((r-1)\gamma)}  \right] \,,
\eeq
where $\langle \ell_1 \rangle$ and $\langle \ell_2 \rangle$ are the expectation values of the number of contractible blobbed and contractible unblobbed loops respectively. We believe that,  as we cross the critical value $r_c$ (or $k-r_c$),  we go from a situation where the ground state is dominated by entropic considerations and $\langle \ell_2 \rangle$ is finite to a situation where the ground state is dominated by energy considerations and $\ell_2$ is zero in the ground state. \\

\subsection{Extended characters}

The generating functions of levels correspond to  traces of powers of the transfer matrix, and as such describe partition functions of the underlying staggered six-vertex model (which will be discussed more in our next paper). These same functions are closely related to, but different from, the physical partition functions of the Potts model itself, which are obtained as certain modified traces. In the loop-model context, this situation corresponds to identifying the top and bottom sides of the lattice, while giving definite weights to the contractible and non-contractible loops obtained under this gluing. This construction has been discussed quite generally in \cite{JS-Blob}. If non-contractible blobbed and unblobbed loops are given the respective weights
\begin{equation}
\ell = 2\cosh\alpha \quad \mbox{and} \quad m={\sinh(\alpha+\beta)\over \sinh \beta} \,,
\end{equation}
the partition function reads
\begin{equation}
Z=\sum_{j=0}^\infty {\sinh(2j\alpha+\beta)\over\sinh\beta}\hbox{Tr}_{W_j^b}-\sum_{j=1}^\infty {\sinh(2j\alpha-\beta)\over\sinh\beta}
\hbox{Tr}_{W_j^u} \,,
\end{equation}
where ${\rm Tr}$ denotes the usual trace within the standard modules of the blob algebra.
Inserting our results for the blobbed and unblobbed partition functions with alt boundary conditions on the left side, we find
\begin{eqnarray}
Z&=&\sum_{j=0}^\infty {\sinh(2j\alpha+\beta)\over\sinh\beta}\lambda_{r,j}-\sum_{j=1}^\infty {\sinh(2j\alpha-\beta)\over\sinh\beta}
\lambda_{k-r,j}\nonumber\\
&=&\sum_{j=0}^\infty {\sinh(\beta+2j\alpha)\over\sinh\beta}\lambda^d_{r/2,j}+\sum_{j=1}^\infty {\sinh(\beta-2j\alpha)\over\sinh\beta}
\lambda^d_{r/2,-j}\nonumber\\
&=&\sum_{j=-\infty}^\infty {\sinh(\beta+2j\alpha)\over\sinh\beta}\lambda^d_{r/2,j} \,,
\end{eqnarray}
where in the second equality we have used the correspondences (\ref{paramcorr}) and (\ref{paramcorr1}) with the discrete $SL(2,\mathbb{R})/U(1)$ characters.
It follows that the generating functions of levels obtained previously are just what is needed to calculate the partition functions of the loop models as well. 
If we now restrict to the Potts model  per se, for which $\ell = \sqrt{Q}=\q+\q^{-1}$, so $\alpha=i\gamma$, and specialise also to the case $k$ integer,  we find that the sum can be split:
\begin{eqnarray}
Z=\sum_{j=0}^{k-1} {\sinh(\beta+2ij\gamma)\over\sinh\beta}\sum_{n=-\infty}^\infty \lambda^d_{r/2,j+nk}\label{physZ} \,.
\end{eqnarray}
Introduce now the  ``extended characters''  \cite{Troostetal} 
\begin{equation}
\Lambda_{J,M}=\sum_{n\in\mathbb{Z}} \lambda^d_{J,M+nk} \,.
\end{equation}
Recall that for $M+nk\geq 0$ we have $\lambda^d_{J,M+nk}=\lambda_{r,j}$, with $J={r\over 2}$ and $M+nk=j$. When $M+nk<0$ we have instead $\lambda^d_{J,M+nk}=\lambda_{k-r,-M-nk}$. Hence, starting with $j\in [0,k[$ we find
for the sum in (\ref{physZ}). 
\begin{equation}
Z=\sum_{M=0}^{k-1} {\sinh(\beta+2i\gamma M)\over\sinh\beta}\Lambda_{J,M}
\end{equation}
with, as usual, $J={r\over 2}$ and $M=j$. Similar identities would hold with free boundary conditions on both sides, or alt on both sides, a case to which we turn now. 

\subsection{Combining alt boundary conditions}\label{altaltloop}

We now wish to consider the case where alt boundary conditions are imposed on both sides of the strip. The general situation is characterised by more parameters than previously. First, the parametrisation (\ref{blobloopweight}) of the alternatingly restricted number of Potts states on the boundary has to made independently for both boundaries. Instead of $r$, we thus have $r_1$ for the left boundary and $r_2$ for the right boundary. Second, the algebraic framework must be similarly extended, so as to have blob and unblob operators---denoted $b_1$, $b_2$ and $u_1=1-b_1$, $u_2=1-b_2$ respectively---for each side. The proper algebraic framework for this situation is called the two-boundary Temperley-Lieb (2BTL) algebra \cite{GierNichols,JS-Blob,DJS}. Third, with $2j > 0$ through-lines, we need to define four different sectors---denoted $bb$, $ub$, $bu$ and $uu$---where the left (resp.\ right) label specifies whether the leftmost (resp.\ rightmost) through-line carries the blob or unblob operator, $b_1$ or $u_1$ (resp.\ $b_2$ or $u_2$). 

Note that even though the lattice model allows continuous values of $r_1$ and $r_2$, the discrete character in equation (\ref{discCharcpos}) (which played the role of the generating function when ``alt'' was imposed on only one side of the system) is only defined for $2J\in\mathbb{N}$. From the correspondences in (\ref{paramcorr}) and (\ref{paramcorr1}) we have then that $r\in\mathbb{N}$ also. As we shall now see, the discrete character also arises when ``alt" is placed on both sides. We hence consider only the case $r_1$ and $r_2$ integer. Note that when $j=0$ the lattice model is more subtle since loops can touch both boundaries.%
\footnote{In particular, one can distinguish between four types of loops touching both boundaries, depending on whether they pick up $b$ or $u$ labels at either boundary.}
We leave the case $j=0$ as a problem for the future and in what follows we instead focus only on $j > 0$.

\begin{figure}[h]
	\centering
\begin{tikzpicture}[scale=0.8]
\draw[black,line width = 1pt](1,1)--(9,9);
\draw[black,line width = 1pt](3,1)--(9,7);
\draw[black,line width = 1pt](5,1)--(9,5);
\draw[black,line width = 1pt](7,1)--(9,3);

\draw[black,line width = 1pt](1,3)--(7,9);
\draw[black,line width = 1pt](1,5)--(5,9);
\draw[black,line width = 1pt](1,7)--(3,9);

\draw[black,line width = 1pt](9,1)--(1,9);
\draw[black,line width = 1pt](7,1)--(1,7);
\draw[black,line width = 1pt](5,1)--(1,5);
\draw[black,line width = 1pt](3,1)--(1,3);

\draw[black,line width = 1pt](9,3)--(3,9);
\draw[black,line width = 1pt](9,5)--(5,9);
\draw[black,line width = 1pt](9,7)--(7,9);

\filldraw[black] (1.32,2) circle (2.5pt);
\filldraw[black] (8.68,2) circle (2.5pt);

\draw[black,line width = 1pt](1.215,3.9)--(1.415,3.9);
\draw[black,line width = 1pt](1.215,3.9)--(1.215,4.1);
\draw[black,line width = 1pt](1.415,3.9)--(1.415,4.1);
\draw[black,line width = 1pt](1.215,4.1)--(1.415,4.1);

\draw[black,line width = 1pt](8.785,3.9)--(8.585,3.9);
\draw[black,line width = 1pt](8.785,3.9)--(8.785,4.1);
\draw[black,line width = 1pt](8.585,3.9)--(8.585,4.1);
\draw[black,line width = 1pt](8.785,4.1)--(8.585,4.1);

\filldraw[black] (1.32,6) circle (2.5pt);
\filldraw[black] (8.68,6) circle (2.5pt);

\draw[black,line width = 1pt](1.215,7.9)--(1.415,7.9);
\draw[black,line width = 1pt](1.215,7.9)--(1.215,8.1);
\draw[black,line width = 1pt](1.415,7.9)--(1.415,8.1);
\draw[black,line width = 1pt](1.215,8.1)--(1.415,8.1);

\draw[black,line width = 1pt](8.785,7.9)--(8.585,7.9);
\draw[black,line width = 1pt](8.785,7.9)--(8.785,8.1);
\draw[black,line width = 1pt](8.585,7.9)--(8.585,8.1);
\draw[black,line width = 1pt](8.785,8.1)--(8.585,8.1);

\filldraw[black] (2,2) circle (4pt);
\filldraw[black] (4,2) circle (4pt);
\filldraw[black] (6,2) circle (4pt);
\filldraw[black] (8,2) circle (4pt);

\filldraw[black] (2,4) circle (4pt);
\filldraw[black] (4,4) circle (4pt);
\filldraw[black] (6,4) circle (4pt);
\filldraw[black] (8,4) circle (4pt);

\filldraw[black] (2,6) circle (4pt);
\filldraw[black] (4,6) circle (4pt);
\filldraw[black] (6,6) circle (4pt);
\filldraw[black] (8,6) circle (4pt);

\filldraw[black] (2,8) circle (4pt);
\filldraw[black] (4,8) circle (4pt);
\filldraw[black] (6,8) circle (4pt);
\filldraw[black] (8,8) circle (4pt);

\draw[black,line width = 1pt] (1.5,1.5) .. controls (1.25,2) .. (1.5,2.5);
\draw[black,line width = 1pt] (1.5,3.5) .. controls (1.25,4) .. (1.5,4.5);
\draw[black,line width = 1pt] (1.5,5.5) .. controls (1.25,6) .. (1.5,6.5);
\draw[black,line width = 1pt] (1.5,7.5) .. controls (1.25,8) .. (1.5,8.5);

\draw[black,line width = 1pt] (8.5,1.5) .. controls (8.75,2) .. (8.5,2.5);
\draw[black,line width = 1pt] (8.5,3.5) .. controls (8.75,4) .. (8.5,4.5);
\draw[black,line width = 1pt] (8.5,5.5) .. controls (8.75,6) .. (8.5,6.5);
\draw[black,line width = 1pt] (8.5,7.5) .. controls (8.75,8) .. (8.5,8.5);

\draw[black,line width = 2pt](4,4)--(6,4);
\draw[black,line width = 2pt](2,6)--(4,6);
\draw[black,line width = 2pt](4,2)--(4,8);
\draw[black,line width = 2pt](6,2)--(6,4);

\draw[black,line width = 1pt] (1.5,1.5) .. controls (2,1.25) .. (2.5,1.5);
\draw[black,line width = 1pt] (3.5,1.5) .. controls (3.65,1.25) .. (3.75,1);
\draw[black,line width = 1pt] (4.5,1.5) .. controls (4.35,1.25) .. (4.25,1);
\draw[black,line width = 1pt] (5.5,1.5) .. controls (6,1.25) .. (6.5,1.5);
\draw[black,line width = 1pt] (7.5,1.5) .. controls (8,1.25) .. (8.5,1.5);

\draw[black,line width = 1pt] (2.5,1.5) .. controls (2.75,2) .. (2.5,2.5);
\draw[black,line width = 1pt] (3.5,1.5) .. controls (3.25,2) .. (3.5,2.5);
\draw[black,line width = 1pt] (4.5,1.5) .. controls (4.75,2) .. (4.5,2.5);
\draw[black,line width = 1pt] (5.5,1.5) .. controls (5.25,2) .. (5.5,2.5);
\draw[black,line width = 1pt] (6.5,1.5) .. controls (6.75,2) .. (6.5,2.5);
\draw[black,line width = 1pt] (7.5,1.5) .. controls (7.25,2) .. (7.5,2.5);

\draw[black,line width = 1pt] (1.5,2.5) .. controls (2,2.75) .. (2.5,2.5);
\draw[black,line width = 1pt] (1.5,3.5) .. controls (2,3.25) .. (2.5,3.5);
\draw[black,line width = 1pt] (3.5,2.5) .. controls (3.75,3) .. (3.5,3.5);
\draw[black,line width = 1pt] (4.5,2.5) .. controls (4.25,3) .. (4.5,3.5);
\draw[black,line width = 1pt] (5.5,2.5) .. controls (5.75,3) .. (5.5,3.5);
\draw[black,line width = 1pt] (6.5,2.5) .. controls (6.25,3) .. (6.5,3.5);
\draw[black,line width = 1pt] (7.5,2.5) .. controls (8,2.75) .. (8.5,2.5);
\draw[black,line width = 1pt] (7.5,3.5) .. controls (8,3.25) .. (8.5,3.5);

\draw[black,line width = 1pt] (2.5,3.5) .. controls (2.75,4) .. (2.5,4.5);
\draw[black,line width = 1pt] (3.5,3.5) .. controls (3.25,4) .. (3.5,4.5);
\draw[black,line width = 1pt] (4.5,3.5) .. controls (5,3.75) .. (5.5,3.5);
\draw[black,line width = 1pt] (4.5,4.5) .. controls (5,4.25) .. (5.5,4.5);
\draw[black,line width = 1pt] (6.5,3.5) .. controls (6.75,4) .. (6.5,4.5);
\draw[black,line width = 1pt] (7.5,3.5) .. controls (7.25,4) .. (7.5,4.5);

\draw[black,line width = 1pt] (1.5,4.5) .. controls (2,4.75) .. (2.5,4.5);
\draw[black,line width = 1pt] (1.5,5.5) .. controls (2,5.25) .. (2.5,5.5);
\draw[black,line width = 1pt] (3.5,4.5) .. controls (3.75,5) .. (3.5,5.5);
\draw[black,line width = 1pt] (4.5,4.5) .. controls (4.25,5) .. (4.5,5.5);
\draw[black,line width = 1pt] (5.5,4.5) .. controls (6,4.75) .. (6.5,4.5);
\draw[black,line width = 1pt] (5.5,5.5) .. controls (6,5.25) .. (6.5,5.5);
\draw[black,line width = 1pt] (7.5,4.5) .. controls (8,4.75) .. (8.5,4.5);
\draw[black,line width = 1pt] (7.5,5.5) .. controls (8,5.25) .. (8.5,5.5);

\draw[black,line width = 1pt] (2.5,6.5) .. controls (3,6.25) .. (3.5,6.5);
\draw[black,line width = 1pt] (2.5,5.5) .. controls (3,5.75) .. (3.5,5.5);
\draw[black,line width = 1pt] (4.5,5.5) .. controls (4.75,6) .. (4.5,6.5);
\draw[black,line width = 1pt] (5.5,5.5) .. controls (5.25,6) .. (5.5,6.5);
\draw[black,line width = 1pt] (6.5,5.5) .. controls (6.75,6) .. (6.5,6.5);
\draw[black,line width = 1pt] (7.5,5.5) .. controls (7.25,6) .. (7.5,6.5);

\draw[black,line width = 1pt] (1.5,6.5) .. controls (2,6.75) .. (2.5,6.5);
\draw[black,line width = 1pt] (1.5,7.5) .. controls (2,7.25) .. (2.5,7.5);
\draw[black,line width = 1pt] (3.5,6.5) .. controls (3.75,7) .. (3.5,7.5);
\draw[black,line width = 1pt] (4.5,6.5) .. controls (4.25,7) .. (4.5,7.5);
\draw[black,line width = 1pt] (5.5,6.5) .. controls (6,6.75) .. (6.5,6.5);
\draw[black,line width = 1pt] (5.5,7.5) .. controls (6,7.25) .. (6.5,7.5);
\draw[black,line width = 1pt] (7.5,6.5) .. controls (8,6.75) .. (8.5,6.5);
\draw[black,line width = 1pt] (7.5,7.5) .. controls (8,7.25) .. (8.5,7.5);

\draw[black,line width = 1pt] (2.5,7.5) .. controls (2.75,8) .. (2.5,8.5);
\draw[black,line width = 1pt] (3.5,7.5) .. controls (3.25,8) .. (3.5,8.5);
\draw[black,line width = 1pt] (4.5,7.5) .. controls (4.75,8) .. (4.5,8.5);
\draw[black,line width = 1pt] (5.5,7.5) .. controls (5.25,8) .. (5.5,8.5);
\draw[black,line width = 1pt] (6.5,7.5) .. controls (6.75,8) .. (6.5,8.5);
\draw[black,line width = 1pt] (7.5,7.5) .. controls (7.25,8) .. (7.5,8.5);

\draw[black,line width = 1pt] (1.5,8.5) .. controls (2,8.75) .. (2.5,8.5);
\draw[black,line width = 1pt] (3.5,8.5) .. controls (3.65,8.75) .. (3.75,9);
\draw[black,line width = 1pt] (4.5,8.5) .. controls (4.35,8.75) .. (4.25,9);
\draw[black,line width = 1pt] (5.5,8.5) .. controls (6,8.75) .. (6.5,8.5);
\draw[black,line width = 1pt] (7.5,8.5) .. controls (8,8.75) .. (8.5,8.5);

\end{tikzpicture}

\caption{Alt on both sides: correlated boundary conditions in the bu sector with $j=1$.}\label{altaltcorr}
\end{figure}
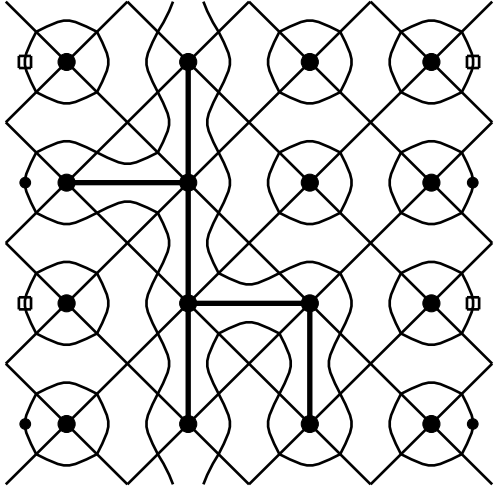

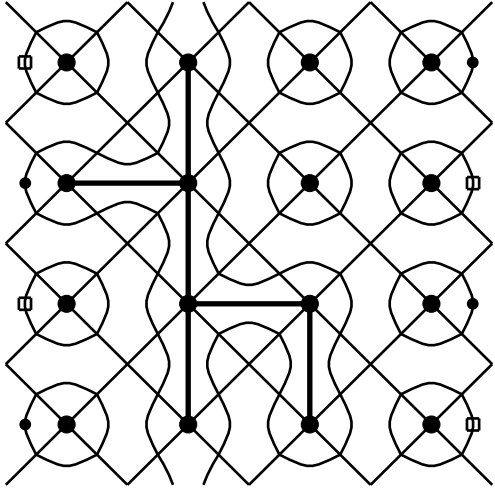
\begin{figure}[h]
	\centering
\begin{tikzpicture}[scale=0.8]
\draw[black,line width = 1pt](1,1)--(9,9);
\draw[black,line width = 1pt](3,1)--(9,7);
\draw[black,line width = 1pt](5,1)--(9,5);
\draw[black,line width = 1pt](7,1)--(9,3);

\draw[black,line width = 1pt](1,3)--(7,9);
\draw[black,line width = 1pt](1,5)--(5,9);
\draw[black,line width = 1pt](1,7)--(3,9);

\draw[black,line width = 1pt](9,1)--(1,9);
\draw[black,line width = 1pt](7,1)--(1,7);
\draw[black,line width = 1pt](5,1)--(1,5);
\draw[black,line width = 1pt](3,1)--(1,3);

\draw[black,line width = 1pt](9,3)--(3,9);
\draw[black,line width = 1pt](9,5)--(5,9);
\draw[black,line width = 1pt](9,7)--(7,9);

\filldraw[black] (1.32,2) circle (2.5pt);
\filldraw[black] (8.68,4) circle (2.5pt);

\draw[black,line width = 1pt](1.215,3.9)--(1.415,3.9);
\draw[black,line width = 1pt](1.215,3.9)--(1.215,4.1);
\draw[black,line width = 1pt](1.415,3.9)--(1.415,4.1);
\draw[black,line width = 1pt](1.215,4.1)--(1.415,4.1);

\draw[black,line width = 1pt](8.785,5.9)--(8.585,5.9);
\draw[black,line width = 1pt](8.785,5.9)--(8.785,6.1);
\draw[black,line width = 1pt](8.585,5.9)--(8.585,6.1);
\draw[black,line width = 1pt](8.785,6.1)--(8.585,6.1);

\filldraw[black] (1.32,6) circle (2.5pt);
\filldraw[black] (8.68,8) circle (2.5pt);

\draw[black,line width = 1pt](1.215,7.9)--(1.415,7.9);
\draw[black,line width = 1pt](1.215,7.9)--(1.215,8.1);
\draw[black,line width = 1pt](1.415,7.9)--(1.415,8.1);
\draw[black,line width = 1pt](1.215,8.1)--(1.415,8.1);

\draw[black,line width = 1pt](8.785,1.9)--(8.585,1.9);
\draw[black,line width = 1pt](8.785,1.9)--(8.785,2.1);
\draw[black,line width = 1pt](8.585,1.9)--(8.585,2.1);
\draw[black,line width = 1pt](8.785,2.1)--(8.585,2.1);

\filldraw[black] (2,2) circle (4pt);
\filldraw[black] (4,2) circle (4pt);
\filldraw[black] (6,2) circle (4pt);
\filldraw[black] (8,2) circle (4pt);

\filldraw[black] (2,4) circle (4pt);
\filldraw[black] (4,4) circle (4pt);
\filldraw[black] (6,4) circle (4pt);
\filldraw[black] (8,4) circle (4pt);

\filldraw[black] (2,6) circle (4pt);
\filldraw[black] (4,6) circle (4pt);
\filldraw[black] (6,6) circle (4pt);
\filldraw[black] (8,6) circle (4pt);

\filldraw[black] (2,8) circle (4pt);
\filldraw[black] (4,8) circle (4pt);
\filldraw[black] (6,8) circle (4pt);
\filldraw[black] (8,8) circle (4pt);

\draw[black,line width = 1pt] (1.5,1.5) .. controls (1.25,2) .. (1.5,2.5);
\draw[black,line width = 1pt] (1.5,3.5) .. controls (1.25,4) .. (1.5,4.5);
\draw[black,line width = 1pt] (1.5,5.5) .. controls (1.25,6) .. (1.5,6.5);
\draw[black,line width = 1pt] (1.5,7.5) .. controls (1.25,8) .. (1.5,8.5);

\draw[black,line width = 1pt] (8.5,1.5) .. controls (8.75,2) .. (8.5,2.5);
\draw[black,line width = 1pt] (8.5,3.5) .. controls (8.75,4) .. (8.5,4.5);
\draw[black,line width = 1pt] (8.5,5.5) .. controls (8.75,6) .. (8.5,6.5);
\draw[black,line width = 1pt] (8.5,7.5) .. controls (8.75,8) .. (8.5,8.5);

\draw[black,line width = 2pt](4,4)--(6,4);
\draw[black,line width = 2pt](2,6)--(4,6);
\draw[black,line width = 2pt](4,2)--(4,8);
\draw[black,line width = 2pt](6,2)--(6,4);

\draw[black,line width = 1pt] (1.5,1.5) .. controls (2,1.25) .. (2.5,1.5);
\draw[black,line width = 1pt] (3.5,1.5) .. controls (3.65,1.25) .. (3.75,1);
\draw[black,line width = 1pt] (4.5,1.5) .. controls (4.35,1.25) .. (4.25,1);
\draw[black,line width = 1pt] (5.5,1.5) .. controls (6,1.25) .. (6.5,1.5);
\draw[black,line width = 1pt] (7.5,1.5) .. controls (8,1.25) .. (8.5,1.5);

\draw[black,line width = 1pt] (2.5,1.5) .. controls (2.75,2) .. (2.5,2.5);
\draw[black,line width = 1pt] (3.5,1.5) .. controls (3.25,2) .. (3.5,2.5);
\draw[black,line width = 1pt] (4.5,1.5) .. controls (4.75,2) .. (4.5,2.5);
\draw[black,line width = 1pt] (5.5,1.5) .. controls (5.25,2) .. (5.5,2.5);
\draw[black,line width = 1pt] (6.5,1.5) .. controls (6.75,2) .. (6.5,2.5);
\draw[black,line width = 1pt] (7.5,1.5) .. controls (7.25,2) .. (7.5,2.5);

\draw[black,line width = 1pt] (1.5,2.5) .. controls (2,2.75) .. (2.5,2.5);
\draw[black,line width = 1pt] (1.5,3.5) .. controls (2,3.25) .. (2.5,3.5);
\draw[black,line width = 1pt] (3.5,2.5) .. controls (3.75,3) .. (3.5,3.5);
\draw[black,line width = 1pt] (4.5,2.5) .. controls (4.25,3) .. (4.5,3.5);
\draw[black,line width = 1pt] (5.5,2.5) .. controls (5.75,3) .. (5.5,3.5);
\draw[black,line width = 1pt] (6.5,2.5) .. controls (6.25,3) .. (6.5,3.5);
\draw[black,line width = 1pt] (7.5,2.5) .. controls (8,2.75) .. (8.5,2.5);
\draw[black,line width = 1pt] (7.5,3.5) .. controls (8,3.25) .. (8.5,3.5);

\draw[black,line width = 1pt] (2.5,3.5) .. controls (2.75,4) .. (2.5,4.5);
\draw[black,line width = 1pt] (3.5,3.5) .. controls (3.25,4) .. (3.5,4.5);
\draw[black,line width = 1pt] (4.5,3.5) .. controls (5,3.75) .. (5.5,3.5);
\draw[black,line width = 1pt] (4.5,4.5) .. controls (5,4.25) .. (5.5,4.5);
\draw[black,line width = 1pt] (6.5,3.5) .. controls (6.75,4) .. (6.5,4.5);
\draw[black,line width = 1pt] (7.5,3.5) .. controls (7.25,4) .. (7.5,4.5);

\draw[black,line width = 1pt] (1.5,4.5) .. controls (2,4.75) .. (2.5,4.5);
\draw[black,line width = 1pt] (1.5,5.5) .. controls (2,5.25) .. (2.5,5.5);
\draw[black,line width = 1pt] (3.5,4.5) .. controls (3.75,5) .. (3.5,5.5);
\draw[black,line width = 1pt] (4.5,4.5) .. controls (4.25,5) .. (4.5,5.5);
\draw[black,line width = 1pt] (5.5,4.5) .. controls (6,4.75) .. (6.5,4.5);
\draw[black,line width = 1pt] (5.5,5.5) .. controls (6,5.25) .. (6.5,5.5);
\draw[black,line width = 1pt] (7.5,4.5) .. controls (8,4.75) .. (8.5,4.5);
\draw[black,line width = 1pt] (7.5,5.5) .. controls (8,5.25) .. (8.5,5.5);

\draw[black,line width = 1pt] (2.5,6.5) .. controls (3,6.25) .. (3.5,6.5);
\draw[black,line width = 1pt] (2.5,5.5) .. controls (3,5.75) .. (3.5,5.5);
\draw[black,line width = 1pt] (4.5,5.5) .. controls (4.75,6) .. (4.5,6.5);
\draw[black,line width = 1pt] (5.5,5.5) .. controls (5.25,6) .. (5.5,6.5);
\draw[black,line width = 1pt] (6.5,5.5) .. controls (6.75,6) .. (6.5,6.5);
\draw[black,line width = 1pt] (7.5,5.5) .. controls (7.25,6) .. (7.5,6.5);

\draw[black,line width = 1pt] (1.5,6.5) .. controls (2,6.75) .. (2.5,6.5);
\draw[black,line width = 1pt] (1.5,7.5) .. controls (2,7.25) .. (2.5,7.5);
\draw[black,line width = 1pt] (3.5,6.5) .. controls (3.75,7) .. (3.5,7.5);
\draw[black,line width = 1pt] (4.5,6.5) .. controls (4.25,7) .. (4.5,7.5);
\draw[black,line width = 1pt] (5.5,6.5) .. controls (6,6.75) .. (6.5,6.5);
\draw[black,line width = 1pt] (5.5,7.5) .. controls (6,7.25) .. (6.5,7.5);
\draw[black,line width = 1pt] (7.5,6.5) .. controls (8,6.75) .. (8.5,6.5);
\draw[black,line width = 1pt] (7.5,7.5) .. controls (8,7.25) .. (8.5,7.5);

\draw[black,line width = 1pt] (2.5,7.5) .. controls (2.75,8) .. (2.5,8.5);
\draw[black,line width = 1pt] (3.5,7.5) .. controls (3.25,8) .. (3.5,8.5);
\draw[black,line width = 1pt] (4.5,7.5) .. controls (4.75,8) .. (4.5,8.5);
\draw[black,line width = 1pt] (5.5,7.5) .. controls (5.25,8) .. (5.5,8.5);
\draw[black,line width = 1pt] (6.5,7.5) .. controls (6.75,8) .. (6.5,8.5);
\draw[black,line width = 1pt] (7.5,7.5) .. controls (7.25,8) .. (7.5,8.5);

\draw[black,line width = 1pt] (1.5,8.5) .. controls (2,8.75) .. (2.5,8.5);
\draw[black,line width = 1pt] (3.5,8.5) .. controls (3.65,8.75) .. (3.75,9);
\draw[black,line width = 1pt] (4.5,8.5) .. controls (4.35,8.75) .. (4.25,9);
\draw[black,line width = 1pt] (5.5,8.5) .. controls (6,8.75) .. (6.5,8.5);
\draw[black,line width = 1pt] (7.5,8.5) .. controls (8,8.75) .. (8.5,8.5);

\end{tikzpicture}

\caption{Alt on both sides: anti-correlated boundary conditions in the bu sector with $j=1$.}\label{altaltanticorr}
\end{figure}

Interestingly, in this framework there are two distinct ways to implement alt boundary conditions on both sides of the system. In Figure~\ref{altaltcorr} the blob operator on both the left and right boundaries acts on odd-numbered rows, while the unblob operator acts on even-numbered rows on both boundaries. We call this setup ``correlated boundary conditions''. The alternative to this is shown in Figure \ref{altaltanticorr}. Here the blob operator acts on odd rows on the left, and on even rows on the right---and vice versa for the unblob operators; we call this setup ``anti-correlated boundary conditions''.
In order to make sense of the continuum limit in terms of conformal field theory, we must consider correlated boundary conditions when the width of the lattice $L$ is even and anti-correlated boundary conditions when $L$ is odd, or anti-correlated boundary conditions when the width of the lattice $L$ is even and correlated boundary conditions when the width of the lattice $L$ is odd.
(Note that in Figures~\ref{altaltcorr}--\ref{altaltanticorr} our conventions are such that $L=4$). 

These two ways of implementing the alt boundary conditions on both sides give rise to two different generating functions in the continuum limit. The continuum limit must include the two generating functions found, as we have explained, by treating seperately the correlated and anti-correlated boundary conditions. Redefining now $r=\min(r_1,r_2)$,%
\footnote{That is, $r$ here has a different meaning than in the one-boundary case.}
extensive numerical studies and consistency arguments similar to those in \cite{DJS} lead to the following conjectures in the case $r_1,r_2$ integer and $j\neq 0$:
\begin{eqnarray}
\label{bb}
\hbox{Tr}_{\BW_j^{bb}} &\mapsto & \sum\limits_{n=0}^{\infty}\lambda_{r_1+r_2-1-2nr,j+nr}+\lambda_{|r_2-r_1|+1-2nr,j+(n+1)r-1} \,, \\
\label{ub}
\hbox{Tr}_{\BW_j^{ub}} &\mapsto& \sum\limits_{n=0}^{\infty}\lambda_{-r_1+r_2-1-2nr,j+nr}+\lambda_{-r_1+r_2+1-2r(n+1),j+(n+1)r-1} \,, \\
\label{bu}
\hbox{Tr}_{\BW_j^{bu}} &\mapsto& \sum\limits_{n=0}^{\infty}\lambda_{r_1-r_2-1-2nr,j+nr}+\lambda_{r_1-r_2+1-2r(n+1),j+(n+1)r-1} \,, \\
\label{uu}
\hbox{Tr}_{\BW_j^{uu}} &\mapsto& \sum\limits_{n=0}^{\infty}\lambda_{-r_1-r_2-1-2nr,j+nr}+\lambda_{-r_1-r_2+1-2r(n+1),j+(n+1)r-1} \,.
\end{eqnarray}

\section{From Potts AF to $Z_{k-2}$ parafermions for  $k$ integer}\label{RSOSkint}

\subsection{The case of free boundary conditions. RSOS truncation.}\label{rsostrunc}

It was also observed in \cite{S-AF} that when $\q$ is a primitive root of unity, so that $k$ is integer, the generating functions of levels belonging to simple representations (so-called type II in \cite{PasquierSaleurQG}) of $U_qsl(2)$, which can be expressed as usual using an alternating sum
\begin{equation} \label{strfct2}
c_l^0 = \sum\limits_{n=0}^{\infty} \left( K_{l+2nk}-K_{2(n+1)k-l-2} \right) = K_l-K_{2k-l-2}+K_{l+2k}-K_{4k-l-2}+\ldots \,,
\end{equation}
exactly coincide with the string function of the $Z_{k-2}$ symmetric CFT. The string function $c_l^0$ in (\ref{strfct2}) is a special case of the more general object \cite{Jayaraman}:
\beq
 c_l^m = \frac{1}{\eta(q)^2}\sum_{\substack{n_1, n_2 \in \mathbb{Z}/2  \\ n_1-n_2\in \mathbb{Z} \\ n_1 \geq |n_2|, -n_1 > |n_2| }}(-1)^{2n_1}\text{sign}(n_1)q^{\frac{(l+1+2n_1k)^2}{4k}-\frac{(m+2n_2(k-2))^2}{4(k-2)}} \,.
\eeq
Meanwhile, the alternating sum in (\ref{strfct2}) can be interpreted as the partition function of a staggered RSOS model where heights live on an $A_{k-1}$ Dynkin diagram, with heights on the left side of the strip fixed to 1, and those on the right side fixed to $l+1$ \cite{Pasquier6j}.
(Note that this partition function can also be interpreted as the trace over the simple submodule of the Temperley-Lieb algebra module $\StTL{j}$.)
This result, which will be discussed in more details below, strongly suggests that the RSOS version of the AF Potts model for $k$ integer coincides with the $Z_{k-2}$ parafermion theory. 
Our aim is now to return to the alt boundary conditions defined in section~\ref{altbcsec}, in order to explore their potential relationship with the string functions
missing in (\ref{strfct2}), viz., all those with $m\neq 0$. 

First, however, we must clarify what we mean by the ``RSOS version" of the AF Potts model. These RSOS models are not the same models as the RSOS models introduced by Andrews, Baxter and Forrester (ABF) \cite{ABF}. The AF Potts RSOS models are defined by assigning integer heights---or more precisely, nodes in
the associated Dynkin diagram---to the union of Potts spin vertices (solid circles in Figure~\ref{fulllattice}) and their duals. The RSOS constraint amounts, as usual,
to imposing that nearest-neighbour vertices (i.e., an adjacent pair of a Potts vertex and a dual vertex) carry nearest-neighbour heights. But unlike the ABF models,
the RSOS model associated with the AF Potts model has  Boltzmann weights that are ``staggered'', i.e., even and odd numbered tiles will have different
Boltzmann weights. These staggered weights ultimately stem from the alternating local weights $x_1$ and $x_2$ defined in (\ref{looppartition}).

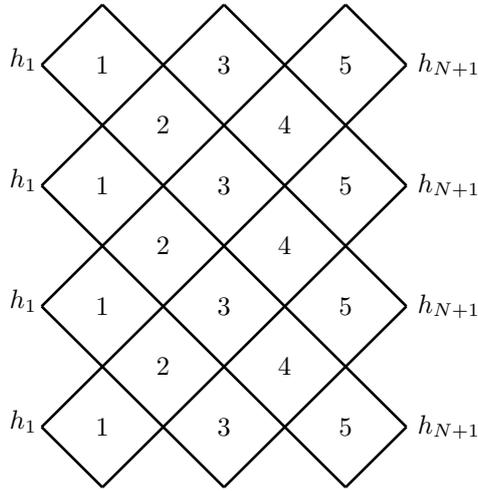
\begin{figure}[ht]
	\centering
\begin{tikzpicture}[scale=0.8]
\draw[black,line width = 1pt](2,2)--(8,8);
\draw[black,line width = 1pt](3,1)--(8,6);
\draw[black,line width = 1pt](5,1)--(8,4);
\draw[black,line width = 1pt](7,1)--(8,2);

\draw[black,line width = 1pt](2,4)--(7,9);
\draw[black,line width = 1pt](2,6)--(5,9);
\draw[black,line width = 1pt](2,8)--(3,9);

\draw[black,line width = 1pt](8,2)--(2,8);
\draw[black,line width = 1pt](7,1)--(2,6);
\draw[black,line width = 1pt](5,1)--(2,4);
\draw[black,line width = 1pt](3,1)--(2,2);

\draw[black,line width = 1pt](8,4)--(3,9);
\draw[black,line width = 1pt](8,6)--(5,9);
\draw[black,line width = 1pt](8,8)--(7,9);

\node at (3,2) {1};
\node at (5,2) {3};
\node at (7,2) {5};
\node at (4,3) {2};
\node at (6,3) {4};

\node at (3,4) {1};
\node at (5,4) {3};
\node at (7,4) {5};
\node at (4,5) {2};
\node at (6,5) {4};

\node at (3,6) {1};
\node at (5,6) {3};
\node at (7,6) {5};
\node at (4,7) {2};
\node at (6,7) {4};

\node at (3,8) {1};
\node at (5,8) {3};
\node at (7,8) {5};

\node at (1.7,2.1) {$h_1$};
\node at (1.7,4.1) {$h_1$};
\node at (1.7,6.1) {$h_1$};
\node at (1.7,8.1) {$h_1$};

\node at (8.7,2) {$h_{N+1}$};
\node at (8.7,4) {$h_{N+1}$};
\node at (8.7,6) {$h_{N+1}$};
\node at (8.7,8) {$h_{N+1}$};

\end{tikzpicture}
\caption{The staggered RSOS model. An integer heights lives on each vertex, here shown for a lattice of width $N=2L=6$. The numbers on the centre of the tiles are labels; in the staggered model, even and odd numbered tiles get different Boltzmann weights. When we pick the boundary heights $h_1=1$ and $h_{N+1}=1+l$, this RSOS model produces the string function $c^0_{l}$ in the continuum limit.}\label{rsoslattice}
\end{figure}

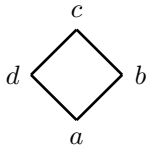
\begin{figure}[ht]
	\centering
\begin{tikzpicture}[scale=1.2]
\draw[black,line width = 1pt](1.5,2.5)--(2,3);
\draw[black,line width = 1pt](1.5,2.5)--(1,3);
\draw[black,line width = 1pt](1,3)--(1.5,3.5);
\draw[black,line width = 1pt](2,3)--(1.5,3.5);

\node at (1.5,2.3) {$a$};
\node at (2.2,3) {$b$};
\node at (1.5,3.7) {$c$};
\node at (0.8,3) {$d$};

\end{tikzpicture}

\caption{Four heights around a tile in the RSOS model}\label{RSOStile}
\end{figure}

To understand this in detail, consider the lattice in Figure~\ref{rsoslattice}. The degrees of freedom which live on the vertices of the lattice are integer heights
from the set $\{1,2,\ldots,k-1\}$---i.e., the Dynkin diagram $A_{k-1}$---where the parameter $k$ will be related to the loop weight $\ell = 2 \cos \gamma$ via
the relation $\gamma=\frac{\pi}{k}$. The heights attributed neighbouring vertices differ by $\pm 1$. Boltzmann weights are then
associated to each tile and depend on the heights at each of the tile's vertices; see Figure~\ref{RSOStile}.

Each tile in the lattice is labelled by an integer giving its position along the horizontal axis, as shown in Figure~\ref{rsoslattice}.
The staggering comes about by allowing the Boltzmann weights to depend on the parity of the tile label.
Specifically, to tiles with an even label we give the weight
\beq\label{ABFBoltzmann}
W(a,b,c,d)=\delta(a,c)+x_2\frac{\sqrt{S_a S_c}}{\sqrt{S_b S_d}}\delta(b,d) \,,
\eeq
whereas tiles with an odd label get the weight
\beq\label{ABFBoltzmannstag}
W(a,b,c,d)=x_1\delta(a,c)+\frac{\sqrt{S_a S_c}}{\sqrt{S_b S_d}}\delta(b,d) \,.
\eeq
In these expressions, the height values $a,b,c,d$ are associated with the vertices bordering the tile as shown in Figure~\ref{RSOStile},
and we have defined
\beq\label{Sadef}
S_a=\frac{\sin(\frac{a\pi}{k})}{\sin(\frac{\pi}{k})} \,.
\eeq
It is easy to see that the weights $x_1$ and $x_2$ are analogous to those appearing in Figure~\ref{fourtiles}. Moreover, the Kronecker deltas
are related with the line expansion in Figure~\ref{fulllattice}.

In the following we will mostly be interested in the isotropic case, $x_1=x_2=x$, where as before $x=\frac{e^K-1}{\sqrt{Q}}$. Note also that $x\leq 0$ along the antiferromagnetic critical line (since $Q\in[0,4]$), and hence some of the tiles will have negative Boltzmann weights.  

\begin{figure}[ht]
	\centering
\begin{tikzpicture}[scale=1.2]
	
\draw[black,line width = 1pt](1,1)--(1.5,1.5);
\draw[black,line width = 1pt](2,1)--(2.5,1.5);
\draw[black,line width = 1pt](3,1)--(3.5,1.5);
\draw[black,line width = 1pt](0.5,1.5)--(1,1);
\draw[black,line width = 1pt](1.5,1.5)--(2,1);
\draw[black,line width = 1pt](2.5,1.5)--(3,1);

\node at (0.5,1.7) {$h_1$};
\node at (1,1.3) {$h_2$};
\node at (1.5,1.7) {$h_3$};
\node at (2,1.3) {$h_4$};
\node at (2.5,1.7) {$h_5$};
\node at (3,1.3) {$h_6$};
\node at (3.5,1.7) {$h_7$};

\end{tikzpicture}

\caption{The state $\left| h_1h_2 h_3 h_4 h_5 h_6 h_7 \right\rangle$.}\label{RSOSrow}
\end{figure}
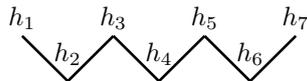

Note that we can write the Boltzmann weights in eqs.~(\ref{ABFBoltzmann})--(\ref{ABFBoltzmannstag}) in terms of a spectral parameter $u$. As in \cite{JS-AF} we have
\beq\label{x1x2}
\begin{aligned}
&x_1=\frac{\sin u}{\sin(\gamma-u)} \,, \\
&x_2=\frac{-\cos (\gamma-u)}{\cos(u)} \,. \\
\end{aligned}
\eeq
Using eqs.~(\ref{x1x2}), we can write the Boltzmann weights as components of an integrable $R$-matrix.  We define
\beq\label{Rmatu}
R(u)=\mathcal{I}+\frac{\sin(\gamma-u)}{\sin(u)}e_i \,,
\eeq
where $e_i$ are the TL generators acting in the RSOS representation, given explicitly by
\beq\label{tlrsos}
e_i \left| h_1,\ldots,h_{i-1},h_i,h_{i+1},\ldots,h_{N+1} \right\rangle =\delta(h_{i-1},h_{i+1})\sum\limits_{h_i'}\frac{\sqrt{S_{h_i}S_{h_i'}}}{S_{h_{i-1}}} \left| h_1,\ldots,h_{i-1},h_i',h_{i+1}, \ldots,h_{N+1} \right\rangle \,,
\eeq
where $\left| h_1,\ldots,h_{i-1},h_i,h_{i+1},\ldots,h_{N+1} \right\rangle$ is a state specifying the heights on the $N+1$ consecutive sites along one row of the lattice (see Figure~\ref{RSOSrow}). Clearly, eq.~(\ref{Rmatu}) recovers the Boltzmann weights (\ref{ABFBoltzmann}) for even-labelled tiles when we consider the $R$-matrix to live on these tiles. The shifted matrix $R(u-\frac{\pi}{2})$ then recovers the weights of odd-labelled tiles in (\ref{ABFBoltzmannstag}). We stress that even though we are using an integrable $R$-matrix to define these Boltzmann weights, the diagonal geometry of Figure \ref{rsoslattice} means that the model is not integrable. This staggering process was applied to the vertex representation in \cite{IJS2008}, the so-called ``staggered six-vertex model''.

\no Equation (\ref{strfct2}) describes the relationship between the AF Potts loop model and the AF Potts RSOS model and suggests that the string function $c^0_l$ can be realised on the lattice via the AF Potts RSOS model. It turns out that fixing the heights on the left boundary of the lattice to be $h_1 = 1$, and those on the right boundary to be $h_{N+1}=1+l$ indeed reproduces the full string function $c^0_l$ in the continuum limit. By symmetry of the Dynkin diagram, and of the weights in (\ref{tlrsos}), obviously the same result would be recovered for $h_1 = 1+l$ and $h_{N+1} = 1$.

We now wish to find other boundary conditions whose continuum limit produces all of the other string functions $c^m_l$, i.e., the cases $m\neq 0$. It will turn out that the ``alt'' boundary conditions in the AF Potts RSOS model will serve this purpose.

\subsection{Missing string functions and alt boundary conditions}\label{stringfunckint}

\color{black}
Of course the discrete character $\lambda^d_{J,M}$ pertains naturally to a theory with the central charge $c_{\rm BH}$ given by (\ref{cbh}) and conformal weights
\begin{equation}
h_{J,M}={(J+M)^2\over k}-{J(J-1)\over k-2} \,.
\end{equation}
Nonetheless it is also possible to interpret it formally within a theory with central charge $c_{\rm PF}$ given by (\ref{cafpotts}). In this case, the corresponding conformal weights become
\begin{equation}
\Delta_{J,M}=h_{J,M}-{1\over 4(k-2)}-{1\over 4k}={(J+M-1/2)(J+M+1/2)\over k}-{(J-1/2)^2\over k-2} \,.
\end{equation}
In the following we shall maintain this distinction, by denoting always the weights pertaining to the black-hole theory by $h$, and those related the parafermionic interpretation of the AF Potts model by $\Delta$. 
Assuming $M\geq 0$, and using the correspondence in (\ref{paramcorr}), this latter expression leads to the weight 
\begin{equation}
\Delta_{r/2,j}=\frac{(r-1+2j)(r+1+2j)}{4k}-\frac{(r-1)^2}{4(k-2)} \,,
\end{equation}
which can be matched with the parafermion exponents  $\Delta_l^m$ defined in eq.~(\ref{PFexp}) if we set 
\begin{eqnarray}
m&=&r-1 \,, \nonumber\\
l&=&r-1+2j \,. \label{identif}
\end{eqnarray}

This suggests we are on the right track to identifying the other string functions. Indeed, suppose now that $k$ is integer. Using the representation theory of the blob algebra, the simple top is formally obtained via
\begin{equation}
\BX_{r,j}^{b}=\bigoplus_{n=0}^\infty \left[\left(\BW_{j+nk}^b/\BW_{j+r+nk}^u\right) -\left(\BW^b_{k-j-r+nk}/\BW_{k-j+nk}^u\right)\right]\label{topstructure}
\end{equation}
The corresponding generating function of levels  for the alt boundary conditions is then
\begin{eqnarray}\label{discretechar}
\hbox{Tr}_{\BX_{r,j}^{b}}q^{L_0-c/24}&=&\sum_{n=0}^\infty\left[ \lambda_{m+1,{l-m\over 2}+nk}-\lambda_{k-m-1, {l+m+2\over 2}+nk}
-\lambda_{m+1,{2k-l-m-2\over 2}+nk}+\lambda_{k-m-1, {2k-l+m\over 2}+nk}\right]\nonumber\\
&=& \sum_{p=-\infty}^\infty\left[ \lambda^d_{{m+1\over 2},{l-m\over 2}+pk}-\lambda^d_{{m+1\over 2},-{l+m+2\over 2}+pk}\right]
\end{eqnarray}
This can be reformulated as (see eq.~(\ref{stringJaya}) in the Appendix)
\begin{equation}
\hbox{Tr}_{\BX_{r,j}^{b}}q^{L_0-c/24}=\sum_{n=0}^\infty\left[ \hbox{Tr}_{\tilde{F}_{l+2nk,m}}q^{L_0-c/24}-\hbox{Tr}_{\tilde{F}_{2k-l-2+2nk,m}}q^{L_0-c/24}\right]=c_l^m \,, \mbox{ for } m\leq l \label{stringidentif}
\end{equation}
with the correspondence (\ref{identif}). As the notation indicates, this expression  coincides with  the string function for the $Z_{k-2}$ parafermion theory (see the Appendix for more detail and formulas).%
\footnote{Sometimes string functions are defined with an extra factor $\eta(q)$, so what we denote $c_l^m$ is referred to as $\eta(q) c_l^m$.}
Note that since $j$ is integer, $l$ and $m$ have the same parity, as required.  Note finally that the generating functions $\tilde{F}_{l,m}$ are invariant under $m\to -m$, so we get as well the string functions $c_l^{-m}=c_l^m$. 

We can now study the range of values of the parameters. For  $\q=e^{i\pi/k}$, the simplest case is  $r=1$, for which we have to consider in fact only the TL algebra. The possible values of $j$ are given by $j=0,\ldots,{k\over 2}-1$, so 
\begin{equation}
l=2j=0,\ldots,k-2 \,. \label{firstrange}
\end{equation}
For $r=2,\ldots k-2$, we see first, from $m=r-1$, that
\begin{equation}
m=1,\ldots,k-3 \,. \label{secondrange}
\end{equation}
As far as $l$ is concerned, we have to invoke the representation theory of the blob algebra. For a given value of $r$, we have now the allowed values of $j$ in the blobbed sector, $j=0,\ldots,{k-1-r\over 2}$,  which correspond to 
\begin{equation}
l=r-1+2j=m,\ldots,k-2\label{thirdrange}
\end{equation}
To get the missing values of $l$ we need to consider the unblobbed sector. We have first the equivalent  of (\ref{topstructure})
\begin{equation}
\BX_{r,j}^{u}=\bigoplus_{n=0}^\infty\left[ \left(\BW_{j+nk}^u/\BW_{r-j+nk}^u\right) -\left(\BW^b_{k-r+j+nk}/\BW_{k-j+nk}^b\right)\right]=c^m_{2m-l} \,, \mbox{ for } m\leq l \,. \label{topstructure}
\end{equation}
Indeed, straightforward calculation first leads to an identity similar to (\ref{stringidentif}) but with $c_{k-2+l-2m}^{k-2-m}$, following from the transformation $r\to k-r$. Using a  standard identity for string functions $c_l^m=c_{k-2-l}^{k-2-m}$ in $Z_{k-2}$ theories \cite{GepnerQiu}, gives instead  the string function $c_{2m-l}^m$. For the unblobbed sector, meanwhile, the allowed values of $j$ are $j=0,\ldots, {r-1\over 2}$, so $l=r-1+2j=r-1,\ldots,2(r-1)=m,\ldots,2m$. It follows that $2m-l=0,\ldots m$, recovering what was missing in (\ref{thirdrange}) to cover the whole set%
\footnote{Note that $l=m$ is common to both sets, since it corresponds to $j=0$ for which there is no distinction between blobbed and unblobbed.}
\begin{equation}
l=0,\ldots,k-2
\end{equation}
as in (\ref{firstrange}).

Equations (\ref{discretechar}) and (\ref{stringidentif}) tell us that the alt boundary conditions in the AF Potts RSOS model should produce the string functions $c^m_l$ in the continuum limit. We now describe what precisely ``alt'' means for the RSOS model.

\medskip

 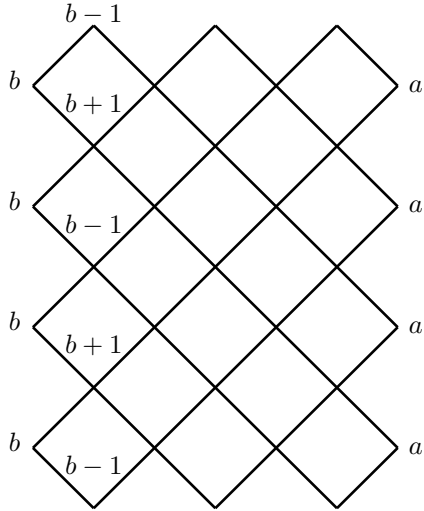
\begin{figure}
 	\centering
 \begin{tikzpicture}[scale=0.8]
 \draw[black,line width = 1pt](2,2)--(8,8);
 \draw[black,line width = 1pt](3,1)--(8,6);
 \draw[black,line width = 1pt](5,1)--(8,4);
 \draw[black,line width = 1pt](7,1)--(8,2);

 \draw[black,line width = 1pt](2,4)--(7,9);
 \draw[black,line width = 1pt](2,6)--(5,9);
 \draw[black,line width = 1pt](2,8)--(3,9);

 \draw[black,line width = 1pt](8,2)--(2,8);
 \draw[black,line width = 1pt](7,1)--(2,6);
 \draw[black,line width = 1pt](5,1)--(2,4);
 \draw[black,line width = 1pt](3,1)--(2,2);

 \draw[black,line width = 1pt](8,4)--(3,9);
 \draw[black,line width = 1pt](8,6)--(5,9);
 \draw[black,line width = 1pt](8,8)--(7,9);

 \node at (1.7,2.1) {$b$};
 \node at (1.7,4.1) {$b$};
 \node at (1.7,6.1) {$b$};
 \node at (1.7,8.1) {$b$};

 \node at (3,1.7) {$b-1$};
 \node at (3,3.7) {$b+1$};
 \node at (3,5.7) {$b-1$};
 \node at (3,7.7) {$b+1$};
 \node at (3,9.2) {$b-1$};

 \node at (8.3,2) {$a$};
 \node at (8.3,4) {$a$};
 \node at (8.3,6) {$a$};
 \node at (8.3,8) {$a$};

 \end{tikzpicture}
 \caption{Alternating boundary conditions in the antiferromagnetic RSOS model. We will write this boundary condition as $b^{\pm},\ldots,a$.}\label{rsosaltlattice}
 \end{figure}
 
Consider the RSOS boundary conditions on the lattice in figure \ref{rsosaltlattice}. Heights are fixed to $h_{N+1} = a$ on the right boundary, to $h_1 = b$ on the left boundary, and heights $h_2$ next to the left boundary alternate between $b\pm 1$ as shown. We will write this boundary condition as $b^{\pm},\ldots,a$. It is found that the boundary condition:
\beq
(m+1)^{\pm},...,(l+1)
\eeq
produces the string function $c^m_l$ in the continuum limit ($m\neq 0$). Similarly, we will write $1^+,\ldots,(l+1)$
to denote the boundary condition in Figure \ref{rsoslattice} in section \ref{rsostrunc}---i.e., with a constant, non-alternating value of $h_2$---that produces the string function $c^0_l$.%
\footnote{Note that since heights are restricted to be between $1$ and $k-1$ we cannot have an ``alt'' condition on a boundary where the boundary height is fixed as $h_1=1$.
The superscript in $1^+,\ldots$ is actually redundant, since $h_2=2$ then follows from the RSOS constraint alone.}
Using this notation, Table \ref{rsosafk6} shows the exact correspondence between string functions and RSOS boundary conditions for the case $k=6$. The generating functions (i.e., the string functions) are written up to the number of terms that we have observed by the numerical study of the lattice model.

\begin{table}
\begin{center}
\begin{tabular}{ c | c | l}
Boundary condition & Exponent & Generating function\\
\hline
$1^+,...,1$ & 0 & $c^{m=0}_{l=0}=q^{h-\frac{c}{24}}(1+q^2+2q^3+\ldots)$ \\[1.1ex]
\hline
$1^+,...,3$ & $\frac{1}{3}$ & $c^{m=0}_{l=2}=q^{h-\frac{c}{24}}(1+2q+3q^2+5q^3+\ldots)$ \\[1.1ex]
\hline
$1^+,...,5$ & 1 & $c^{m=0}_{l=4}=q^{h-\frac{c}{24}}(1+q+3q^2+3q^3+\ldots)$ \\[1.1ex]
\hline
$2^{\pm}$,...,2 & $\frac{1}{16}$ & $c^{m=1}_{l=1}=q^{h-\frac{c}{24}}(1+q+2q^2+4q^3+\ldots)$ \\[1.1ex]
\hline
$2^{\pm}$,...,4  & $\frac{9}{16}$ & $c^{m=1}_{l=3}=q^{h-\frac{c}{24}}(1+2q+3q^2+5q^3+\ldots)$\\[1.1ex]
\hline
$3^{\pm}$,...,1  & $\frac{3}{4}$ & $c^{m=2}_{l=0}=q^{h-\frac{c}{24}}(1+q+2q^2+3q^3+\ldots)$\\[1.1ex]
\hline
$3^{\pm}$,...,3 & $\frac{1}{12}$ & $c^{m=2}_{l=2}=q^{h-\frac{c}{24}}(1+q+3q^2+4q^3+\ldots)$\\[1.1ex]
\hline
$3^{\pm}$,...,5  & $\frac{3}{4}$ & $c^{m=2}_{l=4}=q^{h-\frac{c}{24}}(1+q+2q^2+3q^3+\ldots)$\\[1.1ex]
\hline
$4^{\pm}$,...,2  & $\frac{9}{16}$ & $c^{m=3}_{l=1}=q^{h-\frac{c}{24}}(1+2q+3q^2+5q^3+\ldots)$\\[1.1ex]
\hline
$4^{\pm}$,...,4  & $\frac{1}{16}$ & $c^{m=3}_{l=3}=q^{h-\frac{c}{24}}(1+q+2q^2+4q^3+\ldots)$\\[1.1ex]
\end{tabular}
\caption{String functions in the $k=6$ antiferromagnetic RSOS model. The string functions are expanded up to the terms we can clearly observe on the lattice.}\label{rsosafk6}
\end {center}
\end{table}

\subsection{The case of alt on both sides: fusion of string functions}

We have found that the continuum limit of the AF Potts RSOS model coincides with that of the $Z_{k-2}$ parafermion theory and we have found an ``alt'' boundary condition corresponding to each of the string functions in these models. This prescription, however, was restricted to the case where the alt boundary condition is on one side only with the other side having free boundary conditions. We would expect that putting the alt boundary conditions on both sides of the lattice would correspond to the fusion of fields in the $Z_{k-2}$ parafermion theory. As will be shown below from our numerical results, this is indeed the case.

\medskip

We can however recover this result from knowledge of the generating functions produced in the continuum limit of the loop model with the alt conditions on both sides; see eqs.~(\ref{bb})--(\ref{uu}). Sections \ref{rsostrunc} and \ref{stringfunckint} used the representation theory of the blob algebra to move between the loop model and the RSOS model; the generating function of the irreducible representation of the blob algebra created by the infinite sum in eq.~(\ref{discretechar}) produced the RSOS representation. We can use the same method for the case with the alt condition on both sides, i.e., when there are two blob operators, to calculate the generating functions in the RSOS model produced by putting the alt condition on both sides. The relevant algebra in this case is the two-boundary Temperley-Lieb (2BTL) algebra \cite{GierNichols,JS-Blob,DJS}.

\medskip

Section \ref{rsosfusionnumerics} will present the numerical results of the RSOS model with the alternating boundary condition on both sides. These results will be interpreted in terms of the fusion of fields in $Z_{k-2}$ parafermion theory. Section \ref{loopfusionstring} will recover these results by studying the representation theory of the 2BTL algebra.

\subsubsection{Alt on both sides in the RSOS model: Numerics}\label{rsosfusionnumerics}

 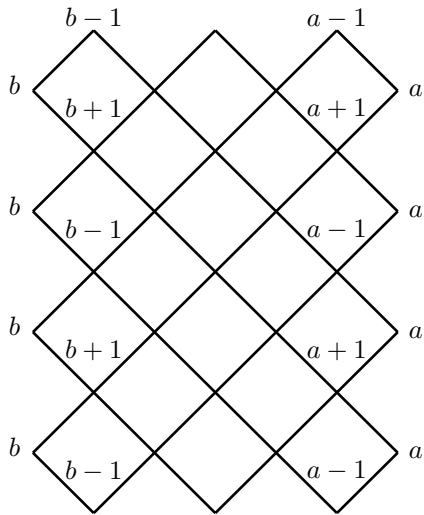
\begin{figure}
 	\centering
 \begin{tikzpicture}[scale=0.8]
 \draw[black,line width = 1pt](2,2)--(8,8);
 \draw[black,line width = 1pt](3,1)--(8,6);
 \draw[black,line width = 1pt](5,1)--(8,4);
 \draw[black,line width = 1pt](7,1)--(8,2);

 \draw[black,line width = 1pt](2,4)--(7,9);
 \draw[black,line width = 1pt](2,6)--(5,9);
 \draw[black,line width = 1pt](2,8)--(3,9);

 \draw[black,line width = 1pt](8,2)--(2,8);
 \draw[black,line width = 1pt](7,1)--(2,6);
 \draw[black,line width = 1pt](5,1)--(2,4);
 \draw[black,line width = 1pt](3,1)--(2,2);

 \draw[black,line width = 1pt](8,4)--(3,9);
 \draw[black,line width = 1pt](8,6)--(5,9);
 \draw[black,line width = 1pt](8,8)--(7,9);

 \node at (1.7,2.1) {$b$};
 \node at (1.7,4.1) {$b$};
 \node at (1.7,6.1) {$b$};
 \node at (1.7,8.1) {$b$};

 \node at (3,1.7) {$b-1$};
 \node at (3,3.7) {$b+1$};
 \node at (3,5.7) {$b-1$};
 \node at (3,7.7) {$b+1$};
 \node at (3,9.2) {$b-1$};

 \node at (7,1.7) {$a-1$};
 \node at (7,3.7) {$a+1$};
 \node at (7,5.7) {$a-1$};
  \node at (7,7.7) {$a+1$};
   \node at (7,9.2) {$a-1$};

 \node at (8.3,2) {$a$};
 \node at (8.3,4) {$a$};
 \node at (8.3,6) {$a$};
 \node at (8.3,8) {$a$};

 \end{tikzpicture}
 \caption{Alternating boundary conditions on both sides. We will write this boundary condition $b^{\pm},...,^{\pm} \! a$  and refer to it as ``correlated'' boundary conditions.}\label{rsosaltaltcorrelated}
 \end{figure}

 \begin{figure}
 	\centering
 \begin{tikzpicture}[scale=0.8]
 \draw[black,line width = 1pt](2,2)--(8,8);
 \draw[black,line width = 1pt](3,1)--(8,6);
 \draw[black,line width = 1pt](5,1)--(8,4);
 \draw[black,line width = 1pt](7,1)--(8,2);

 \draw[black,line width = 1pt](2,4)--(7,9);
 \draw[black,line width = 1pt](2,6)--(5,9);
 \draw[black,line width = 1pt](2,8)--(3,9);

 \draw[black,line width = 1pt](8,2)--(2,8);
 \draw[black,line width = 1pt](7,1)--(2,6);
 \draw[black,line width = 1pt](5,1)--(2,4);
 \draw[black,line width = 1pt](3,1)--(2,2);

 \draw[black,line width = 1pt](8,4)--(3,9);
 \draw[black,line width = 1pt](8,6)--(5,9);
 \draw[black,line width = 1pt](8,8)--(7,9);

 \node at (1.7,2.1) {$b$};
 \node at (1.7,4.1) {$b$};
 \node at (1.7,6.1) {$b$};
 \node at (1.7,8.1) {$b$};

 \node at (3,1.7) {$b-1$};
 \node at (3,3.7) {$b+1$};
 \node at (3,5.7) {$b-1$};
 \node at (3,7.7) {$b+1$};
 \node at (3,9.2) {$b-1$};

 \node at (7,1.7) {$a+1$};
 \node at (7,3.7) {$a-1$};
 \node at (7,5.7) {$a+1$};
  \node at (7,7.7) {$a-1$};
   \node at (7,9.2) {$a+1$};

 \node at (8.3,2) {$a$};
 \node at (8.3,4) {$a$};
 \node at (8.3,6) {$a$};
 \node at (8.3,8) {$a$};

 \end{tikzpicture}
 \caption{Alternating boundary conditions on both sides. We will write this boundary condition $b^{\pm},...,^{\mp} \! a$ and refer to it as ``anti-correlated'' boundary conditions.}\label{rsosaltaltanticorrelated}
 \end{figure}
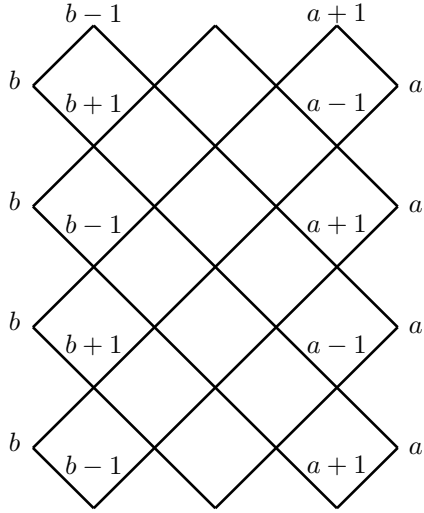

As was the case in the loop model (see section \ref{altaltloop}) there are two ways to put the alt boundary condition on both sides and these two ways will give two different continuum limits.
Consider the boundary conditions in Figures \ref{rsosaltaltcorrelated} and \ref{rsosaltaltanticorrelated}. The heights at the boundaries are fixed to the same, constant values in the two figures, namely $h_1 = b$ on the left boundary and $h_{N+1} = a$ on the right boundary. However, in Figure \ref{rsosaltaltcorrelated} the lattice sites with heights $h_N = a-1$ just next to the right boundary appear on the same rows as lattice sites with heights $h_2 = b-1$ just next the left boundary. Conversely, in Figure \ref{rsosaltaltanticorrelated} the heights $a-1$ appear on the same row as $b+1$. We will refer to these two situations as ``correlated'' and ``anticorrelated'' (alternating) boundary conditions respectively.

We will write a generic correlated boundary condition as $b^{\pm},\ldots,^{\pm} \! a$ and a generic anti-correlated boundary condition as $b^{\pm},\ldots,^{\mp} \! a$. We find that one can obtain two different conformally invariant continuum limits
from these boundary conditions. In the first case, we must consider correlated boundary conditions when the width of the lattice $L$ is even and anti-correlated boundary conditions when the width $L$ is odd. The second case works the other way around: we take anti-correlated boundary conditions for $L$ even and correlated boundary conditions for $L$ odd. (Note that in our conventions $N=2L=6$ in figures \ref{rsosaltaltcorrelated} and \ref{rsosaltaltanticorrelated}). The two different choices correspond to two different continuum limits. The string functions obtained from this prescription for the case $k=6$ are shown in Table \ref{rsosafdoublealtk6}. The boundary conditions reported in the table are the boundary conditions we take for $L$ even. The number of terms we have written in the expansion of the string function correspond to the number of terms that we can clearly observe on the lattice.  We have the general result that for the correlated boundary condition: 
\beq\label{doublealtcorr}
(m_1+1)^{\pm}...^{\pm}(m_2+1)
\eeq
in even sizes (and hence anti-correlated in odd sizes) the continuum limit is given by the string function $c^{m_1+m_2}_0$. Similarly, the anti-correlated boundary condition

\beq\label{doublealtanticorr}
(m_1+1)^{\pm}...^{\mp}(m_2+1)
\eeq
in even sizes (and hence correlated in odd sizes) gives the string function $c^{m_1-m_2}_0$. Compare this with the parafermion fusion rules \cite{GepnerQiu} for fields of the form $\phi^{m}_{l=0}$:

\beq\label{fusionplus}
\phi^{m_1}_{l=0}\times\phi^{m_2}_{l=0}=\phi^{m_1+m_2}_{l=0} 
\eeq
Clearly then, the boundary condition for $L$ even in (\ref{doublealtcorr}) corresponds to the fusion of the fields in equation (\ref{fusionplus}) and the boundary condition in (\ref{doublealtanticorr}) for $L$ even corresponds to the fusion of the fields

\beq\label{fusionminus}
\phi_0^{m_1}\times \phi_{0}^{-m_2}=\phi_0^{m_1-m_2} 
\eeq

Finally, let us notice that when we put correlated and anti-correlated boundary conditions together on the same lattice (i.e., we sum the correlated and anti-correlated configurations) we will clearly get the sum of the two string functions in the continuum limit $c^{m_1+m_2}_0+c^{m_1-m_2}_0$. We can write interpret this as the result of the fusion product: 
\beq\label{fusionplusminus}
\frac{1}{\sqrt{2}}(\phi^{m_1}_{l=0}+\phi^{-m_1}_{l=0})\times\frac{1}{\sqrt{2}}(\phi^{m_2}_{l=0}+\phi^{-m_2}_{l=0}) 
\eeq

\begin{table}
\begin{center}
\begin{tabular}{c | c | l }
Boundary condition & Exponent & Generating function\\
\hline
$2^{\pm}$,...,$^{\pm} 2$ & $\frac{3}{4}$ & $c^{m=2}_{l=0}=q^{h-\frac{c}{24}}(1+q+2q^2+3q^3...)$\\[1.1ex]
\hline
$2^{\pm}$,...,$^{\mp} 2$ & $0$ & $c^{m=0}_{l=0}=q^{h-\frac{c}{24}}(1+q^2+2q^3...)$\\[1.1ex]
\hline
$3^{\pm}$,...,$^{\pm} 3$ & $1$ & $c^{m=4}_{l=0}=q^{h-\frac{c}{24}}(1+q+3q^2+3q^4...)$\\[1.1ex]
\hline
$3^{\pm}$,...,$^{\mp} 3$ & $0$ & $c^{m=0}_{l=0}=q^{h-\frac{c}{24}}(1+q^2+2q^3...)$\\[1.1ex]
\hline
$2^{\pm}$,...,$^{\pm} 4$ & $1$ & $c^{m=4}_{l=0}=q^{h-\frac{c}{24}}(1+q+3q^2+3q^3...)$\\[1.1ex]
\hline
$2^{\pm}$,...,$^{\mp} 4$ & $\frac{3}{4}$ & $c^{m=2}_{l=0}=q^{h-\frac{c}{24}}(1+q+2q^2+3q^3...)$\\[1.1ex]
\hline
$4^{\pm}$,...,$^{\pm} 4$ & $\frac{3}{4}$ & $c^{m=2}_{l=0}=q^{h-\frac{c}{24}}(1+q+2q^2+3q^3...)$\\[1.1ex]
\hline
$4^{\pm}$,...,$^{\mp} 4$ & $0$ & $c^{m=0}_{l=0}=q^{h-\frac{c}{24}}(1+q^2+2q^3...)$\\[1.1ex]
\end{tabular}
\caption{String functions in the $k=6$ antiferromagnetic RSOS model with the alternating boundary condition on both sides. Note that the boundary condition written in the table corresponds to the boundary condition when $L$ is even. When the left and right boundary conditions are  ``correlated'' for $L$ even then we take ``anticorrelated'' boundary conditions for $L$ odd, and vice versa. The string functions are expanded up to the terms we can clearly observe on the lattice. }\label{rsosafdoublealtk6}
\end {center}
\end{table}

\subsubsection{Alt on both sides in the RSOS model: 2BTL representation theory}\label{loopfusionstring}
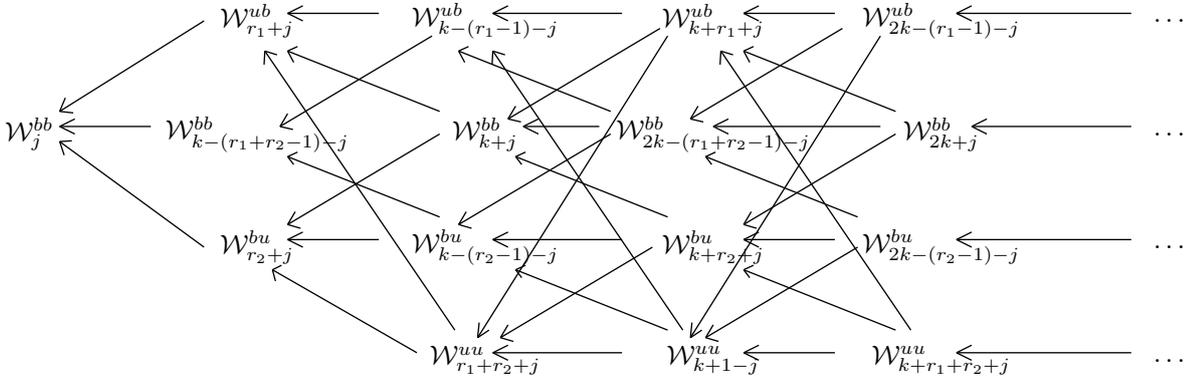
\begin{figure}[h]
	\begin{tikzpicture}
		\node at (0,0) {$\mathcal{W}_j^{bb}$};
		\draw[black,line width = 0.5pt](0.4,0.1)--(1.6,0.1);
		\draw[black,line width = 0.5pt](0.4,0.1)--(0.6,0.2);
		\draw[black,line width = 0.5pt](0.4,0.1)--(0.6,0);
		
		\draw[black,line width = 0.5pt](0.4,0.3)--(2.3,1.5);
		\draw[black,line width = 0.5pt](0.4,0.3)--(0.475,0.48);
		\draw[black,line width = 0.5pt](0.4,0.3)--(0.58,0.28);

		\draw[black,line width = 0.5pt](0.4,-0.1)--(2.3,-1.5);
		\draw[black,line width = 0.5pt](0.4,-0.1)--(0.46,-0.28);
		\draw[black,line width = 0.5pt](0.4,-0.1)--(0.56,-0.09);

		\node at (3,0) {$\mathcal{W}_{k-(r_1+r_2-1)-j}^{bb}$};
		\draw[black,line width = 0.5pt](3.3,0.1)--(5.3,1.3);
		\draw[black,line width = 0.5pt](3.3,0.1)--(3.4,0.27);
		\draw[black,line width = 0.5pt](3.3,0.1)--(3.47,0.06);
		
		\draw[black,line width = 0.5pt](3.4,-0.3)--(5.4,-1.1);
		\draw[black,line width = 0.5pt](3.4,-0.3)--(3.48,-0.405);
		\draw[black,line width = 0.5pt](3.4,-0.3)--(3.53,-0.24);

		\node at (6,0) {$\mathcal{W}_{k+j}^{bb}$};
		\draw[black,line width = 0.5pt](6.5,0.1)--(7.5,0.1);
		\draw[black,line width = 0.5pt](6.5,0.1)--(6.7,0.2);
		\draw[black,line width = 0.5pt](6.5,0.1)--(6.7,0);
		
		\draw[black,line width = 0.5pt](6.3,0.2)--(8.3,1.4);
		\draw[black,line width = 0.5pt](6.3,0.2)--(6.4,0.37);
		\draw[black,line width = 0.5pt](6.3,0.2)--(6.47,0.16);
		
		\draw[black,line width = 0.5pt](6.4,-0.3)--(8.4,-1.1);
		\draw[black,line width = 0.5pt](6.4,-0.3)--(6.48,-0.405);
		\draw[black,line width = 0.5pt](6.4,-0.3)--(6.53,-0.24);

		\node at (9,0) {$\mathcal{W}_{2k-(r_1+r_2-1)-j}^{bb}$};
		\draw[black,line width = 0.5pt](9,0.1)--(11.2,0.1);
		\draw[black,line width = 0.5pt](9,0.1)--(9.2,0.2);
		\draw[black,line width = 0.5pt](9,0.1)--(9.2,0);
		
		\draw[black,line width = 0.5pt](8.7,0.2)--(10.7,1.4);
		\draw[black,line width = 0.5pt](8.7,0.2)--(8.8,0.37);
		\draw[black,line width = 0.5pt](8.7,0.2)--(8.87,0.16);
		
		\draw[black,line width = 0.5pt](8.9,-0.3)--(10.9,-1.1);
		\draw[black,line width = 0.5pt](8.9,-0.3)--(8.98,-0.405);
		\draw[black,line width = 0.5pt](8.9,-0.3)--(9.03,-0.24);
		
		\node at (12,0) {$\mathcal{W}_{2k+j}^{bb}$};
		\draw[black,line width = 0.5pt](12.4,0.1)--(14.5,0.1);
		\draw[black,line width = 0.5pt](12.4,0.1)--(12.6,0.2);
		\draw[black,line width = 0.5pt](12.4,0.1)--(12.6,0);
				
		\node at (15,0) {$\ldots$};
		
		\node at (15,1.5) {$\ldots$};
		
		\node at (15,-1.5) {$\ldots$};
		
		\node at (15,-3) {$\ldots$};
		
		\node at (3,1.5) {$\mathcal{W}_{r_1+j}^{ub}$};
		\draw[black,line width = 0.5pt](3.4,1.6)--(4.6,1.6);
		\draw[black,line width = 0.5pt](3.4,1.6)--(3.6,1.7);
		\draw[black,line width = 0.5pt](3.4,1.6)--(3.6,1.5);
		
		\draw[black,line width = 0.5pt](3.4,1.1)--(5.4,0.3);
		\draw[black,line width = 0.5pt](3.4,1.1)--(3.48,0.95);
		\draw[black,line width = 0.5pt](3.4,1.1)--(3.53,1.16);
		
		\draw[black,line width = 0.5pt](3.1,1.1)--(5.6,-2.6);
		\draw[black,line width = 0.5pt](3.1,1.1)--(3.09,0.92);
		\draw[black,line width = 0.5pt](3.1,1.1)--(3.26,1.05);

		\node at (6,1.5) {$\mathcal{W}_{k-(r_1-1)-j}^{ub}$};
		\draw[black,line width = 0.5pt](6.1,1.6)--(7.8,1.6);
		\draw[black,line width = 0.5pt](6.1,1.6)--(6.3,1.7);
		\draw[black,line width = 0.5pt](6.1,1.6)--(6.3,1.5);
		
		\draw[black,line width = 0.5pt](5.65,1.1)--(7.65,0.3);
		\draw[black,line width = 0.5pt](5.65,1.1)--(5.73,0.95);
		\draw[black,line width = 0.5pt](5.65,1.1)--(5.78,1.16);

		\draw[black,line width = 0.5pt](6.1,1.1)--(8.6,-2.6);
		\draw[black,line width = 0.5pt](6.1,1.1)--(6.09,0.92);
		\draw[black,line width = 0.5pt](6.1,1.1)--(6.26,1.05);
		
		\node at (9,1.5) {$\mathcal{W}_{k+r_1+j}^{ub}$};
		\draw[black,line width = 0.5pt](9.4,1.6)--(10.6,1.6);
		\draw[black,line width = 0.5pt](9.4,1.6)--(9.6,1.7);
		\draw[black,line width = 0.5pt](9.4,1.6)--(9.6,1.5);
		
		\draw[black,line width = 0.5pt](9.4,1.1)--(11.4,0.3);
		\draw[black,line width = 0.5pt](9.4,1.1)--(9.48,0.95);
		\draw[black,line width = 0.5pt](9.4,1.1)--(9.53,1.16);

		\draw[black,line width = 0.5pt](9.1,1.1)--(11.6,-2.6);
		\draw[black,line width = 0.5pt](9.1,1.1)--(9.09,0.92);
		\draw[black,line width = 0.5pt](9.1,1.1)--(9.26,1.05);
		
		\node at (12,1.5) {$\mathcal{W}_{2k-(r_1-1)-j}^{ub}$};
		\draw[black,line width = 0.5pt](12.2,1.6)--(14.5,1.6);
		\draw[black,line width = 0.5pt](12.2,1.6)--(12.4,1.7);
		\draw[black,line width = 0.5pt](12.2,1.6)--(12.4,1.5);
		
		\node at (3,-1.5) {$\mathcal{W}_{r_2+j}^{bu}$};
		\draw[black,line width = 0.5pt](3.4,-1.4)--(4.6,-1.4);
		\draw[black,line width = 0.5pt](3.4,-1.4)--(3.6,-1.5);
		\draw[black,line width = 0.5pt](3.4,-1.4)--(3.6,-1.3);
		
		\draw[black,line width = 0.5pt](3.2,-1.8)--(5.1,-2.9);
		\draw[black,line width = 0.5pt](3.2,-1.8)--(3.24,-1.95);
		\draw[black,line width = 0.5pt](3.2,-1.8)--(3.32,-1.74);
		
		\draw[black,line width = 0.5pt](3.4,-1.2)--(5.4,0);
		\draw[black,line width = 0.5pt](3.4,-1.2)--(3.5,-1.03);
		\draw[black,line width = 0.5pt](3.4,-1.2)--(3.57,-1.25);

		\node at (6,-1.5) {$\mathcal{W}_{k-(r_2-1)-j}^{bu}$};
		\draw[black,line width = 0.5pt](6.1,-1.4)--(7.8,-1.4);
		\draw[black,line width = 0.5pt](6.1,-1.4)--(6.3,-1.3);
		\draw[black,line width = 0.5pt](6.1,-1.4)--(6.3,-1.5);
		
		\draw[black,line width = 0.5pt](5.65,-1.2)--(7.65,0);
		\draw[black,line width = 0.5pt](5.65,-1.2)--(5.75,-1.03);
		\draw[black,line width = 0.5pt](5.65,-1.2)--(5.82,-1.25);
		
		\draw[black,line width = 0.5pt](6.4,-1.8)--(8.4,-2.6);
		\draw[black,line width = 0.5pt](6.4,-1.8)--(6.48,-1.905);
		\draw[black,line width = 0.5pt](6.4,-1.8)--(6.53,-1.74);

		\node at (9,-1.5) {$\mathcal{W}_{k+r_2+j}^{bu}$};
		\draw[black,line width = 0.5pt](9.4,-1.4)--(10.6,-1.4);
		\draw[black,line width = 0.5pt](9.4,-1.4)--(9.6,-1.5);
		\draw[black,line width = 0.5pt](9.4,-1.4)--(9.6,-1.3);
		
		\draw[black,line width = 0.5pt](9.4,-1.2)--(11.4,0);
		\draw[black,line width = 0.5pt](9.4,-1.2)--(9.5,-1.03);
		\draw[black,line width = 0.5pt](9.4,-1.2)--(9.57,-1.25);
		
		\draw[black,line width = 0.5pt](9.4,-1.8)--(11.4,-2.6);
		\draw[black,line width = 0.5pt](9.4,-1.8)--(9.48,-1.905);
		\draw[black,line width = 0.5pt](9.4,-1.8)--(9.53,-1.74);

		\node at (12,-1.5) {$\mathcal{W}_{2k-(r_2-1)-j}^{bu}$};
		\draw[black,line width = 0.5pt](12.2,-1.4)--(14.5,-1.4);
		\draw[black,line width = 0.5pt](12.2,-1.4)--(12.4,-1.3);
		\draw[black,line width = 0.5pt](12.2,-1.4)--(12.4,-1.5);
		
		\node at (6,-3) {$\mathcal{W}_{r_1+r_2+j}^{uu}$};
		\draw[black,line width = 0.5pt](6.1,-2.9)--(7.8,-2.9);
		\draw[black,line width = 0.5pt](6.1,-2.9)--(6.3,-2.8);
		\draw[black,line width = 0.5pt](6.1,-2.9)--(6.3,-3);
		
		\draw[black,line width = 0.5pt](6.2,-2.7)--(8.2,-1.5);
		\draw[black,line width = 0.5pt](6.2,-2.7)--(6.3,-2.53);
		\draw[black,line width = 0.5pt](6.2,-2.7)--(6.37,-2.75);
		
		\draw[black,line width = 0.5pt](5.9,-2.7)--(8.4,1.3);
		\draw[black,line width = 0.5pt](5.9,-2.7)--(5.92,-2.5);
		\draw[black,line width = 0.5pt](5.9,-2.7)--(6.05,-2.6);
		
		\node at (9,-3) {$\mathcal{W}_{k+1-j}^{uu}$};
		\draw[black,line width = 0.5pt](9.4,-2.9)--(10.6,-2.9);
		\draw[black,line width = 0.5pt](9.4,-2.9)--(9.6,-3);
		\draw[black,line width = 0.5pt](9.4,-2.9)--(9.6,-2.8);
		
		\draw[black,line width = 0.5pt](8.9,-2.7)--(10.9,-1.5);
		\draw[black,line width = 0.5pt](8.9,-2.7)--(9,-2.53);
		\draw[black,line width = 0.5pt](8.9,-2.7)--(9.07,-2.75);
		
		\draw[black,line width = 0.5pt](8.7,-2.7)--(11.2,1.3);
		\draw[black,line width = 0.5pt](8.7,-2.7)--(8.72,-2.5);
		\draw[black,line width = 0.5pt](8.7,-2.7)--(8.85,-2.6);

		\node at (12,-3) {$\mathcal{W}_{k+r_1+r_2+j}^{uu}$};
		\draw[black,line width = 0.5pt](12.2,-2.9)--(14.5,-2.9);
		\draw[black,line width = 0.5pt](12.2,-2.9)--(12.4,-2.8);
		\draw[black,line width = 0.5pt](12.2,-2.9)--(12.4,-3);
		
	\end{tikzpicture}
	\caption{The corrected 2BTL inclusion diagram, replacing the one published in \cite{DubailThese}. The repeated part of the diagram is such that all standard modules beyond those in the first three columns each have three out-going arrows.}\label{2btldiagram}
\end{figure}

The representation theory of the 2BTL algebra \cite{GierNichols,JS-Blob,DJS} was further studied from a conformal perspective in \cite{DubailThese}. As was discussed in section \ref{altaltloop}, when there is a blob on both sides of the system there are four sectors to consider, labelled by $bb$, $ub$, $bu$ and $uu$. It was found in \cite{DubailThese} that (when $\q=e^{i\pi/k} $ is a primitive root of unity) the following infinite sum corresponds to the generating function of an irreducible representation:

\beq\label{2btlirrep}
\begin{aligned}
\mathcal{X}_{j}^{bb}=&\sum\limits_{n_1=0}\mathcal{W}_{j+n_1k}^{bb}-\sum\limits_{n_1=0}\mathcal{W}_{k-(r_1+r_2)+1-j+n_1k}^{bb}-\sum\limits_{n_1=0}\mathcal{W}_{j+r_1+n_1k}^{ub}+\sum\limits_{n_1=0}\mathcal{W}_{k-(r_2-1)-j+n_1k}^{ub}\\
-&\sum\limits_{n_1=0}\mathcal{W}_{j+r_2+n_1k}^{bu}+\sum\limits_{n_1=0}\mathcal{W}_{k-(r_1-1)-j+n_1k}^{bu}+\sum\limits_{n_1=0}\mathcal{W}_{r_1+r_2+j+n_1k}^{uu}-\sum\limits_{n_1=0}\mathcal{W}_{k+1-j+n_1k}^{uu} \,.
\end{aligned}
\eeq
We find results in agreement with equation (\ref{2btlirrep}), but we also find that the diagram of inclusions from which this infinite sum can be derived requires modification, i.e., the known 2BTL inclusion diagram does not in fact lead to equation (\ref{2btlirrep}) as previously thought. When we impose $r_1+r_2-1+2j\leq k-1$ and $j > 0$ we find that the inclusion diagram should instead be given by Figure \ref{2btldiagram}. The only difference here to the diagram published in \cite{DubailThese} is that we do not have an arrow from $\mathcal{W}_{k+j}^{bb}$ to $\mathcal{W}_{k-(r_1+r_2-1)-j}^{bb}$. 

\no We will use the expressions in eqs.~(\ref{bb})--(\ref{uu}) to calculate $\mathcal{X}_{j}^{bb}$, and we will show that the resulting generating function is that corresponding to the fusion of parafermion fields. We have from expressions (\ref{bb}) to (\ref{uu}):
\begin{footnotesize}
\begin{eqnarray}
\label{bb1}
\mathcal{W}_{j+n_1k}^{bb} &\rightarrow&\sum\limits_{n_2=0}^{\infty}\lambda_{r_1+r_2-1-2n_2r,j+n_2r+n_1k}+\lambda_{|r_2-r_1|+1-2n_2r,j+(n_2+1)r-1+n_1k} \\
\label{bb2}
\mathcal{W}_{k-(r_1+r_2)+1-j+n_1k}^{bb} &\rightarrow& \sum\limits_{n_2=0}^{\infty}\lambda_{r_1+r_2-1-2n_2r,k-(r_1+r_2)+1-j+n_2r+n_1k}+\lambda_{|r_2-r_1|+1-2n_2r,k-(r_1+r_2)-j+(n_2+1)r+n_1k} \\
\label{ub3}
\mathcal{W}_{j+r_1+n_1k}^{ub} &\rightarrow& \sum\limits_{n_2=0}^{\infty}\lambda_{-r_1+r_2-1-2n_2r,j+r_1+n_2r+n_1k}+\lambda_{-r_1+r_2+1-2r(n_2+1),j+r_1+(n_2+1)r-1+n_1k} \\
\label{ub4}
\mathcal{W}_{k-(r_2-1)-j+n_1k}^{ub} &\rightarrow& \sum\limits_{n_2=0}^{\infty}\lambda_{-r_1+r_2-1-2n_2r,k-(r_2-1)-j+n_2r+n_1k}+\lambda_{-r_1+r_2+1-2r(n_2+1),k-r_2-j+(n_2+1)r+n_1k} \\
\label{bu5}
\mathcal{W}_{j+r_2+n_1k}^{bu} &\rightarrow& \sum\limits_{n_2=0}^{\infty}\lambda_{r_1-r_2-1-2n_2r,j+r_2+n_2r+n_1k}+\lambda_{r_1-r_2+1-2r(n_2+1),j+r_2+(n_2+1)r-1+n_1k} \\
\label{bu6}
\mathcal{W}_{k-(r_1-1)-j+n_1k}^{bu} &\rightarrow& \sum\limits_{n_2=0}^{\infty}\lambda_{r_1-r_2-1-2n_2r,k-(r_1-1)-j+n_2r+n_1k}+\lambda_{r_1-r_2+1-2r(n_2+1),k-r_1-j+(n_2+1)r+n_1k} \\
\label{uu7}
\mathcal{W}_{r_1+r_2+j+n_1k}^{uu} &\rightarrow& \sum\limits_{n_2=0}^{\infty}\lambda_{-r_1-r_2-1-2n_2r,j+r_1+r_2+n_2r+n_1k}+\lambda_{-r_1-r_2+1-2r(n_2+1),j+r_1+r_2+(n_2+1)r-1+n_1k} \\
\label{uu8}
\mathcal{W}_{k+1-j+n_1k}^{uu} &\rightarrow& \sum\limits_{n_2=0}^{\infty}\lambda_{-r_1-r_2-1-2n_2r,k+1-j+n_2r+n_1k}+\lambda_{-r_1-r_2+1-2r(n_2+1),k-j+(n_2+1)r+n_1k}
\end{eqnarray}
\end{footnotesize}
We will consider for now the case $r_1\leq r_2$, so that we have $r=r_1$. After taking the sum $\sum\limits_{n_1=0}^{\infty}$ of the expression in (\ref{bb1}), we can write it as:

\beq
\begin{aligned}
	&\sum\limits_{n_1=0}^{\infty}\lambda_{r_1+r_2-1,j+n_1k}+\lambda_{-r_1+r_2+1,j+r_1-1+n_1k}\\
	+&\sum\limits_{n_1=0}^{\infty} \sum\limits_{n_2=0}^{\infty}\lambda_{-r_1+r_2-1-2n_2r_1,j+r_1+n_2r_1+n_1k}+\lambda_{-r_1+r_2+1-2(n_2+1)r_1,j+r_1+(n_2+1)r_1-1+n_1k} \,,
\end{aligned}
\eeq
where we see that the second term cancels the expression in eq.~(\ref{ub3}) entirely. Similarly, applying the same sum to the expression in (\ref{bb2}) gives:
\beq
\begin{aligned}
	&\sum\limits_{n_1=0}^{\infty}\lambda_{r_1+r_2-1,k-(r_1+r_2)+1-j+n_1k}+\lambda_{-r_1+r_2+1,k-r_2-j+n_1k}\\
	+&\sum\limits_{n_1=0}^{\infty}\sum\limits_{n_2=0}^{\infty}\lambda_{-r_1+r_2-1-2n_2r_1,k-(r_2-1)-j+n_1k+n_2r_1}+\lambda_{-r_1+r_2+1-2(n_2+1)r_1,k-r_2-j+n_1k+(n_2+1)r_1} \,,
\end{aligned}
\eeq
where the second term now cancels the expression in (\ref{ub4}) entirely. In a similar way, we can make the same cancellations for the terms in (\ref{bu5}) and (\ref{uu7}) as well as (\ref{bu6}) and (\ref{uu8}). In the end, we are left with the terms:
\beq\label{total1}
\begin{aligned}
\mathcal{X}_{j}^{bb}=&\sum\limits_{n_1=0}^{\infty}\lambda_{r_1+r_2-1,j+n_1k}-\sum\limits_{n_1=0}^{\infty}\lambda_{k-r_1-r_2+1,j+r_1+r_2-1+n_1k}\\
-&\sum\limits_{n_1=0}^{\infty}\lambda_{r_1+r_2-1,k-r_1-r_2+1-j+n_1k}+\sum\limits_{n_1=0}^{\infty}\lambda_{k-r_1-r_2+1,k-j+n_1k}\\
+&\sum\limits_{n_1=0}^{\infty}\lambda_{-r_1+r_2+1,j+r_1-1+n_1k}-\sum\limits_{n1=0}^{\infty}\lambda_{k+r_1-r_2-1,j+r_2+n_1k}\\
-&\sum\limits_{n_1=0}^{\infty}\lambda_{-r_1+r_2+1,k-r_2-j+n_1k}+\sum\limits_{n1=0}^{\infty}\lambda_{k+r_1-r_2-1,k-r_1+1-j+n_1k}\\
\end{aligned}
\eeq
where we have dealt with the negative values of $r$ in $\lambda_{r,j}$ by defining $r$ modulo $k$. We further have the relationship
\beq
2j=k-r_2-r_1
\eeq
when we want the loop model parameters to correspond to the RSOS model. Using equations (\ref{stringidentif}) and (\ref{topstructure}), equation (\ref{total1}) then becomes
\beq\label{total2}
\mathcal{X}_{j}^{bb}=c_{l=k-2}^{m=r_2-r_1}+c_{l=k-2}^{m=r_1+r_2-2} \,,
\eeq
which, after using the identities $r_1=m_1+1$ and $r_2=m_2+1$, yields
\beq\label{total3}
\mathcal{X}_{j}^{bb}=c_{l=k-2}^{m=m_2-m_1}+c_{l=k-2}^{m=m_1+m_2} \,.
\eeq
But using the string function identities $c^m_l=c^{-m}_l$ and $c^{k-2-m}_{k-2-l}=c^m_l$, we can rewrite this as 
\beq\label{total4}
\mathcal{X}_{j}^{bb}=c_{l=k-2}^{m=m_1-m_2}+c_{l=0}^{m=k-2-m_1-m_2} \,,
\eeq
which, when we compare with eq.~(\ref{fusionplusminus}), is nothing but the fusion product
\beq
\frac{1}{\sqrt{2}}(\phi^{m_1}_{l=0}+\phi^{-m_1}_{l=0})\times\frac{1}{\sqrt{2}}(\phi^{k-2-m_2}_{l=0}+\phi^{-k+2+m_2}_{l=0}) \,.
\eeq

In summary, then, we have recovered the numerical results found in section \ref{rsosfusionnumerics} from studying the representation theory of the 2BTL algebra.

\section{Special cases: the two and three-state Potts models}\label{3statepotts}

It is well known that the two-state  ferromagnetic Potts model---alias the Ising model---is equivalent to an antiferromagnetic one in the bulk. Indeed, the mapping between these models is easily obtained by switching the sign of the couplings and flipping the spins at the same time. Under this transformation, the critical ferromagnetic coupling $e^{K}=1+\sqrt{2}$ becomes $e^{-K}= (1+\sqrt{2})^{-1} =\sqrt{2}-1$, which is the critical coupling for the $Q=2$ AF Potts model indeed.

This mapping becomes more interesting in the presence of a boundary. While free boundary conditions map onto free, fixed boundary conditions map onto alt boundary conditions (that is, alternating $+$ and $-$ spins on the boundary for $Q=2$). This suggests that the natural ``equivalent'' of fixed boundary conditions in the AF case is alt, an observation we have confirmed in detail earlier. 
 
\medskip

Coming now to the three-state  critical AF Potts model, it is also well known that it  can be reformulated as a colouring problem (see the discussion in \cite{S-AF}, for example). As was discussed in section \ref{PottsQgen}, the antiferromagnetic critical point of the Potts model is defined by \cite{B-AF}
\beq
\exp(K)=-1+\sqrt{4-Q} \,.
\eeq
We see  that setting $Q=3$ sends $K\rightarrow -\infty$. Considering then the classical Potts Hamiltonian in eq.~(\ref{PottsHam}), the only allowed configurations of Potts spins are those where no two neighbouring spins can be identical. This is the ``colouring problem''; the partition function is given by the number of configurations that allow $Q=3$ possible colours on each site, but no two neighbouring sites can have the same colour.

\begin{figure}
\centering
\begin{minipage}{.35\textwidth}
	\centering
\begin{tikzpicture}[scale=0.4]
	\filldraw[black] (2,2) circle (4pt);
	\filldraw[black] (4,2) circle (4pt);
	\filldraw[black] (6,2) circle (4pt);
	\filldraw[black] (8,2) circle (4pt);
	\filldraw[black] (10,2) circle (4pt);

	\filldraw[black] (2,4) circle (4pt);
	\filldraw[black] (4,4) circle (4pt);
	\filldraw[black] (6,4) circle (4pt);
	\filldraw[black] (8,4) circle (4pt);
	\filldraw[black] (10,4) circle (4pt);

	\filldraw[black] (2,6) circle (4pt);
	\filldraw[black] (4,6) circle (4pt);
	\filldraw[black] (6,6) circle (4pt);
	\filldraw[black] (8,6) circle (4pt);
	\filldraw[black] (10,6) circle (4pt);

	\filldraw[black] (2,8) circle (4pt);
	\filldraw[black] (4,8) circle (4pt);
	\filldraw[black] (6,8) circle (4pt);
	\filldraw[black] (8,8) circle (4pt);
	\filldraw[black] (10,8) circle (4pt);

	\filldraw[black] (2,10) circle (4pt);
	\filldraw[black] (4,10) circle (4pt);
	\filldraw[black] (6,10) circle (4pt);
	\filldraw[black] (8,10) circle (4pt);
	\filldraw[black] (10,10) circle (4pt);

	\draw[black,line width = 1pt](3,1)--(3,11);
	\draw[black,line width = 1pt](5,1)--(5,11);
	\draw[black,line width = 1pt](7,1)--(7,11);
	\draw[black,line width = 1pt](9,1)--(9,11);

	\draw[black,line width = 1pt](1,3)--(11,3);
	\draw[black,line width = 1pt](1,5)--(11,5);
	\draw[black,line width = 1pt](1,7)--(11,7);
	\draw[black,line width = 1pt](1,9)--(11,9);

	\node at (1.5,2) {\tiny 2};
	\node at (1.5,4) {\tiny 1,3};
	\node at (1.5,6) {\tiny 2};
	\node at (1.5,8) {\tiny 1,3};
	\node at (1.5,10) {\tiny 2};
	
	\node at (10.5,2) {\tiny 1,3};
	\node at (10.5,4) {\tiny 2};
	\node at (10.5,6) {\tiny 1,3};
	\node at (10.5,8) {\tiny 2};
	\node at (10.5,10) {\tiny 1,3};

\end{tikzpicture}
 \end{minipage}%
\begin{minipage}{.35\textwidth}
	\centering
\begin{tikzpicture}[scale=0.4]

	\filldraw[black] (2,2) circle (4pt);
	\filldraw[black] (4,2) circle (4pt);
	\filldraw[black] (6,2) circle (4pt);
	\filldraw[black] (8,2) circle (4pt);
	\filldraw[black] (10,2) circle (4pt);

	\filldraw[black] (2,4) circle (4pt);
	\filldraw[black] (4,4) circle (4pt);
	\filldraw[black] (6,4) circle (4pt);
	\filldraw[black] (8,4) circle (4pt);
	\filldraw[black] (10,4) circle (4pt);

	\filldraw[black] (2,6) circle (4pt);
	\filldraw[black] (4,6) circle (4pt);
	\filldraw[black] (6,6) circle (4pt);
	\filldraw[black] (8,6) circle (4pt);
	\filldraw[black] (10,6) circle (4pt);

	\filldraw[black] (2,8) circle (4pt);
	\filldraw[black] (4,8) circle (4pt);
	\filldraw[black] (6,8) circle (4pt);
	\filldraw[black] (8,8) circle (4pt);
	\filldraw[black] (10,8) circle (4pt);

	\filldraw[black] (2,10) circle (4pt);
	\filldraw[black] (4,10) circle (4pt);
	\filldraw[black] (6,10) circle (4pt);
	\filldraw[black] (8,10) circle (4pt);
	\filldraw[black] (10,10) circle (4pt);

	\draw[black,line width = 1pt](3,1)--(3,11);
	\draw[black,line width = 1pt](5,1)--(5,11);
	\draw[black,line width = 1pt](7,1)--(7,11);
	\draw[black,line width = 1pt](9,1)--(9,11);

	\draw[black,line width = 1pt](1,3)--(11,3);
	\draw[black,line width = 1pt](1,5)--(11,5);
	\draw[black,line width = 1pt](1,7)--(11,7);
	\draw[black,line width = 1pt](1,9)--(11,9);

	\node at (1.5,2) {\tiny 2};
	\node at (1.5,4) {\tiny 1,3};
	\node at (1.5,6) {\tiny 2};
	\node at (1.5,8) {\tiny 1,3};
	\node at (1.5,10) {\tiny 2};
	
	\node at (10.5,2) {\tiny 2};
	\node at (10.5,4) {\tiny 1,3};
	\node at (10.5,6) {\tiny 2};
	\node at (10.5,8) {\tiny 1,3};
	\node at (10.5,10) {\tiny 2};

\end{tikzpicture}
\end{minipage}%
\caption{Left panel: Anti-correlated boundary condition $A^=$ . Right panel: Correlated boundary condition $C^=$.}\label{3StatePottsAlt1}
\end{figure}
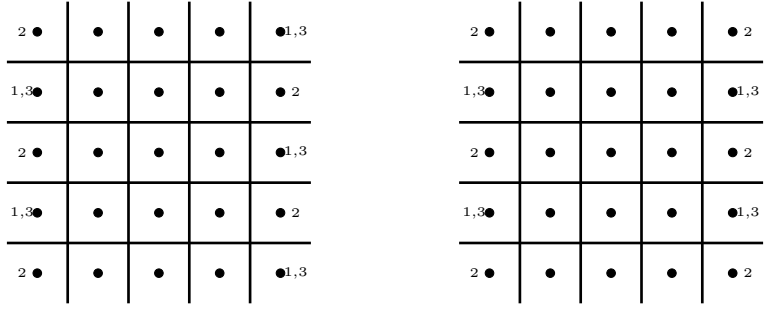

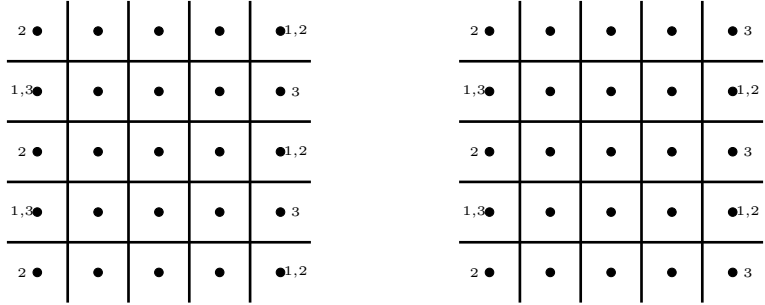
\begin{figure}
\centering
\begin{minipage}{.35\textwidth}
	\centering
\begin{tikzpicture}[scale=0.4]
	\filldraw[black] (2,2) circle (4pt);
	\filldraw[black] (4,2) circle (4pt);
	\filldraw[black] (6,2) circle (4pt);
	\filldraw[black] (8,2) circle (4pt);
	\filldraw[black] (10,2) circle (4pt);

	\filldraw[black] (2,4) circle (4pt);
	\filldraw[black] (4,4) circle (4pt);
	\filldraw[black] (6,4) circle (4pt);
	\filldraw[black] (8,4) circle (4pt);
	\filldraw[black] (10,4) circle (4pt);

	\filldraw[black] (2,6) circle (4pt);
	\filldraw[black] (4,6) circle (4pt);
	\filldraw[black] (6,6) circle (4pt);
	\filldraw[black] (8,6) circle (4pt);
	\filldraw[black] (10,6) circle (4pt);

	\filldraw[black] (2,8) circle (4pt);
	\filldraw[black] (4,8) circle (4pt);
	\filldraw[black] (6,8) circle (4pt);
	\filldraw[black] (8,8) circle (4pt);
	\filldraw[black] (10,8) circle (4pt);

	\filldraw[black] (2,10) circle (4pt);
	\filldraw[black] (4,10) circle (4pt);
	\filldraw[black] (6,10) circle (4pt);
	\filldraw[black] (8,10) circle (4pt);
	\filldraw[black] (10,10) circle (4pt);

	\draw[black,line width = 1pt](3,1)--(3,11);
	\draw[black,line width = 1pt](5,1)--(5,11);
	\draw[black,line width = 1pt](7,1)--(7,11);
	\draw[black,line width = 1pt](9,1)--(9,11);

	\draw[black,line width = 1pt](1,3)--(11,3);
	\draw[black,line width = 1pt](1,5)--(11,5);
	\draw[black,line width = 1pt](1,7)--(11,7);
	\draw[black,line width = 1pt](1,9)--(11,9);

	\node at (1.5,2) {\tiny 2};
	\node at (1.5,4) {\tiny 1,3};
	\node at (1.5,6) {\tiny 2};
	\node at (1.5,8) {\tiny 1,3};
	\node at (1.5,10) {\tiny 2};
	
	\node at (10.5,2) {\tiny 1,2};
	\node at (10.5,4) {\tiny 3};
	\node at (10.5,6) {\tiny 1,2};
	\node at (10.5,8) {\tiny 3};
	\node at (10.5,10) {\tiny 1,2};

\end{tikzpicture}
 \end{minipage}%
\begin{minipage}{.35\textwidth}
	\centering
\begin{tikzpicture}[scale=0.4]

	\filldraw[black] (2,2) circle (4pt);
	\filldraw[black] (4,2) circle (4pt);
	\filldraw[black] (6,2) circle (4pt);
	\filldraw[black] (8,2) circle (4pt);
	\filldraw[black] (10,2) circle (4pt);

	\filldraw[black] (2,4) circle (4pt);
	\filldraw[black] (4,4) circle (4pt);
	\filldraw[black] (6,4) circle (4pt);
	\filldraw[black] (8,4) circle (4pt);
	\filldraw[black] (10,4) circle (4pt);

	\filldraw[black] (2,6) circle (4pt);
	\filldraw[black] (4,6) circle (4pt);
	\filldraw[black] (6,6) circle (4pt);
	\filldraw[black] (8,6) circle (4pt);
	\filldraw[black] (10,6) circle (4pt);

	\filldraw[black] (2,8) circle (4pt);
	\filldraw[black] (4,8) circle (4pt);
	\filldraw[black] (6,8) circle (4pt);
	\filldraw[black] (8,8) circle (4pt);
	\filldraw[black] (10,8) circle (4pt);

	\filldraw[black] (2,10) circle (4pt);
	\filldraw[black] (4,10) circle (4pt);
	\filldraw[black] (6,10) circle (4pt);
	\filldraw[black] (8,10) circle (4pt);
	\filldraw[black] (10,10) circle (4pt);

	\draw[black,line width = 1pt](3,1)--(3,11);
	\draw[black,line width = 1pt](5,1)--(5,11);
	\draw[black,line width = 1pt](7,1)--(7,11);
	\draw[black,line width = 1pt](9,1)--(9,11);

	\draw[black,line width = 1pt](1,3)--(11,3);
	\draw[black,line width = 1pt](1,5)--(11,5);
	\draw[black,line width = 1pt](1,7)--(11,7);
	\draw[black,line width = 1pt](1,9)--(11,9);

	\node at (1.5,2) {\tiny 2};
	\node at (1.5,4) {\tiny 1,3};
	\node at (1.5,6) {\tiny 2};
	\node at (1.5,8) {\tiny 1,3};
	\node at (1.5,10) {\tiny 2};
	
	\node at (10.5,2) {\tiny 3};
	\node at (10.5,4) {\tiny 1,2};
	\node at (10.5,6) {\tiny 3};
	\node at (10.5,8) {\tiny 1,2};
	\node at (10.5,10) {\tiny 3};

\end{tikzpicture}
\end{minipage}%
\caption{Left panel: Anti-correlated boundary condition $A^{\neq}$. Right panel: Correlated boundary condition $C^{\neq}$.}\label{3StatePottsAlt2}
\end{figure}

The conformal boundary conditions found for the AF Potts model in the general case are relevant for $Q=3$ as well, of course. Free boundary conditions mean just that for $Q=3$ as well. With alt boundary conditions, 
even boundary sites can  take two particular values, and odd boundary sites are fixed to the remaining value (or vice versa). In the language of the colouring problem, we can fix for example all odd boundary sites to have the colour labelled by $2$, and even boundary sites can take either the colours $1$ or $3$; see Figures \ref{3StatePottsAlt1} and \ref{3StatePottsAlt2}. We see that this boundary condition automatically satisfies the colouring constraint that no neighbouring sites can have the same colour. (We will in fact show below that the $Q=3$ AF Potts model with alt boundary conditions is equivalent to the XXZ spin chain with $\Delta=\frac{1}{2}$ and free boundary conditions.)

Note however that just as in the RSOS and loop models, there are different ways to implement the alt condition when it is applied to both the left and right boundaries at the same time. Once we fix, for example, odd boundary sites on the left to take only one colour (in figure \ref{3StatePottsAlt1} this colour is 2) we have a choice between fixing odd boundary sites on the right to one specific colour or to allow odd boundary sites on the right to take two colours. If we choose the former, (i.e. odd boundary sites on both the right and left are fixed to take one colour) we refer to this as ``correlated''. This is the case in the right panel of Figure \ref{3StatePottsAlt1}. If we choose the inverse, i.e., we say that odd boundary sites on the left are fixed to one specific colour but odd boundary sites on the right are allowed to take two colours (and hence \textit{even} sites on the right boundary are fixed to the one remaining colour), we refer to this as ``anti-correlated''. This is the case in the left panel of Figure \ref{3StatePottsAlt1}.

But even after we have specified ``correlated'' or ``anti-correlated'' we still have an additional choice in relation to which particular colours we choose. In Figure \ref{3StatePottsAlt1} the boundary site that is fixed to \textit{one} particular value is equal to 2 on both sides. In Figure \ref{3StatePottsAlt2}, however, this is not the case; we again have ``anti-correlated'' boundary conditions in the left image and correlated boundary conditions in the right image, but the fixed height on the left boundary is equal to $2$ whereas on the right boundary the fixed height is equal to $3$. Since everything in the colouring problem is defined modulo $3$, there are only four independent cases, corresponding to those in Figures \ref{3StatePottsAlt1} and \ref{3StatePottsAlt2}. We will denote these four cases by by $C^=$, $C^{\neq}$, $A^=$ and $A^{\neq}$; $C$ and $A$ refer to ``correlated'' and ``anti-correlated'' respectively, while the $=$ and $\neq$ signs refer to the cases where the left boundary colour that is fixed is equal (resp. not equal) to the right boundary colour that is fixed. (See Figures \ref{3StatePottsAlt1} and \ref{3StatePottsAlt2}).

As we saw in sections \ref{rsosfusionnumerics} and \ref{altaltloop} when considering the RSOS and loop models respectively, making sense of the continuum limit of the alt condition on both sides of the boundary requires much care. In particular one must consider the two different scenarios:
\begin{enumerate}
 \item Correlated boundary conditions when the lattice width $L$ is even and anti-correlated boundary conditions when the lattice width $L$ is odd.
 \item Anti-correlated boundary conditions when the lattice width $L$ is even and correlated boundary conditions when the lattice width $L$ is odd.
\end{enumerate}

(Note that in our notations $L=5$ in Figures \ref{3StatePottsAlt1} and \ref{3StatePottsAlt2}). These two scenarios give two different continuum limits. The results are shown in Table \ref{d4stringfunc}.

\begin{table}[ht]
\begin{center}
\begin{tabular}{c | l }
Boundary condition (for odd lattice size $L$) & Generating function\\
\hline
Free/Free & $c^{m=0}_{l=0}+c^{m=0}_{l=4}$ \\[1.1ex]
\hline
Alt/Free & $c^{m=1}_{l=1}+c^{m=1}_{l=3}$ \\[1.1ex]
\hline
Alt/Alt: $C^=$ & $c^{m=0}_{l=0}+c^{m=4}_{l=0}$ \\[1.1ex]
\hline
Alt/Alt: $A^=$& $2c^{m=2}_{l=0}$  \\[1.1ex]
\hline
Alt/Alt: $C^{\neq}$& $c^{m=0}_{l=2}$ \\[1.1ex]
\hline
Alt/Alt: $A^{\neq}$& $c^{m=2}_{l=2}$  \\[1.1ex]
\end{tabular}
\caption{Boundary conditions and their continuum limits. As explained in the text there are four independent ``alt/alt" boundary conditions. The precise meaning of the labels of these four types are explained in Figures \ref{3StatePottsAlt1} and \ref{3StatePottsAlt2}. The boundary conditions written above are those taken for $L$ odd. See the main text for a discussion on the relevance of the parity of $L$.}\label{d4stringfunc}
\end {center}
\end{table}

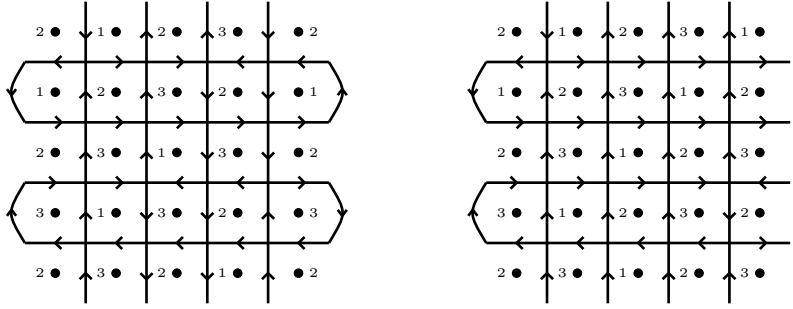
\begin{figure}
\centering
\begin{minipage}{.35\textwidth}
	\centering
\begin{tikzpicture}[scale=0.4]

\filldraw[black] (2,2) circle (4pt);
\filldraw[black] (4,2) circle (4pt);
\filldraw[black] (6,2) circle (4pt);
\filldraw[black] (8,2) circle (4pt);
\filldraw[black] (10,2) circle (4pt);

\filldraw[black] (2,4) circle (4pt);
\filldraw[black] (4,4) circle (4pt);
\filldraw[black] (6,4) circle (4pt);
\filldraw[black] (8,4) circle (4pt);
\filldraw[black] (10,4) circle (4pt);

\filldraw[black] (2,6) circle (4pt);
\filldraw[black] (4,6) circle (4pt);
\filldraw[black] (6,6) circle (4pt);
\filldraw[black] (8,6) circle (4pt);
\filldraw[black] (10,6) circle (4pt);

\filldraw[black] (2,8) circle (4pt);
\filldraw[black] (4,8) circle (4pt);
\filldraw[black] (6,8) circle (4pt);
\filldraw[black] (8,8) circle (4pt);
\filldraw[black] (10,8) circle (4pt);

\filldraw[black] (2,10) circle (4pt);
\filldraw[black] (4,10) circle (4pt);
\filldraw[black] (6,10) circle (4pt);
\filldraw[black] (8,10) circle (4pt);
\filldraw[black] (10,10) circle (4pt);

\draw[black,line width = 1pt](3,1)--(3,11);
\draw[black,line width = 1pt](5,1)--(5,11);
\draw[black,line width = 1pt](7,1)--(7,11);
\draw[black,line width = 1pt](9,1)--(9,11);

\draw[black,line width = 1pt](1,3)--(11,3);
\draw[black,line width = 1pt](1,5)--(11,5);
\draw[black,line width = 1pt](1,7)--(11,7);
\draw[black,line width = 1pt](1,9)--(11,9);

\node at (1.5,2) {\tiny 2 };
\node at (1.5,4) {\tiny 3 };
\node at (1.5,6) {\tiny 2 };
\node at (1.5,8) {\tiny 1 };
\node at (1.5,10) {\tiny 2 };

\node at (3.5,2) {\tiny 3 };
\node at (3.5,4) {\tiny 1 };
\node at (3.5,6) {\tiny 3 };
\node at (3.5,8) {\tiny 2 };
\node at (3.5,10) {\tiny 1 };

\node at (5.5,2) {\tiny 2 };
\node at (5.5,4) {\tiny 3 };
\node at (5.5,6) {\tiny 1 };
\node at (5.5,8) {\tiny 3 };
\node at (5.5,10) {\tiny 2 };

\node at (7.5,2) {\tiny 1 };
\node at (7.5,4) {\tiny 2 };
\node at (7.5,6) {\tiny 3 };
\node at (7.5,8) {\tiny 2 };
\node at (7.5,10) {\tiny 3 };

\node at (10.5,2) {\tiny 2 };
\node at (10.5,4) {\tiny 3 };
\node at (10.5,6) {\tiny 2 };
\node at (10.5,8) {\tiny 1 };
\node at (10.5,10) {\tiny 2 };

\draw[black,line width = 1pt](2.2,3.2)--(2,3);
\draw[black,line width = 1pt](2.2,2.8)--(2,3);

\draw[black,line width = 1pt](1.8,5.2)--(2,5);
\draw[black,line width = 1pt](1.8,4.8)--(2,5);

\draw[black,line width = 1pt](2,7.2)--(2.2,7);
\draw[black,line width = 1pt](2,6.8)--(2.2,7);

\draw[black,line width = 1pt](2.2,9.2)--(2,9);
\draw[black,line width = 1pt](2.2,8.8)--(2,9);

\draw[black,line width = 1pt] (1,3) .. controls (0.4,4) .. (1,5);
\draw[black,line width = 1pt](0.41,3.9)--(0.56,4.1);
\draw[black,line width = 1pt](0.71,3.9)--(0.56,4.1);

\draw[black,line width = 1pt] (1,7) .. controls (0.4,8) .. (1,9);
\draw[black,line width = 1pt](0.41,8.1)--(0.56,7.9);
\draw[black,line width = 1pt](0.71,8.1)--(0.56,7.9);

\draw[black,line width = 1pt] (11,9) .. controls (11.6,8) .. (11,7);
\draw[black,line width = 1pt](11.44,8.1)--(11.59,7.9);
\draw[black,line width = 1pt](11.44,8.1)--(11.29,7.9);

\draw[black,line width = 1pt] (11,5) .. controls (11.6,4) .. (11,3);
\draw[black,line width = 1pt](11.44,3.9)--(11.59,4.1);
\draw[black,line width = 1pt](11.44,3.9)--(11.29,4.1);

\draw[black,line width = 1pt](2.8,1.8)--(3,2);
\draw[black,line width = 1pt](3.2,1.8)--(3,2);
\draw[black,line width = 1pt](5,1.8)--(4.8,2);
\draw[black,line width = 1pt](5.2,2)--(5,1.8);
\draw[black,line width = 1pt](7,1.8)--(6.8,2);
\draw[black,line width = 1pt](7,1.8)--(7.2,2);
\draw[black,line width = 1pt](8.8,1.8)--(9,2);
\draw[black,line width = 1pt](9.2,1.8)--(9,2);

\draw[black,line width = 1pt](2.8,3.8)--(3,4);
\draw[black,line width = 1pt](3.2,3.8)--(3,4);
\draw[black,line width = 1pt](5,3.8)--(4.8,4);
\draw[black,line width = 1pt](5,3.8)--(5.2,4);
\draw[black,line width = 1pt](7,3.8)--(6.8,4);
\draw[black,line width = 1pt](7,3.8)--(7.2,4);
\draw[black,line width = 1pt](9,4)--(8.8,3.8);
\draw[black,line width = 1pt](9,4)--(9.2,3.8);

\draw[black,line width = 1pt](2.8,5.8)--(3,6);
\draw[black,line width = 1pt](3.2,5.8)--(3,6);
\draw[black,line width = 1pt](4.8,5.8)--(5,6);
\draw[black,line width = 1pt](5.2,5.8)--(5,6);
\draw[black,line width = 1pt](7,5.8)--(6.8,6);
\draw[black,line width = 1pt](7,5.8)--(7.2,6);
\draw[black,line width = 1pt](9,5.8)--(8.8,6);
\draw[black,line width = 1pt](9,5.8)--(9.2,6);

\draw[black,line width = 1pt](2.8,7.8)--(3,8);
\draw[black,line width = 1pt](3.2,7.8)--(3,8);
\draw[black,line width = 1pt](4.8,7.8)--(5,8);
\draw[black,line width = 1pt](5.2,7.8)--(5,8);
\draw[black,line width = 1pt](7,7.8)--(6.8,8);
\draw[black,line width = 1pt](7,7.8)--(7.2,8);
\draw[black,line width = 1pt](9,7.8)--(8.8,8);
\draw[black,line width = 1pt](9,7.8)--(9.2,8);

\draw[black,line width = 1pt](2.8,10)--(3,9.8);
\draw[black,line width = 1pt](3.2,10)--(3,9.8);
\draw[black,line width = 1pt](4.8,9.8)--(5,10);
\draw[black,line width = 1pt](5.2,9.8)--(5,10);
\draw[black,line width = 1pt](6.8,9.8)--(7,10);
\draw[black,line width = 1pt](7.2,9.8)--(7,10);
\draw[black,line width = 1pt](9,9.8)--(8.8,10);
\draw[black,line width = 1pt](9,9.8)--(9.2,10);

\draw[black,line width = 1pt](4.2,3.2)--(4,3);
\draw[black,line width = 1pt](4.2,2.8)--(4,3);
\draw[black,line width = 1pt](6.2,3.2)--(6,3);
\draw[black,line width = 1pt](6.2,2.8)--(6,3);
\draw[black,line width = 1pt](8.2,3.2)--(8,3);
\draw[black,line width = 1pt](8.2,2.8)--(8,3);
\draw[black,line width = 1pt](10.2,3.2)--(10,3);
\draw[black,line width = 1pt](10.2,2.8)--(10,3);

\draw[black,line width = 1pt](4,5.2)--(4.2,5);
\draw[black,line width = 1pt](4,4.8)--(4.2,5);
\draw[black,line width = 1pt](6.2,5.2)--(6,5);
\draw[black,line width = 1pt](6.2,4.8)--(6,5);
\draw[black,line width = 1pt](8.2,5.2)--(8,5);
\draw[black,line width = 1pt](8.2,4.8)--(8,5);
\draw[black,line width = 1pt](10,5.2)--(10.2,5);
\draw[black,line width = 1pt](10,4.8)--(10.2,5);

\draw[black,line width = 1pt](4,7.2)--(4.2,7);
\draw[black,line width = 1pt](4,6.8)--(4.2,7);
\draw[black,line width = 1pt](6,7.2)--(6.2,7);
\draw[black,line width = 1pt](6,6.8)--(6.2,7);
\draw[black,line width = 1pt](8,7.2)--(8.2,7);
\draw[black,line width = 1pt](8,6.8)--(8.2,7);
\draw[black,line width = 1pt](10,7.2)--(10.2,7);
\draw[black,line width = 1pt](10,6.8)--(10.2,7);

\draw[black,line width = 1pt](4,9.2)--(4.2,9);
\draw[black,line width = 1pt](4,8.8)--(4.2,9);
\draw[black,line width = 1pt](6,9.2)--(6.2,9);
\draw[black,line width = 1pt](6,8.8)--(6.2,9);
\draw[black,line width = 1pt](8,9)--(8.2,9.2);
\draw[black,line width = 1pt](8,9)--(8.2,8.8);
\draw[black,line width = 1pt](10.2,9.2)--(10,9);
\draw[black,line width = 1pt](10.2,8.8)--(10,9);

\end{tikzpicture}
 \end{minipage}%
\begin{minipage}{.35\textwidth}
	\centering
\begin{tikzpicture}[scale=0.4]
\filldraw[black] (2,2) circle (4pt);
\filldraw[black] (4,2) circle (4pt);
\filldraw[black] (6,2) circle (4pt);
\filldraw[black] (8,2) circle (4pt);
\filldraw[black] (10,2) circle (4pt);

\filldraw[black] (2,4) circle (4pt);
\filldraw[black] (4,4) circle (4pt);
\filldraw[black] (6,4) circle (4pt);
\filldraw[black] (8,4) circle (4pt);
\filldraw[black] (10,4) circle (4pt);

\filldraw[black] (2,6) circle (4pt);
\filldraw[black] (4,6) circle (4pt);
\filldraw[black] (6,6) circle (4pt);
\filldraw[black] (8,6) circle (4pt);
\filldraw[black] (10,6) circle (4pt);

\filldraw[black] (2,8) circle (4pt);
\filldraw[black] (4,8) circle (4pt);
\filldraw[black] (6,8) circle (4pt);
\filldraw[black] (8,8) circle (4pt);
\filldraw[black] (10,8) circle (4pt);

\filldraw[black] (2,10) circle (4pt);
\filldraw[black] (4,10) circle (4pt);
\filldraw[black] (6,10) circle (4pt);
\filldraw[black] (8,10) circle (4pt);
\filldraw[black] (10,10) circle (4pt);

\draw[black,line width = 1pt](3,1)--(3,11);
\draw[black,line width = 1pt](5,1)--(5,11);
\draw[black,line width = 1pt](7,1)--(7,11);
\draw[black,line width = 1pt](9,1)--(9,11);

\draw[black,line width = 1pt](1,3)--(11,3);
\draw[black,line width = 1pt](1,5)--(11,5);
\draw[black,line width = 1pt](1,7)--(11,7);
\draw[black,line width = 1pt](1,9)--(11,9);

\node at (1.5,2) {\tiny 2};
\node at (1.5,4) {\tiny 3};
\node at (1.5,6) {\tiny 2};
\node at (1.5,8) {\tiny 1};
\node at (1.5,10) {\tiny 2};

\node at (3.5,2) {\tiny 3};
\node at (3.5,4) {\tiny 1};
\node at (3.5,6) {\tiny 3};
\node at (3.5,8) {\tiny 2};
\node at (3.5,10) {\tiny 1};

\node at (5.5,2) {\tiny 1};
\node at (5.5,4) {\tiny 2};
\node at (5.5,6) {\tiny 1};
\node at (5.5,8) {\tiny 3};
\node at (5.5,10) {\tiny 2};

\node at (7.5,2) {\tiny 2};
\node at (7.5,4) {\tiny 3};
\node at (7.5,6) {\tiny 2};
\node at (7.5,8) {\tiny 1};
\node at (7.5,10) {\tiny 3};

\node at (9.5,2) {\tiny 3};
\node at (9.5,4) {\tiny 2};
\node at (9.5,6) {\tiny 3};
\node at (9.5,8) {\tiny 2};
\node at (9.5,10) {\tiny 1};

\draw[black,line width = 1pt](2.2,3.2)--(2,3);
\draw[black,line width = 1pt](2.2,2.8)--(2,3);

\draw[black,line width = 1pt](1.8,5.2)--(2,5);
\draw[black,line width = 1pt](1.8,4.8)--(2,5);

\draw[black,line width = 1pt](2,7.2)--(2.2,7);
\draw[black,line width = 1pt](2,6.8)--(2.2,7);

\draw[black,line width = 1pt](2.2,9.2)--(2,9);
\draw[black,line width = 1pt](2.2,8.8)--(2,9);

\draw[black,line width = 1pt] (1,3) .. controls (0.4,4) .. (1,5);
\draw[black,line width = 1pt](0.41,3.9)--(0.56,4.1);
\draw[black,line width = 1pt](0.71,3.9)--(0.56,4.1);

\draw[black,line width = 1pt] (1,7) .. controls (0.4,8) .. (1,9);
\draw[black,line width = 1pt](0.41,8.1)--(0.56,7.9);
\draw[black,line width = 1pt](0.71,8.1)--(0.56,7.9);

\draw[black,line width = 1pt](2.8,1.8)--(3,2);
\draw[black,line width = 1pt](3.2,1.8)--(3,2);
\draw[black,line width = 1pt](4.8,1.8)--(5,2);
\draw[black,line width = 1pt](5.2,1.8)--(5,2);
\draw[black,line width = 1pt](6.8,1.8)--(7,2);
\draw[black,line width = 1pt](7.2,1.8)--(7,2);
\draw[black,line width = 1pt](8.8,1.8)--(9,2);
\draw[black,line width = 1pt](9.2,1.8)--(9,2);

\draw[black,line width = 1pt](2.8,3.8)--(3,4);
\draw[black,line width = 1pt](3.2,3.8)--(3,4);
\draw[black,line width = 1pt](4.8,3.8)--(5,4);
\draw[black,line width = 1pt](5.2,3.8)--(5,4);
\draw[black,line width = 1pt](6.8,3.8)--(7,4);
\draw[black,line width = 1pt](7.2,3.8)--(7,4);
\draw[black,line width = 1pt](8.8,4)--(9,3.8);
\draw[black,line width = 1pt](9.2,4)--(9,3.8);

\draw[black,line width = 1pt](2.8,5.8)--(3,6);
\draw[black,line width = 1pt](3.2,5.8)--(3,6);
\draw[black,line width = 1pt](4.8,5.8)--(5,6);
\draw[black,line width = 1pt](5.2,5.8)--(5,6);
\draw[black,line width = 1pt](6.8,5.8)--(7,6);
\draw[black,line width = 1pt](7.2,5.8)--(7,6);
\draw[black,line width = 1pt](8.8,5.8)--(9,6);
\draw[black,line width = 1pt](9.2,5.8)--(9,6);

\draw[black,line width = 1pt](2.8,7.8)--(3,8);
\draw[black,line width = 1pt](3.2,7.8)--(3,8);
\draw[black,line width = 1pt](4.8,7.8)--(5,8);
\draw[black,line width = 1pt](5.2,7.8)--(5,8);
\draw[black,line width = 1pt](6.8,7.8)--(7,8);
\draw[black,line width = 1pt](7.2,7.8)--(7,8);
\draw[black,line width = 1pt](8.8,7.8)--(9,8);
\draw[black,line width = 1pt](9.2,7.8)--(9,8);

\draw[black,line width = 1pt](2.8,10)--(3,9.8);
\draw[black,line width = 1pt](3.2,10)--(3,9.8);
\draw[black,line width = 1pt](4.8,9.8)--(5,10);
\draw[black,line width = 1pt](5.2,9.8)--(5,10);
\draw[black,line width = 1pt](6.8,9.8)--(7,10);
\draw[black,line width = 1pt](7.2,9.8)--(7,10);
\draw[black,line width = 1pt](8.8,9.8)--(9,10);
\draw[black,line width = 1pt](9.2,9.8)--(9,10);

\draw[black,line width = 1pt](4.2,3.2)--(4,3);
\draw[black,line width = 1pt](4.2,2.8)--(4,3);
\draw[black,line width = 1pt](6.2,3.2)--(6,3);
\draw[black,line width = 1pt](6.2,2.8)--(6,3);
\draw[black,line width = 1pt](8.2,3.2)--(8,3);
\draw[black,line width = 1pt](8.2,2.8)--(8,3);
\draw[black,line width = 1pt](10,3.2)--(10.2,3);
\draw[black,line width = 1pt](10,2.8)--(10.2,3);

\draw[black,line width = 1pt](4,5.2)--(4.2,5);
\draw[black,line width = 1pt](4,4.8)--(4.2,5);
\draw[black,line width = 1pt](6,5.2)--(6.2,5);
\draw[black,line width = 1pt](6,4.8)--(6.2,5);
\draw[black,line width = 1pt](8,5.2)--(8.2,5);
\draw[black,line width = 1pt](8,4.8)--(8.2,5);
\draw[black,line width = 1pt](10.2,5.2)--(10,5);
\draw[black,line width = 1pt](10.2,4.8)--(10,5);

\draw[black,line width = 1pt](4,7.2)--(4.2,7);
\draw[black,line width = 1pt](4,6.8)--(4.2,7);
\draw[black,line width = 1pt](6,7.2)--(6.2,7);
\draw[black,line width = 1pt](6,6.8)--(6.2,7);
\draw[black,line width = 1pt](8,7.2)--(8.2,7);
\draw[black,line width = 1pt](8,6.8)--(8.2,7);
\draw[black,line width = 1pt](10,7.2)--(10.2,7);
\draw[black,line width = 1pt](10,6.8)--(10.2,7);

\draw[black,line width = 1pt](4,9.2)--(4.2,9);
\draw[black,line width = 1pt](4,8.8)--(4.2,9);
\draw[black,line width = 1pt](6,9.2)--(6.2,9);
\draw[black,line width = 1pt](6,8.8)--(6.2,9);
\draw[black,line width = 1pt](8,9.2)--(8.2,9);
\draw[black,line width = 1pt](8,8.8)--(8.2,9);
\draw[black,line width = 1pt](10,9.2)--(10.2,9);
\draw[black,line width = 1pt](10,8.8)--(10.2,9);

\end{tikzpicture}
\end{minipage}%
\caption{The partition function of the three-state Potts model is equal to that of the six-vertex model, where the six vertices all have the same Boltzmann weight. Each of the circles can take one of three colours (represented by $1$,$2$ and $3$) in any given configuration, and no two nearest neighbours can have the same colour. Each configuration of the three-state Potts model is mapped to the six vertex model. An example of this mapping with alt boundary conditions on both the left and right boundary is shown in the image on the left. Alt boundary conditions can be seen to correspond to reflecting boundary conditions in the six-vertex model. An example of the mapping for alt boundary conditions on the left boundary and free boundary conditions on the right boundary is shown in the figure on the right. With free boundary conditions there is no constraint on arrows, as illustrated on the right boundary (in the figure on the right).}\label{AFPottsSixVertex}
\end{figure}

The central charge for $Q=3$ becomes simply $c_{\rm AF}=1$, corresponding to a free boson (see, e.g., \cite{CJS01}). It is well known that the AF Potts model at this point is in fact equivalent to a compactified free boson, which is itself equivalent to the (diagonal) $\mathbb{Z}_4$ parafermionic theory. These identifications are encoded in the torus partition function
\beq\label{zpotts}
Z_{Q=3}=\frac{1}{\eta(q)\eta(\bar{q})}\sum_{\substack{e \in \frac{\mathbb{Z}}{3}\\ m\in 3\mathbb{Z}}}q^{\frac{1}{4}(\sqrt{3}e+\frac{m}{\sqrt{3}})^2}\bar{q}^{\frac{1}{4}(\sqrt{3}e-\frac{m}{\sqrt{3}})^2}
\eeq
and the identity \cite{GepnerQiu,YangZ4paraf}
\beq
Z_{Q=3}=\frac{1}{2}\sum\limits_{l,m}|c^m_l|^2.
\eeq
The expansions of the string functions $c^m_l$ worked out in \cite{YangZ4paraf} are repeated here. (Our notations for $c^m_l$ however differ from \cite{YangZ4paraf} by a factor of $1/\eta(q)$). Define the objects:

\beq
\begin{aligned}
	W=&\frac{1}{\eta(q)}\sum\limits_{k\in\mathbb{Z}}(-1)^k q^{k^2} \,, \\
	W_{\pm}=&\frac{1}{\eta(q)}\sum\limits_{k\in\mathbb{Z}}(\pm 1)^k q^{(k+\frac{1}{4})^2} \,, \\
	Y_{n}(q)=&\frac{1}{\eta(q)}\sum\limits_{k\in\mathbb{Z}}q^{3(k+\frac{n}{6})^2} \,, \mbox{ for } n=0,1,2,3 \,.
\end{aligned}
\eeq
We then have:
\beq\label{yangstringfunc}
\begin{aligned}
c^{m=0}_{l=0}= & \frac{1}{2}(Y_0+W) \,,
 & c^{m=2}_{l=0}= & \frac{1}{2}Y_3 \,, 
 & c^{m=4}_{l=0}= & \frac{1}{2}(Y_0-W) \,, \\
c^{m=1}_{l=1}= & \frac{1}{2}(W_+ +W_-) \,,
 & c^{m=3}_{l=1}= & \frac{1}{2}(W_+ -W_-) \,, \\
c^{m=0}_{l=2}= & Y_2 \,,
 & c^{m=2}_{l=2}= & Y_1 \,.
\end{aligned}
\eeq
The relationship with the free boson corresponds, on the lattice, to the fact that the critical AF three-state Potts model is equivalent to the six-vertex model \cite{ExSolvMod}. The idea is the following: pick any lattice site and choose a particular nearest neighbour. We will orient ourselves such that we are facing the nearest neighbour of interest. If this nearest neighbour has a colour with a label succeding (modulo 3) that of the site we are standing at, we will draw a left-pointing arrow between the two lattice sites.%
\footnote{Note that left and right are well-defined because we have chosen to stand facing a particular direction}
If on the contrary this nearest neighbour has a colour with a label preceeding (modulo 3) that of the site we are standing at, then we will assign a right-pointing arrow. There is then a one-to-one mapping between allowed configurations in the colouring problem and allowed configurations in the six-vertex model. See Figure~\ref{AFPottsSixVertex} for an application of this mapping.

We can see then that the alternating boundary conditions in the Potts model translate into ``reflecting'' boundary conditions in the six vertex model: arrows on the boundary are divided into pairs as shown in Figure~\ref{AFPottsSixVertex} and each pair must include one left and one right-pointing arrow, so any outgoing arrow is indeed reflected back into the system. Meanwhile, free boundary conditions for the AF Potts model correspond to free boundary conditions for the six-vertex model, meaning that arrows can freely go into or emerge from the boundary. Note then that in the context of integrable boundary conditions \cite{SklyaninBdry}, the alt/alt condition can be interpreted as a boundary condition described by a $K$-matrix equal to the identity on both the left and the right boundary. Let us consider now the implications of this correspondence. We will parameterise the six-vertex model $R$-matrix by:

\beq
R(u)=
\begin{pmatrix}
	\sinh(u+i\gamma) & 0 & 0 & 0 \\
	0 & \sinh(u) & \sinh(i\gamma) & 0 \\
	0 & \sinh(i\gamma) & \sinh(u) & 0 \\
	0 & 0 & 0 & \sinh(u+i\gamma)
\end{pmatrix}
\eeq
(which is the same as that in \cite{SklyaninBdry} when we relate the parameter $\eta$ appearing in \cite{SklyaninBdry} to the parameter $\gamma$ by $\eta=i\gamma$). The diagonal $K$-matrices then read
\beq\label{kmat}
K(u)=
\begin{pmatrix}
	\sinh(u+\xi) & 0 \\
	0 & -\sinh(u-\xi) 
\end{pmatrix} \,.
\eeq
As was shown in \cite{SklyaninBdry} the corresponding Hamiltonian is given by
\beq\label{HXXZbdry}
H=-\sum\limits_{n=1}^{N-1}(\sigma_j^x\sigma_{j+1}^x+\sigma_{j}^y\sigma_{j+1}^y-\cos\gamma\sigma_j^z\sigma_{j+1}^z)+\sinh i\gamma(\sigma_1^3\coth\xi+\sigma_{N}^3\coth\xi) \,,
\eeq
which is the XXZ Hamiltonian with some extra boundary terms. We have that for the three-state AF Potts model $\gamma=\frac{2\pi}{3}$, and for the isotropic case that we are considering $u=-\frac{i\pi}{3}$. Then setting the free parameter $\xi=\frac{i\pi}{2}$ ensures that the $K$-matrix in (\ref{kmat}) becomes proportional to the identity, hence corresponding to the reflecting boundary conditions of Figure \ref{AFPottsSixVertex}. With these values of the parameters the boundary term in (\ref{HXXZbdry}) disappears and we are left with
\beq\label{HXXZbdry2}
H=-\sum\limits_{n=1}^{N-1}(\sigma_j^x\sigma_{j+1}^x+\sigma_{j}^y\sigma_{j+1}^y+\frac{1}{2}\sigma_j^z\sigma_{j+1}^z) \,.
\eeq
The continuum limit of this Hamiltonian has been studied in \cite{SALEURBauer}; comparing this work to our results in Table \ref{d4stringfunc} allows us, as we will now show, to recover eqs.~(\ref{yangstringfunc}). In \cite{SALEURBauer} it was found that the generating function of the spectrum of the Hamiltonian (\ref{HXXZbdry}) (in the sector with spin $S_z$) in the continuum limit is given by
\beq\label{ZXXZ}
Z(S_z)=\frac{q^{gS_z^2}}{\eta(q)}=\frac{q^{\frac{S_z^2}{3}}}{\eta(q)} \,.
\eeq
The second equality comes from the definition of $g$: we have $g=1-\frac{\gamma}{\pi}$ and $\gamma=\frac{2\pi}{3}$. Consider then the three-state AF Potts model with the boundary condition $C^=$  (defined in Figure \ref{3StatePottsAlt1}) in odd sizes. From the relationship with the six-vertex model (Figure \ref{AFPottsSixVertex}) we can see that this boundary condition corresponds to the spin sectors $S_z=0,\pm 3, \pm 6, \pm 9,\ldots$ Then from (\ref{ZXXZ}) we have that the generating function of the boundary condition $C^=$ should be given by
\beq
\sum\limits_{S_z \in \mathbb{Z}}Z(3S_z)=\frac{1}{\eta(q)}\sum\limits_{S_z \in \mathbb{Z}}q^{3S_z^2} \,.
\eeq
However, by comparing this to Table \ref{d4stringfunc} we also have that the generating function arising from the boundary condition $C^=$ gives $c^{m=0}_{l=0}+c^{m=4}_{l=0}$. We can see from eqs.~(\ref{yangstringfunc}) that this is indeed consistent and we have
\beq
\frac{1}{\eta(q)}\sum\limits_{S_z \in \mathbb{Z}}q^{3S_z^2}= Y_0 = c^{m=0}_{l=0}+c^{m=4}_{l=0} \,.
\eeq
Similarly, consider the the three-state AF Potts model with the boundary condition $C^{\neq}$  (defined in Figure \ref{3StatePottsAlt2}) in odd sizes. In this case the correspondence can be seen to be with the six-vertex spin sectors: $S_z=\ldots -7,-4,-1, 2, 5, 8,... \ldots$ So according to (\ref{ZXXZ}) we get for the generating function
\beq
\frac{1}{\eta(q)}\sum\limits_{k \in \mathbb{Z}}q^{\frac{1}{3}(3k-2)^2} \,,
\eeq
which when comparing with the result in table \ref{d4stringfunc} for the generating function of this boundary condition we find:
\beq
c^{m=0}_{l=2}= Y_2 = \frac{1}{\eta(q)}\sum\limits_{k \in \mathbb{Z}}q^{\frac{1}{3}(3k-2)^2}
\eeq
which is consistent with eqs.~(\ref{yangstringfunc}). Eqs.~(\ref{yangstringfunc}) can therefore be seen as the continuum version of the equivalence between the XXZ chain with free boundary conditions and the three-state AF Potts model with alt boundary conditions.

\medskip

By considering the two remaining alt/alt boundary conditions in table \ref{d4stringfunc} (i.e. $A^=$ and $A^{\neq}$) we can derive two more identities appearing in (\ref{yangstringfunc}). They are
\beq
c^{m=2}_{l=0}= \frac12 Y_3 = \frac{1}{2\eta(q)}\sum\limits_{k \in \mathbb{Z}}q^{3(k+1/2)}
\eeq
and
\beq
c^2_2= Y_1 = \frac{1}{\eta(q)}\sum\limits_{k \in \mathbb{Z}}q^{3(k+\frac{1}{6})^2} \,.
\eeq

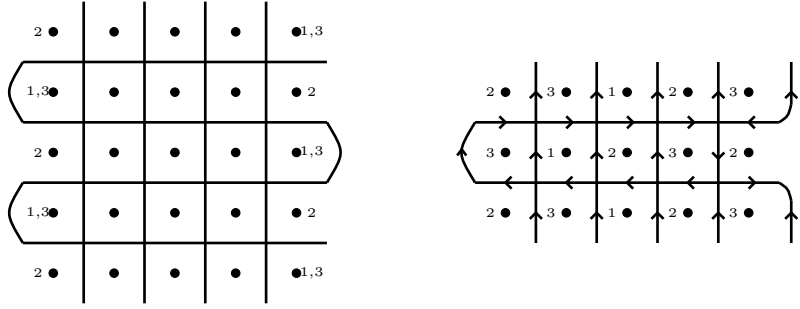
\begin{figure}
\centering
\begin{minipage}{.35\textwidth}
	\centering
\begin{tikzpicture}[scale=0.4]

\filldraw[black] (2,2) circle (4pt);
\filldraw[black] (4,2) circle (4pt);
\filldraw[black] (6,2) circle (4pt);
\filldraw[black] (8,2) circle (4pt);
\filldraw[black] (10,2) circle (4pt);

\filldraw[black] (2,4) circle (4pt);
\filldraw[black] (4,4) circle (4pt);
\filldraw[black] (6,4) circle (4pt);
\filldraw[black] (8,4) circle (4pt);
\filldraw[black] (10,4) circle (4pt);

\filldraw[black] (2,6) circle (4pt);
\filldraw[black] (4,6) circle (4pt);
\filldraw[black] (6,6) circle (4pt);
\filldraw[black] (8,6) circle (4pt);
\filldraw[black] (10,6) circle (4pt);

\filldraw[black] (2,8) circle (4pt);
\filldraw[black] (4,8) circle (4pt);
\filldraw[black] (6,8) circle (4pt);
\filldraw[black] (8,8) circle (4pt);
\filldraw[black] (10,8) circle (4pt);

\filldraw[black] (2,10) circle (4pt);
\filldraw[black] (4,10) circle (4pt);
\filldraw[black] (6,10) circle (4pt);
\filldraw[black] (8,10) circle (4pt);
\filldraw[black] (10,10) circle (4pt);

\draw[black,line width = 1pt](3,1)--(3,11);
\draw[black,line width = 1pt](5,1)--(5,11);
\draw[black,line width = 1pt](7,1)--(7,11);
\draw[black,line width = 1pt](9,1)--(9,11);

\draw[black,line width = 1pt](1,3)--(11,3);
\draw[black,line width = 1pt](1,5)--(11,5);
\draw[black,line width = 1pt](1,7)--(11,7);
\draw[black,line width = 1pt](1,9)--(11,9);

\node at (1.5,2) {\tiny 2 };
\node at (1.5,4) {\tiny 1,3 };
\node at (1.5,6) {\tiny 2 };
\node at (1.5,8) {\tiny 1,3 };
\node at (1.5,10) {\tiny 2 };

\node at (10.5,2) {\tiny 1,3 };
\node at (10.5,4) {\tiny 2 };
\node at (10.5,6) {\tiny 1,3 };
\node at (10.5,8) {\tiny 2 };
\node at (10.5,10) {\tiny 1,3 };

\draw[black,line width = 1pt] (1,3) .. controls (0.4,4) .. (1,5);

\draw[black,line width = 1pt] (1,7) .. controls (0.4,8) .. (1,9);

\draw[black,line width = 1pt] (11,7) .. controls (11.6,6) .. (11,5);

\end{tikzpicture}
 \end{minipage}%
\begin{minipage}{.35\textwidth}
	\centering
\begin{tikzpicture}[scale=0.4]
\filldraw[black] (2,2) circle (4pt);
\filldraw[black] (4,2) circle (4pt);
\filldraw[black] (6,2) circle (4pt);
\filldraw[black] (8,2) circle (4pt);
\filldraw[black] (10,2) circle (4pt);

\filldraw[black] (2,4) circle (4pt);
\filldraw[black] (4,4) circle (4pt);
\filldraw[black] (6,4) circle (4pt);
\filldraw[black] (8,4) circle (4pt);
\filldraw[black] (10,4) circle (4pt);

\filldraw[black] (2,6) circle (4pt);
\filldraw[black] (4,6) circle (4pt);
\filldraw[black] (6,6) circle (4pt);
\filldraw[black] (8,6) circle (4pt);
\filldraw[black] (10,6) circle (4pt);

\draw[black,line width = 1pt](3,1)--(3,7);
\draw[black,line width = 1pt](5,1)--(5,7);
\draw[black,line width = 1pt](7,1)--(7,7);
\draw[black,line width = 1pt](9,1)--(9,7);

\draw[black,line width = 1pt](1,3)--(11,3);
\draw[black,line width = 1pt](1,5)--(11,5);

\node at (1.5,2) {\tiny 2};
\node at (1.5,4) {\tiny 3};
\node at (1.5,6) {\tiny 2};

\node at (3.5,2) {\tiny 3};
\node at (3.5,4) {\tiny 1};
\node at (3.5,6) {\tiny 3};

\node at (5.5,2) {\tiny 1};
\node at (5.5,4) {\tiny 2};
\node at (5.5,6) {\tiny 1};

\node at (7.5,2) {\tiny 2};
\node at (7.5,4) {\tiny 3};
\node at (7.5,6) {\tiny 2};

\node at (9.5,2) {\tiny 3};
\node at (9.5,4) {\tiny 2};
\node at (9.5,6) {\tiny 3};

\draw[black,line width = 1pt](2.2,3.2)--(2,3);
\draw[black,line width = 1pt](2.2,2.8)--(2,3);

\draw[black,line width = 1pt](1.8,5.2)--(2,5);
\draw[black,line width = 1pt](1.8,4.8)--(2,5);

\draw[black,line width = 1pt] (1,3) .. controls (0.4,4) .. (1,5);
\draw[black,line width = 1pt](0.41,3.9)--(0.56,4.1);
\draw[black,line width = 1pt](0.71,3.9)--(0.56,4.1);

\draw[black,line width = 1pt] (11,3) .. controls (11.3,2.8) .. (11.4,2.4);
\draw[black,line width = 1pt] (11.4,2.4)--(11.4,1);
\draw[black,line width = 1pt](11.2,1.8)--(11.4,2);
\draw[black,line width = 1pt](11.6,1.8)--(11.4,2);

\draw[black,line width = 1pt] (11,5) .. controls (11.3,5.2) .. (11.4,5.6);
\draw[black,line width = 1pt] (11.4,5.6)--(11.4,7);
\draw[black,line width = 1pt](11.2,5.8)--(11.4,6);
\draw[black,line width = 1pt](11.6,5.8)--(11.4,6);

\draw[black,line width = 1pt](2.8,1.8)--(3,2);
\draw[black,line width = 1pt](3.2,1.8)--(3,2);
\draw[black,line width = 1pt](4.8,1.8)--(5,2);
\draw[black,line width = 1pt](5.2,1.8)--(5,2);
\draw[black,line width = 1pt](6.8,1.8)--(7,2);
\draw[black,line width = 1pt](7.2,1.8)--(7,2);
\draw[black,line width = 1pt](8.8,1.8)--(9,2);
\draw[black,line width = 1pt](9.2,1.8)--(9,2);

\draw[black,line width = 1pt](2.8,3.8)--(3,4);
\draw[black,line width = 1pt](3.2,3.8)--(3,4);
\draw[black,line width = 1pt](4.8,3.8)--(5,4);
\draw[black,line width = 1pt](5.2,3.8)--(5,4);
\draw[black,line width = 1pt](6.8,3.8)--(7,4);
\draw[black,line width = 1pt](7.2,3.8)--(7,4);
\draw[black,line width = 1pt](8.8,4)--(9,3.8);
\draw[black,line width = 1pt](9.2,4)--(9,3.8);

\draw[black,line width = 1pt](2.8,5.8)--(3,6);
\draw[black,line width = 1pt](3.2,5.8)--(3,6);
\draw[black,line width = 1pt](4.8,5.8)--(5,6);
\draw[black,line width = 1pt](5.2,5.8)--(5,6);
\draw[black,line width = 1pt](6.8,5.8)--(7,6);
\draw[black,line width = 1pt](7.2,5.8)--(7,6);
\draw[black,line width = 1pt](8.8,5.8)--(9,6);
\draw[black,line width = 1pt](9.2,5.8)--(9,6);

\draw[black,line width = 1pt](4.2,3.2)--(4,3);
\draw[black,line width = 1pt](4.2,2.8)--(4,3);
\draw[black,line width = 1pt](6.2,3.2)--(6,3);
\draw[black,line width = 1pt](6.2,2.8)--(6,3);
\draw[black,line width = 1pt](8.2,3.2)--(8,3);
\draw[black,line width = 1pt](8.2,2.8)--(8,3);
\draw[black,line width = 1pt](10,3.2)--(10.2,3);
\draw[black,line width = 1pt](10,2.8)--(10.2,3);

\draw[black,line width = 1pt](4,5.2)--(4.2,5);
\draw[black,line width = 1pt](4,4.8)--(4.2,5);
\draw[black,line width = 1pt](6,5.2)--(6.2,5);
\draw[black,line width = 1pt](6,4.8)--(6.2,5);
\draw[black,line width = 1pt](8,5.2)--(8.2,5);
\draw[black,line width = 1pt](8,4.8)--(8.2,5);
\draw[black,line width = 1pt](10.2,5.2)--(10,5);
\draw[black,line width = 1pt](10.2,4.8)--(10,5);

\end{tikzpicture}
\end{minipage}%
\caption{Left panel: Anti-Correlated alt boundary conditions correspond to reflecting boundary conditions in the six vertex model, but where the pairing of arrows on the left and right boundary are anti-correlated. Right panel: The Hamiltonian/Transfer Matrix that describes the vertex model acts on a chain with an extra spin that is decoupled from the rest of the system. This extra spin has the effect of adding $\pm$ to the total spin. }\label{ACVertex}
\end{figure}
The two ``anti-correlated'' boundary conditions $A^=$ and $A^{\neq}$ correspond to half integer spin sectors of the XXZ chain. We can see this in two ways. The first way is to recall that $A^=$ for $L$ odd has the same continuum limit as $C^{\neq}$ for $L$ even. We have seen from the mapping from 3-state Potts to the XXZ chain that $L$ even in the Potts model corresponds to an XXZ chain of \textit{odd} length, and hence necessarily with half integer spin. In particular, one can see that $A^=$ produces the generating function associated with the spin sectors $S_z=\ldots-\frac{9}{2},-\frac{3}{2},\frac{3}{2},\frac{9}{2}\ldots$, and $A^{\neq}$ to the sectors $S_z=\ldots -\frac{7}{2},-\frac{1}{2},\frac{5}{2},\frac{11}{2}\ldots$\\

We can recover the same result from a different point of view. We can see from Figure \ref{ACVertex} that the ``anti-correlated'' boundary conditions correspond to reflecting boundary conditions in the vertex model but where the pairs of arrows on the left and right boundaries are not correlated to occur on the same rows of the lattice. We see from the right panel in Figure \ref{ACVertex} that this is equivalent to a situation where the Transfer Matrix/Hamiltonian acts on a system with an extra spin that is decoupled from the system. This extra spin contributes $\pm \frac{1}{2}$ to the total spin of the system; we hence go from a system with integer spin to a system with half integer spin.

\section{Odd number of sites}\label{oddsites}

\begin{figure}
	\centering
\begin{tikzpicture}[scale=0.8]
\draw[black,line width = 1pt](1,1)--(8,8);
\draw[black,line width = 1pt](3,1)--(8,6);
\draw[black,line width = 1pt](5,1)--(8,4);
\draw[black,line width = 1pt](7,1)--(8,2);

\draw[black,line width = 1pt](1,3)--(7,9);
\draw[black,line width = 1pt](1,5)--(5,9);
\draw[black,line width = 1pt](1,7)--(3,9);

\draw[black,line width = 1pt](8,2)--(1,9);
\draw[black,line width = 1pt](7,1)--(1,7);
\draw[black,line width = 1pt](5,1)--(1,5);
\draw[black,line width = 1pt](3,1)--(1,3);

\draw[black,line width = 1pt](8,4)--(3,9);
\draw[black,line width = 1pt](8,6)--(5,9);
\draw[black,line width = 1pt](8,8)--(7,9);

\filldraw[black] (2,2) circle (4pt);
\filldraw[black] (4,2) circle (4pt);
\filldraw[black] (6,2) circle (4pt);
\filldraw[black] (8,2) circle (4pt);

\filldraw[black] (2,4) circle (4pt);
\filldraw[black] (4,4) circle (4pt);
\filldraw[black] (6,4) circle (4pt);
\filldraw[black] (8,4) circle (4pt);

\filldraw[black] (2,6) circle (4pt);
\filldraw[black] (4,6) circle (4pt);
\filldraw[black] (6,6) circle (4pt);
\filldraw[black] (8,6) circle (4pt);

\filldraw[black] (2,8) circle (4pt);
\filldraw[black] (4,8) circle (4pt);
\filldraw[black] (6,8) circle (4pt);
\filldraw[black] (8,8) circle (4pt);

\draw[black,line width = 1pt] (1.5,1.5) .. controls (1.25,2) .. (1.5,2.5);
\draw[black,line width = 1pt] (1.5,3.5) .. controls (1.25,4) .. (1.5,4.5);
\draw[black,line width = 1pt] (1.5,5.5) .. controls (1.25,6) .. (1.5,6.5);
\draw[black,line width = 1pt] (1.5,7.5) .. controls (1.25,8) .. (1.5,8.5);

\draw[black,line width = 2pt](2,2)--(2,4);
\draw[black,line width = 2pt](2,4)--(6,4);
\draw[black,line width = 2pt](4,4)--(4,2);
\draw[black,line width = 2pt](6,4)--(6,2);
\draw[black,line width = 2pt](2,6)--(4,6);
\draw[black,line width = 2pt](2,8)--(4,8);

\draw[black,line width = 1pt] (1.5,1.5) .. controls (2,1.25) .. (2.5,1.5);
\draw[black,line width = 1pt] (3.5,1.5) .. controls (4,1.25) .. (4.5,1.5);
\draw[black,line width = 1pt] (5.5,1.5) .. controls (6,1.25) .. (6.5,1.5);

\draw[black,line width = 1pt] (2.5,1.5) .. controls (2.75,2) .. (2.5,2.5);
\draw[black,line width = 1pt] (3.5,1.5) .. controls (3.25,2) .. (3.5,2.5);
\draw[black,line width = 1pt] (4.5,1.5) .. controls (4.75,2) .. (4.5,2.5);
\draw[black,line width = 1pt] (5.5,1.5) .. controls (5.25,2) .. (5.5,2.5);
\draw[black,line width = 1pt] (6.5,1.5) .. controls (6.75,2) .. (6.5,2.5);
\draw[black,line width = 1pt] (7.5,1.5) .. controls (7.25,2) .. (7.5,2.5);

\draw[black,line width = 1pt] (1.5,2.5) .. controls (1.75,3) .. (1.5,3.5);
\draw[black,line width = 1pt] (2.5,2.5) .. controls (2.25,3) .. (2.5,3.5);
\draw[black,line width = 1pt] (3.5,2.5) .. controls (3.75,3) .. (3.5,3.5);
\draw[black,line width = 1pt] (4.5,2.5) .. controls (4.25,3) .. (4.5,3.5);
\draw[black,line width = 1pt] (5.5,2.5) .. controls (5.75,3) .. (5.5,3.5);
\draw[black,line width = 1pt] (6.5,2.5) .. controls (6.25,3) .. (6.5,3.5);
\draw[black,line width = 1pt] (7.5,2.5) .. controls (7.75,3) .. (7.5,3.5);

\draw[black,line width = 1pt] (2.5,3.5) .. controls (3,3.75) .. (3.5,3.5);
\draw[black,line width = 1pt] (2.5,4.5) .. controls (3,4.25) .. (3.5,4.5);
\draw[black,line width = 1pt] (4.5,3.5) .. controls (5,3.75) .. (5.5,3.5);
\draw[black,line width = 1pt] (4.5,4.5) .. controls (5,4.25) .. (5.5,4.5);
\draw[black,line width = 1pt] (6.5,3.5) .. controls (6.75,4) .. (6.5,4.5);
\draw[black,line width = 1pt] (7.5,3.5) .. controls (7.25,4) .. (7.5,4.5);

\draw[black,line width = 1pt] (1.5,4.5) .. controls (2,4.75) .. (2.5,4.5);
\draw[black,line width = 1pt] (1.5,5.5) .. controls (2,5.25) .. (2.5,5.5);
\draw[black,line width = 1pt] (3.5,4.5) .. controls (4,4.75) .. (4.5,4.5);
\draw[black,line width = 1pt] (3.5,5.5) .. controls (4,5.25) .. (4.5,5.5);
\draw[black,line width = 1pt] (5.5,4.5) .. controls (6,4.75) .. (6.5,4.5);
\draw[black,line width = 1pt] (5.5,5.5) .. controls (6,5.25) .. (6.5,5.5);
\draw[black,line width = 1pt] (7.5,4.5) .. controls (7.75,5) .. (7.5,5.5);

\draw[black,line width = 1pt] (2.5,5.5) .. controls (3,5.75) .. (3.5,5.5);
\draw[black,line width = 1pt] (2.5,6.5) .. controls (3,6.25) .. (3.5,6.5);
\draw[black,line width = 1pt] (4.5,5.5) .. controls (4.75,6) .. (4.5,6.5);
\draw[black,line width = 1pt] (5.5,5.5) .. controls (5.25,6) .. (5.5,6.5);
\draw[black,line width = 1pt] (6.5,5.5) .. controls (7,5.75) .. (7.5,5.5);
\draw[black,line width = 1pt] (6.5,6.5) .. controls (7,6.25) .. (7.5,6.5);

\draw[black,line width = 1pt] (1.5,6.5) .. controls (2,6.75) .. (2.5,6.5);
\draw[black,line width = 1pt] (1.5,7.5) .. controls (2,7.25) .. (2.5,7.5);
\draw[black,line width = 1pt] (3.5,6.5) .. controls (4,6.75) .. (4.5,6.5);
\draw[black,line width = 1pt] (3.5,7.5) .. controls (4,7.25) .. (4.5,7.5);
\draw[black,line width = 1pt] (5.5,6.5) .. controls (6,6.75) .. (6.5,6.5);
\draw[black,line width = 1pt] (5.5,7.5) .. controls (6,7.25) .. (6.5,7.5);
\draw[black,line width = 1pt] (7.5,6.5) .. controls (7.75,7) .. (7.5,7.5);

\draw[black,line width = 1pt] (2.5,7.5) .. controls (3,7.75) .. (3.5,7.5);
\draw[black,line width = 1pt] (2.5,8.5) .. controls (3,8.25) .. (3.5,8.5);
\draw[black,line width = 1pt] (4.5,7.5) .. controls (4.75,8) .. (4.5,8.5);
\draw[black,line width = 1pt] (5.5,7.5) .. controls (5.25,8) .. (5.5,8.5);
\draw[black,line width = 1pt] (6.5,7.5) .. controls (6.75,8) .. (6.5,8.5);
\draw[black,line width = 1pt] (7.5,7.5) .. controls (7.25,8) .. (7.5,8.5);

\draw[black,line width = 1pt] (1.5,8.5) .. controls (2,8.75) .. (2.5,8.5);
\draw[black,line width = 1pt] (3.5,8.5) .. controls (4,8.75) .. (4.5,8.5);
\draw[black,line width = 1pt] (5.5,8.5) .. controls (6,8.75) .. (6.5,8.5);

\end{tikzpicture}

\caption{The loop model with and odd number of strands. There must be at least one defect line running through the system when $N$ is odd. We have $N=7$ and $j=\frac{1}{2}$ in this example. An odd number of strands in the loop model corresponds to wired boundary conditions on one boundary in the Potts model: in this figure the wired boundary conditions are imposed on the right boundary: all Potts spins on the right boundary are identified, since they cannot be separated by any loop.}\label{oddlattice}
\end{figure}
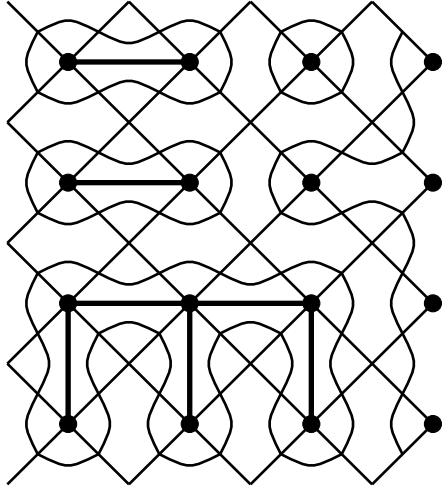

All the discussion so far was restricted to spin chains where the number of sites $N=2L$ was even. It is also possible  to consider instead the case $N$ odd, which in the  Potts model formulation corresponds to having wired boundary conditions on one boundary. See Figure \ref{oddlattice} and the discussion of wired boundary conditions in section \ref{ferrpotts}.

Introduce first $Z_{\rm ND}$,  the partition function of the free boson with Neumann/Dirichlet boundary conditions:
\begin{equation}
Z_{\rm ND}={1\over \eta(q)}\sum_{n\in\mathbb{Z}} q^{(n-{1\over 4})^2} \,.
\end{equation}
From extensive numerical studies, our first result is the generating function of levels in the loop model for half-odd integer spin $j$---the equivalent of (\ref{Kfct}): 
\begin{equation}\label{Kjodd}
\tilde{K}_j=Z_{\rm ND}\times {1\over \eta(q)}\left(q^{k\left[{1\over 4}-{(j+{1\over 2})\over k}\right]^2}-q^{k\left[{1\over 4}+{(j+{1\over 2})\over k}\right]^2}\right) \,.
\end{equation}
We can give a few examples to support this result. When we fix $k=4.2$ and analyse the first 40 eigenvalues of the loop model transfer matrix (with free boundary conditions) up to size $N=25$ we find
\beq
\tilde{K}_{j=\frac{1}{2}}=q^{\Delta-\frac{c}{24}}(1+q^{\frac{1}{2}}+q+2q^{\frac{3}{2}}+3q^2+4q^{\frac{5}{2}}+6q^3+\ldots) \,,
\eeq
which is consistent with eq.~(\ref{Kjodd}) up to the level written. Similarly, we have
\beq
\tilde{K}_{j=\frac{3}{2}}=q^{\Delta-\frac{c}{24}}(1+q^{\frac{1}{2}}+2q+3q^{\frac{3}{2}}+4q^2+6q^{\frac{5}{2}}+\ldots) \,.
\eeq

When $k$ is an integer, we can again ask about the potential relationship between the RSOS model and parafermions. To start,  recall 
the expression for characters of the C-disorder fields in $Z_{k-2}$ parafermionic CFT \cite{YangRavanini}. Setting $l=k+2r$, with $1\leq r\leq {k\over 2}$, we have
\begin{equation}
\xi_l=q^{-1/48} \prod_{n=1}^\infty \left(1-q^{n-1/2}\right)\prod_{n=1}^\infty {1+q^{n/2}\over 1-q^{n/2}}\left[\Theta_{l,2k}(\tau/2)-\Theta_{2k-l,2k}(\tau/2)\right] \,.
\end{equation}
The pre-factor can be massaged to read
\begin{equation}
q^{-1/48} \prod_{n=1}^\infty \left(1-q^{n-1/2}\right)\prod_{n=1}^\infty {1+q^{n/2}\over 1-q^{n/2}}={Z_{ \rm ND}\over \eta(q)} \,,
\end{equation}
while it is easy to check that  
\begin{equation}
\xi_l= \sum_{n=0}^\infty\left[ \tilde{K}_{j+nk}-\tilde{K}_{k-1-j+nk}\right]\label{characrsos} \,,
\end{equation}
using also $l=k-2-4j$ or $r=2j+1$. On the other hand, we know by general quantum group arguments (see above) that such an alternating sum is, when $k$ is integer,  the result for the corresponding RSOS model with heights $1$ on the left boundary and even height $r=2j+1$ (recall that $j$ is half an odd integer). These boundary conditions correspond therefore to  the character of a C-disorder field in the $Z_{k-2}$ theory with dimension
\begin{equation}
\Delta={k-4+(k-2-4j)^2\over 16k} \,.
\end{equation}
Observe that  we can write
\begin{equation}
\Delta={1\over 16}-{1\over 4k} +{(2j-{k-2\over 2})^2\over 4k} \,,
\end{equation}
that is,
\begin{equation}
\Delta={1\over 16}+{\tilde{l}(\tilde{l}+2)\over 4k} \,,
\end{equation}
where $\tilde{l}={k-2\over 2}-2j-1$. This is formally similar to the form of the conformal weights for the vertex model or spin chain and even lengths, up to the addition of  ${1\over 16}$---the dimension of the ``twist-field'' in a free-boson theory.

We compare the numerical results obtained from the lattice model with the CFT quantities defined in (\ref{characrsos}) in Table \ref{rsostabodd}.

\begin{table}
\begin{center}
\begin{tabular}{c | c | c | l}
Boundary condition & k & Exponent & Generating function\\
\hline
1,...,2 & $4$ & $0$ & $1+q^{\frac{1}{2}}+q^{\frac{3}{2}}+q^2+q^{\frac{5}{2}}+q^3+q^{\frac{7}{2}}+2q^4+2q^{\frac{5}{2}}+...$\\[1.1ex]
\hline
1,...,2 & $5$ & $\frac{1}{40}$ & $1+q^{\frac{1}{2}}+q+q^{\frac{3}{2}}+2q^2+2q^{\frac{5}{2}}+3q^3+3q^{\frac{7}{2}}+4q^4+5q^{\frac{9}{2}}+6q^5+7q^{\frac{11}{2}}+...$\\[1.1ex]
\hline
1,...,4 & $5$ & $\frac{1}{8}$ & $1+q+q^{\frac{3}{2}}+q^2+q^{\frac{5}{2}}+2q^3+2q^{\frac{7}{2}}+3q^4+3q^{\frac{9}{2}}+4q^5+4q^{\frac{11}{2}}+...$\\[1.1ex]
\hline
1,...,2 & $6$ & $\frac{1}{16}$ & $1+q^{\frac{1}{2}}+q+2q^{\frac{3}{2}}+2q^2+3q^{\frac{5}{2}}+4q^3+...$\\[1.1ex]
\hline
1,...,4 & $6$ & $\frac{1}{16}$ & $1+q^{\frac{1}{2}}+q+2q^{\frac{3}{2}}+2q^2+3q^{\frac{5}{2}}+4q^3+...$\\[1.1ex]
\hline
1,...,2 & $7$ & $\frac{3}{28}$ & $1+q^{\frac{1}{2}}+q+2q^{\frac{3}{2}}+3q^2+3q^{\frac{5}{2}}+5q^3+...$\\[1.1ex]
\hline
1,...,4 & $7$ & $\frac{1}{28}$ & $1+q^{\frac{1}{2}}+2q+2q^{\frac{3}{2}}+3q^2+4q^{\frac{5}{2}}+6q^3+...$\\[1.1ex]
\hline
1,...,6 & $7$ & $\frac{1}{4}$ & $1+q+q^{\frac{3}{2}}+2q^2+2q^{\frac{5}{2}}+3q^3+...$\\[1.1ex]
\end{tabular}
\caption{Generating functions in the RSOS model with an odd numbers of sites. The generating functions are written up to the number of terms that we have observed on the lattice.}\label{rsostabodd}
\end {center}
\end{table}

\medskip

In conclusion, we see that the correspondence, for $k$ integer, between RSOS restrictions of the critical antiferromagnetic Potts model and parafermions extends to the case of disorder operators for odd numbers of sites. It is not clear to us what this means for the $SL(2,\mathbb{R})/U(1)$ theory. In particular, it is not clear what eq.~(\ref{Kjodd}) means  from the point of view of (deformed) $W_\infty$ algebra. 

\section{Conclusion}

Our analysis of conformal boundary conditions for the AF Potts model confirms the close relationship of the CFT for this model with the $SL(2,\mathbb{R})/U(1)$ coset sigma model, since, for all types of boundary conditions we have found, the generating functions of levels are expressed in terms of discrete characters for the sigma model. Puzzles remain however. The most obvious one is that we have not been able to find conformal boundary conditions leading to a continuous spectrum. This may well have to do with the subtle (and not fully understood) difference between the critical theory for the Potts model and its ``untwisted'' staggered six-vertex model version. Nonetheless, our results confirm that the bulk theory possesses a spectrum with a continuous component: this is a simple consequence, for instance, of the observation of discrete characters, together with the general formulas for their modular transforms \cite{Israel}. 

 We should further highlight that the geometry of the lattice on which we considered our model, while natural in the formulation of the Potts model, does not allow us to write an integrable transfer matrix that commutes for all values of the spectral parameter $u$. (Except, that is, for the special case of the three-state Potts model discussed in section \ref{3statepotts} where the six-vertex model to which the Potts model was mapped does indeed have the required geometry to be integrable). We therefore cannot directly use the tools of the Bethe Ansatz to study the model in this geometry. One would hope, however, that changing the geometry to one in which we can take advantage of integrability, will not change the universality class of the model. Doing so should allow us to further explore the conformally invariant boundary conditions and their continuum limits: this will be explored in our next paper. A related aspect is that all of our analysis was related to isotropic transfer matrices and not Hamiltonians; what the `alt' boundary conditions correspond to in the anisotropic limit is not so clear.

Another troubling aspect is that our results do not have much overlap with those  in \cite{RS}. In this paper, the authors 
have identified a single set of boundary conditions leading to a discrete spectrum, the case of strings stretching between D0-branes. The corresponding partition functions are written in this reference as
\begin{equation}
Z_{mm'}^{D0}=\sum_{2J+1=\rm max(m,m')}^{m+m'-1}\sum_{l\in \mathbb{Z}}\left[\chi^d_{(J,l-J)}-\chi^d_{(-J-1,l+J+1)}\right] \,.
\end{equation}
This corresponds, in our notations, to
\begin{equation}
Z_{mm'}^{D0}=\sum_{2J+1=\rm max(m,m')}^{m+m'-1}\sum_{l\in \mathbb{Z}}\left[\lambda^d_{-J,l}-\lambda^d_{J+1,l}\right] \,.
\end{equation}
For $m=m'=1$ for instance we have 
\begin{equation}
Z_{11}^{D0}=\sum_{l\in \mathbb{Z}}\left[\lambda^d_{0,l}-\lambda^d_{1,l}\right]\label{ZD0} \,.
\end{equation}
A remarkable aspect of this expression is that it expands at small $q$ with the central charge $c_{\rm BH}$
\begin{equation}
Z_{11}^{D0}=q^{-c_{\rm BH}/24}\left(1+2q^{1+{1\over k}}+q^2+\ldots\right) \,,
\end{equation}
Since  none of the  boundary conditions (at least, in the context of the Potts model) we have discussed  exhibit the true central charge of the black-hole sigma model,  the D0-branes of \cite{RS} remain unidentified in lattice terms  %
\footnote{Note that while the states in the representation with spin $J=0$ are not normalisable, the corresponding character can still appear in the partition functions in \cite{RS}.}.

On the other hand,  we have found boundary conditions not discussed in \cite{RS}, for which the generating functions expand in terms of discrete characters of $SL(2,\mathbb{R})/U(1)$, and thus which should respect all the symmetries of the theory. What this means with respect to the analysis in \cite{RS} is not clear to us at this point. We note however that the results in \cite{RS} are not supposed to be a complete classification: see \cite{AddtlBdr},\cite{AddtlBdr1} for more on this. 

In conclusion, we finally note  that putting together the fixed and the alt boundary conditions for the RSOS version of the AF Potts model for rational values of $k$ gives rise to a set of generating functions that seem to correspond formally to  ``rational parafermions'' string functions in a $SU(2)_{k=P/Q}/U(1)$ theory. This aspect will be discussed further in an upcoming paper. 

\subsection*{Acknowledgments}

This work was supported by the ERC Advanced Grant NuQFT. We thank M. Pawelkiewicz, S. Ribault, J. Troost and V. Schomerus for many useful discussions, and especially S. Ribault for his careful reading of the manuscript, and for pointing out the references \cite{AddtlBdr},\cite{AddtlBdr1}.

\appendix
\section{Coulomb gas}\label{coulombgas}

We have no clear understanding of the relationship between the $SU(2)$ parafermions and the $SL(2,\mathbb{R})/U(1)$ coset CFT, apart from the fact that this seems to involve discrete representations. It is interesting however to observe that both theories can be described quite naturally in terms of a ``Coulomb gas'', which should give one additional intuition on the continuum limit of our lattice model. 

We follow here the work of  \cite{Jayaraman}.%
\footnote{Note that we use $k$ instead of $k+2$ in that reference.} 
The starting point is  a pair of bosonic fields 
   $\phi_{1}$ and $\phi_{2}$, with 
  propagators
  \begin{eqnarray}
      \langle\phi_{1}(z)\phi_{1}(w)\rangle&=&-2\ln(z-w)\nonumber\\
      \langle\phi_{2}(z)\phi_{2}(w)\rangle&=&-2\ln(z-w)\nonumber\\
  \end{eqnarray}
  and a  stress tensor 
  \begin{equation}
      T=-{1\over 4}\left(\partial\phi_{1}\right)^{2}+{1\over 
      4}\left(\partial\phi_{2}\right)^{2}+i\alpha_{0}\partial^{2}\phi_{1} \,.
  \end{equation}
  With a charge at infinity 
  $\alpha_{0}= {1\over 2\sqrt{k}}$ for the first boson 
  $\phi_{1}$,
  the central charge is the parafermionic central charge $c=c_{\rm  {PF}}=2-{6\over k}$.  There are various ways to introduce screening operators in this theory. 
  The choice made in \cite{Jayaraman}, uses both 
  bosons $\phi_{1},\phi_{2}$, and leads to the three currents:
  \begin{equation}
      J_{1}=\partial\phi_{2}\exp\left[2i\alpha_{0}\phi_{1}\right]
  \end{equation}
  together with
  \begin{equation}
      J_{\pm}=\exp\left[-{i\over 2}\sqrt{k}\:\phi_{1}\pm{1\over 
      2}\sqrt{k-2}\:\phi_{2}\right] \,.
  \end{equation}
 Thsese screening operators, having conformal dimension one, commute with 
 the Virasoro algebra. It turns out that they also commute with the 
 ``parafermionic'' currents
 \begin{eqnarray}
     \Psi &=& -{i\over 2}\left(\sqrt{k\over 
     k-2}\:\partial\phi_{1}+i\partial\phi_{2}\right)
     \exp\left[{1\over\sqrt{k-2}}\phi_{2}\right] \,, \nonumber\\
     \Psi^{\dagger} &=& -{i\over 2}\left(\sqrt{k\over 
	 k-2}\:\partial\phi_{1}-i\partial\phi_{2}\right)
	 \exp\left[-{1\over\sqrt{k-2}}\phi_{2}\right]
	 \end{eqnarray}
 of conformal dimension ${k-1\over k}$ in the theory with $c_{\rm PF}$. The 
 vertex operators 
 \begin{equation}
     V_{lm}=\exp\left[-i{l\over 2\sqrt{k+2}}\:\phi_{1}+{m\over 
     2\sqrt{k}}\:\phi_{2}\right]
 \end{equation}
with conformal weight 
\begin{equation}
\Delta_l^m={l(l+2)\over 4k}-{m^2\over 4(k-2)}
\end{equation}
then play a special role:
 the action of $J_{\pm}$ on $V_{lm}$ is only well defined for $l,m$ 
 integer and $l\pm m$ even.  $V_{lm}$ is annihilated by the 
 corresponding charges  $Q_{\pm}$ iff $-l\leq m\leq l$.

 We start with  the Fock space $F_{lm}$ with 
\begin{equation}
\hbox{Tr}_{F_{lm}}q^{L_0-c/24}={1\over \eta(q)^2}q^{{(l+1)^2\over 4k}-{m^2\over 4(k-2)}} \,. \label{basicF}
\end{equation}
 We haven't found a realization of this Fock space in the lattice model. However, within  $F_{lm}$ we have  the trace in $\hbox{Ker}\,Q_\pm$:
\begin{eqnarray}
\hbox{Tr}_{F_{lm}\cap\hbox{\tiny{Ker}}\,Q_+}q^{L_0-c/24}={1\over \eta(q)^2}\sum_{n=0}^\infty (-1)^n q^{k\left({n\over 2}+{l+1\over 2k}\right)^2-(k-2)\left({n\over 2}+{m\over 2(k-2)}\right)^2} \nonumber\\
={1\over \eta(q)^2}q^{{(l+1)^2\over 4k}-{m^2\over 4(k-2)}} \sum_{n=0}^\infty (-1)^n q^{{1\over 2}n(n+l+1-m)} \,,
\end{eqnarray}
and 
\begin{eqnarray}
\hbox{Tr}_{F_{lm}\cap \hbox{\tiny{Ker}}\,Q_-}q^{L_0-c/24}={1\over \eta(q)^2}\sum_{n=0}^\infty (-1)^n q^{k\left({n\over 2}+{l+1\over 2k}\right)^2-(k-2)\left({n\over 2}-{m\over 2(k-2)}\right)^2} \nonumber\\
={1\over \eta(q)^2}q^{{(l+1)^2\over 4k}-{m^2\over 4(k-2)}} \sum_{n=0}^\infty (-1)^n q^{{1\over 2}n(n+l+1+m)} \,.
\end{eqnarray}
(note that the exchange of $Q_\pm$ corresponds to switching the sign of $m$.) The discrete characters of the $SL(2,\mathbb{R})/U(1)$ theory are obtained as 
\begin{equation}
\lambda^d_{J={m+1\over 2},M={l-m\over 2}}=\hbox{Tr}_{F_{lm}\cap\hbox{\tiny{Ker}}\,Q_+}q^{L_0-c/24} \,.
\end{equation}
so we see that with alt boundary conditions, the lattice model reproduces the content of $F_{lm}\cap\hbox{{Ker}}\,Q_+$. 

Moreover, we can now  take the trace in $F_{lm}$  over states in $\hbox{Ker}\,Q_+\cap \hbox{Ker}\,Q_-$ and $F_{lm}/
\hbox{Ker}\,Q_+\cap \hbox{Ker}\,Q_-$, corresponding to the space 
 $\tilde{F}_{lm}$ in \cite{Jayaraman}. We have 
\begin{eqnarray}
\hbox{Tr }_{\!\!\tilde{F}_{lm}}q^{L_0-c/24}={1\over \eta(q)^2}\left[\sum_{n=0}^\infty (-1)^n q^{k\left({n\over 2}+{l+1\over 2k}\right)^2-(k-2)\left({n\over 2}+{m\over 2(k-2)}\right)^2} \right.\nonumber\\
\left.+\sum_{n=1}^\infty (-1)^n q^{k\left({n\over 2}+{l+1\over 2k}\right)^2-(k-2)\left({n\over 2}-{m\over 2(k-2)}\right)^2}\right] \,.
\end{eqnarray}
or
\begin{eqnarray}
\hbox{Tr }_{\!\!\tilde{F}_{lm}}q^{L_0-c/24}={q^{{(l+1)^2\over 4k}-{m^2\over 4(k-2)}}\over \eta(q)^2}\left[\sum_{n=0}^\infty (-1)^n q^{{n^2\over 2}+{n(l+1-m)\over 2}} +\sum_{n=1}^\infty (-1)^n q^{{n^2\over 2}+ {n(l+1+m)\over 2}}\right] \,.
\end{eqnarray}
We observe now the identity
\begin{equation}
K_l=\hbox{Tr}_{\tilde{F}_{l0}}q^{L_0-c/24}={q^{(l+1)^2/4k}\over \eta(q)^2}\left[1+2\sum_{n=1}^\infty (-1)^n q^{n(n+l+1)/2}\right] \,.
\end{equation}
and more generally, 
\begin{equation}
\hbox{Tr}_{\BW^{b}_{j}/\BW^{u}_{j+r}}q^{L_0-c/24}=\hbox{Tr }_{\!\!\tilde{F}_{lm}}q^{L_0-c/24}
\end{equation}
with the usual correspondence (\ref{identif}). Hence the ``tops'' of the blob algebra modules in the lattice model with alt boundary conditions reproduce the content of $\tilde{F}_{lm}$. Moreover the Temperley-Lieb modules with free boundary conditions reproduce the content of $\tilde{F}_{l0}$. 

 When $k$ is rational, we can consider also  the action of 
 powers of $Q_{1}$, it is well defined only in the case 
 $Q_{1}^{l+1(\;{\rm mod}\;Q)}$ 
 acting on $F_{lm}$ where   we parametrized  $k=P/Q$. Then $Q_{1}^{l+1}V_{lm}=0$, and it is possible to focus\footnote{One can get 
 outside the range $-l\leq m\leq l$ by acting with the parafermionic 
 fields, which are also annihilated by $Q_{\pm}$ and $Q_{1}$.} on the cohomology 
 of $Q_1$ as in \cite{Jayaraman} (see in particular Table 1 in that reference). 
 
 In the particular case when  $k$ is an integer, the string functions can be obtained as
\begin{equation}
c_l^m=\sum_{n=0}^\infty \hbox{Tr}_{\tilde{F}_{l+2nk,m}}q^{L_0-c/24}-\hbox{Tr}_{\tilde{F}_{2k-l-2+2nk,m}}q^{L_0-c/24}\label{stringJaya}
\end{equation}
The corresponding alternating sum coincides with the sum on the lattice to extract the simple tops of the blob algebra modules, suggesting in particular that $Q_1$ is represented on the lattice by the $U_qsl(2)$ raising (or lowering) operator.

\end{document}